\documentclass[twocolumn]{aastex631}

\usepackage{xspace}

\newcommand{\VCAR}{V658\,Car\xspace}

\newcommand{\Ha}{$\rm H_{\alpha}$\xspace}

\usepackage{ulem}

\usepackage{ragged2e}
\usepackage{hyperref}
\usepackage{multirow}
\usepackage{comment}
\usepackage{amsmath}

\begin{document}

\title{X-Raying a Be star disk: fundamental parameters of the eclipsing binary Be star V658\,Car.}

\author[0000-0001-5563-6629]{Tajan H. de Amorim}
\affiliation{University of S\~ao Paulo
Institute of Astronomy, Geophysics and Atmospherical Sciences 05508-900 S\~ao Paulo, SP, Brazil}
\affiliation{Department of Physics and Astronomy, University of Toledo, Toledo, OH 43606, USA}
\email{tajan.amorim@usp.br}

\author[0000-0002-9369-574X]{Alex C. Carciofi}
\affiliation{University of S\~ao Paulo
Institute of Astronomy, Geophysics and Atmospherical Sciences 05508-900 S\~ao Paulo, SP, Brazil}

\author[0000-0002-2919-6786]{Jonathan Labadie-Bartz}
\affiliation{LESIA, Paris Observatory, PSL University, CNRS, Sorbonne University, Université Paris Cité, F-92195 Meudon, France}
\affiliation{DTU Space, Technical University of Denmark, Elektrovej 327, Kgs., Lyngby 2800, Denmark}

\author[0000-0002-6073-3906]{Ariane C. Silva}
\affiliation{University of S\~ao Paulo
Institute of Astronomy, Geophysics and Atmospherical Sciences 05508-900 S\~ao Paulo, SP, Brazil}

\author[0000-0002-0284-0578]{Felipe Navarete}
\affiliation{National Astrophysics Laboratory (LNA/MCTI), 37504-364 Itajubá, MG, Brazil}

\author[0000-0003-3182-5569]{Saul A. Rappaport}
\affiliation{Department of Physics, Kavli Institute for Astrophysics and Space Research, M.I.T., Cambridge, MA 02139, USA}

\author[0009-0002-2926-0036]{Pamela Querido}
\affiliation{University of S\~ao Paulo
Institute of Astronomy, Geophysics and Atmospherical Sciences 05508-900 S\~ao Paulo, SP, Brazil}

\author[0000-0002-2490-1562]{Amanda C. Rubio}
\affiliation{University of S\~ao Paulo
Institute of Astronomy, Geophysics and Atmospherical Sciences 05508-900 S\~ao Paulo, SP, Brazil}
\affiliation{Max-Planck-Institut für Astrophysik, Karl-Schwarzschild-Str. 1, 85748 Garching b.\ M\"unchen, Germany}

\author[0009-0004-8104-8619]{Jon Bjorkman}
\affiliation{Department of Physics and Astronomy, University of Toledo, Toledo, OH 43606, USA}

\author{Robert Gagliano}
\affiliation{Amateur Astronomer, Glendale, AZ 85308, USA}

\author{Ivan Terentev}
\affiliation{Citizen Scientist, c/o Zooniverse, Department of Physics, University of Oxford, Denys Wilkinson Building, Keble Road, Oxford OX1 3RH, UK}







\begin{abstract}

With its two stellar eclipses and two disk attenuations per binary orbit, {V658~Carinae} stands out as the first clear eclipsing Be + sdOB system. This rare alignment offers a unique opportunity to probe the structure and dynamics of a Be star disk with unprecedented detail. In this study, we present the most comprehensive observational dataset and modeling effort for {this system} to date, including {optical, near-infrared, and ultraviolet spectroscopy}, space-based photometry, and optical polarization. Using a new ray-tracing code, we apply a three-component model, consisting of an oblate, rapidly rotating Be star, a symmetric circumstellar disk, and a compact stripped companion, to reproduce the system's light curve, polarization, and spectral features. Our analysis yields precise constraints on the stellar and disk parameters, determining its status as the second known late-type Be + stripped star, and also provides strong spectroscopic evidence for a tenuous circumsecondary envelope. Despite the model’s overall success, several key observables, such as the \Ha equivalent width and the secondary attenuation, remain poorly reproduced, pointing to the need for more sophisticated {modelling}. In particular, future improvements should incorporate the companion's radiative feedback on the disk and account for asymmetric disk structures {expected by the gravitational interaction with the companion}.
Owing to its unique geometry and rich diagnostics, \VCAR establishes itself as a {benchmark system for Be stars (and rapid rotators in general), stripped stars, post-RLOF massive binaries, and circumstellar disk structures.} 



\end{abstract}


\section{Introduction} \label{sec:intro}

The only direct means of determining the {mass of a star} is by observing its gravitational effects {on other bodies}. {Aside from rare observational techniques that exploit phenomena such as gravitational lensing}, the most reliable method involves binary systems. By measuring {the orbital motions of the two stars}, particularly through radial velocity curves and astrometry, we can derive accurate stellar masses \citep{Andersen1991Binaries, Halm1911binaries}. 
However, not all binary systems offer the same level of diagnostic power. A key distinction is made between single-lined (SB1) and double-lined (SB2) spectroscopic binaries. In SB2 systems, both stars exhibit distinct absorption lines in their spectra, enabling us to trace the radial velocity of each component and directly measure the mass ratio. In contrast, SB1 systems display spectral lines from only one star, typically because the companion contributes {little in the observed wavelength rage}. This complicates the analysis, as the mass function derived from the single velocity curve depends on the sum of the masses. To overcome this limitation in SB1 systems, additional observational techniques are employed. {For instance}, interferometry can spatially resolve the binary components {(e.g., \citealt{klement2024interfsdo})}, while observations in specific wavelength {ranges} can sometimes {identify} the companion's spectral features, e.g., in the ultraviolet (UV) if it is much hotter than the primary, or infrared if it is significantly cooler.

Once precise masses were determined for a statistically meaningful sample of binary systems, empirical relations between mass, luminosity, and temperature could be established, such as the Hertzsprung–Russell Diagram. These relationships became the foundation for stellar evolutionary models and made it possible to estimate the masses of single stars through {spectroscopy alone}. Large-scale surveys of stellar populations have revealed a strong correlation between binarity and stellar mass (e.g., \citealt{offner2023binaries}). Among massive stars, binarity is the rule rather than the exception, {with over 50\% of O stars exchanging mass during the star lifespan} \citep{sana2012binaries, Sana2013}, while B-type stars appear as isolated objects in fewer than 20\% of cases \citep{offner2023binaries}. Recent studies {(e.g., \citealt{offner2023binaries})} have also identified an increasing number of triple and higher-order multiple systems, highlighting the complexity of stellar formation processes and the need to consider multiplicity in models of stellar evolution {\citep[e.g.,][]{navarete2025quadruple}}.

{Eclipsing binaries are particularly useful to refine binary evolution models, as their specific orientation} not only allows for precise mass determinations via orbital motion, but also enables highly accurate measurements of stellar radii. Beyond their role in fundamental stellar parameter estimation, eclipsing binaries offer a unique and largely unexplored opportunity: the light from the companion star can serve as a probe of circumstellar structures, when present.
{As the light passes through the surrounding circumstellar material, it {carries} information about the structure, composition, and dynamics of the environment, analogous to how X-ray imaging reveals internal structures in medical diagnostics.} 
To date, only a few dozen systems with circumstellar disks that are in eclipsing binaries have been identified. 
Algol \citep[$\beta$\,Per][]{Goodricke1783BetPer}, V4142\,Sgr \citep{rosales2023V4142Sgr}, RS~Sgr \citep{Bakis2025RSsgr}, V367\,Cygni \citep{Davidge2022AV367Cygni}, {MWC~882 \citep{Zhou2018Sauleclipse}} and VV Cephei \citep{Pollmann2018VVcep} are a few examples of systems classified as being in an Algol-like phase, where {one} star is likely accreting matter from an evolved star. There are also examples of B[e] \citep[GG Carinae,][]{Pereyra2009GGCarinae}, pre-main sequence stars \citep[V928 Tau, ][]{vanDam2020Ttauri} and {Be/X-ray binaries \citep{Gaudin2024Bexray}}. 

{V658~Carinae (\VCAR)} is the first known example of a new class of {eclipsing binaries with disks}. Despite its relatively high brightness, \VCAR has been surprisingly little studied {in the literature}. {Based on 17 radial velocity measurements from \citet{Gieseking1981v658Car} and a multi-band light curve, the system} was described by \citet{hauck2018eclipsing} as an eclipsing SB2 with a 32~days orbital period consisting of an A0 primary with a circumstellar disk and a less massive B-type companion star where both stars have similar radii, $R_1 = 1.64 (4) \rm \, R_{\odot}$ and $R_2 = 1.46 (4) \, \rm R_{\odot}$, and temperatures, $T_1 = 9.7 (5) \, \rm kK$ and $T_2 = 12.7 (7) \, \rm kK${, respectively}. In this paper, using a more comprehensive dataset and a new analysis, we present a picture of \VCAR that significantly differs from this previous interpretation. We argue that \VCAR is, in fact, a Be + sdOB (subdwarf\footnote{In this paper, we define a subdwarf as a hot B/O star with a radius much larger than a white dwarf and up to a solar radius.} of type O or B) . 

Classical Be/Oe stars are {rapidly}-rotating main sequence stars \citep{rivinius2013classical}. The hallmark of these stars is the strong double-peaked hydrogen emission lines (hence their name, Be/Oe = B/O stars with Balmer emission lines) formed in gaseous circumstellar disks. These are built up from the mechanical ejection of mass (and angular momentum) off the stellar surface, a process that is greatly assisted by their {rapid} rotation. The exact mass loss mechanism is still under scrutiny, with the most prominent explanations suggesting a link with pulsation \citep{Baade1982pulsation}, which seems ubiquitous in Be stars \citep[][and references therein]{rivi2003NRP, labadie-bartz2022classifying}.

The {origin} of Be stars is likely driven by processes that also give rise to their rapid rotation. 
{One hypothesis suggests that a massive star may emerge onto the main sequence (MS) with a relatively high rotation rate, and that during its evolution some mechanism(s) transport angular momentum (AM) from the contracting core outwards, thus spinning up the outermost layers \citep{martayan2006effects,georgy2013grids}. Without some AM transport mechanism (i.e., if AM is locally conserved), the expanding envelope would spin down with time.}
This evolutionary scenario has gained support from, e.g., cluster studies that suggest that the fraction of Be stars relative to normal B stars increases with cluster age \citep[e.g.,][]{mcswain2005}. A second hypothesis suggests acceleration by binary interaction \citep{kriz1975hypothesis, pols1991formation, Rappaport1982xray, Dallas2022binaries, Dodd2024Gaia}. Here, {two stars} with different masses are in a binary system. 
When {the more} massive star evolves past the main sequence and expands, Roche lobe overflow occurs, transferring mass and angular momentum to its companion. When the mass transfer ceases, the systems are first seen as a fast rotator + inflated star \citep{rivi2025bloated, Gabitova2025}, until the companion has enough time to shrink and become a helium subdwarf \citep{Staritsin2024binaries}. 

{Be stars may acquire their fast rotation through either of these two scenarios or a combination of both.} To understand which one is the most prominent, detailed studies of binary incidence {in B and Be stars} are required. However, these studies face a major caveat: the surviving remnant of the mass donor is often very small and difficult to detect, making the binary properties of the rapidly rotating OB population poorly constrained \citep{wang2021detection}.
With the exception of the numerous Be/X-ray binaries, most of the Be + stripped star (typically a hot sdOB) binaries end up being classified as such not by direct observations but by secondary effects, such as: 
1) Low-amplitude radial velocity variations {of the Be star in SB1 systems} \citep[e.g.,][]{Miroshnichenko2023SB1}; %
2) The phenomenon of the spectral energy distribution (SED) turndown \citep{Klement2019sedturndown}, where the presence of the companion limits the growth of the disk \citep{panoglou2016SPH}; 
3) Astrometric signatures {from interferometric observations} \citep[e.g.,][]{Klement2022dynamical}, which are observationally challenging due to the fact that modern interferometers usually operate in the IR, where the companion, being hotter than the Be star, is much dimmer.

Numerous recent studies have explored the effects of a binary companion on the disk, revealing various impacts such as truncation, density waves \citep{okazaki2002,panoglou2016SPH}, tilting \citep{cyr2017}, warping, and even tearing \citep{suffak2002}. {These effects have observational consequences}, for example, \citet{Zharikov2013} tracked V/R variations\footnote{{In a double-peaked profile, V/R is defined as the ratio of the Violet (blue shifted) to Red (red shifted) peaks}} locked with the orbital phase, associated with binary-induced spiral density waves, in the disk of $\pi$~Aqr. {Disk truncation has been linked to the SED turndown of Be stars in radio frequencies} \citep{Klement2019sedturndown}. Disk tilting or precession is suggested as the cause of the transitions between Be and shell\footnote{{Shell stars are defined by the presence of narrow and deep absorption lines superimposed on the standard broad and shallow stellar absorption profile. This phenomenon is directly caused by a circumstellar disk obscuring a rapidly rotating star viewed edge-on}} phases seen in Pleione \citep{marr2022}, $\gamma$~Cas, and 59~Cyg \citep{baade2023gammaCas}. Additionally, disk tearing is believed to occur periodically in Pleione's disk \citep{hirata2007,marr2022,suffak2024,martin2022}. 

{The presence of eclipses in \VCAR offers a rare opportunity to probe the circumstellar disk signatures arising from binary interaction, as well as the fundamental properties of both stars,  with unprecedented clarity. }
{The paper is structured as follows. Section~\ref{sec:Data} describes the observational and data reduction procedures used to construct a consistent data set for \VCAR. Section~\ref{sec:NoModel} presents an observational overview of the system, laying the groundwork for the modeling efforts discussed in Sects.~\ref{sec:Modeling} and \ref{sec:mod_sol}. Finally, Sect.~\ref{sec:final_discussion} outlines our main conclusions and offers a critical discussion of the model's strengths and limitations.}

\section{Observational Data} \label{sec:Data}


We secured a large dataset for \VCAR that includes ultraviolet, optical, and near-infrared (NIR) spectroscopy, optical polarization, as well as space-based optical photometry. A significant portion of the data is original.


\subsection{{Differential Photometry}}
TESS \citep[Transiting Exoplanet Survey Satellite,][]{ricker2015TESS} has acquired single-band photometric data of \VCAR, in the range {$\rm 600 $} -- {$\rm 1000 \, nm$}, with high cadence (2 min), in 5 {observing sectors\footnote{An observing sector of TESS corresponds to a $24^{\circ}$ x $96^{\circ}$ area of the sky observed for about 27 days.}}: sector 10 in 2019, sectors 36 and 37 in 2021, and sectors 63 and 64 in 2023. The typical signal-to-noise ratio (SNR) of the data is 1300. Data from the five sectors are plotted in Fig.~\ref{fig:masterplot} (panel A), phase-folded to the orbital period of 32.1847 days (see Sect.~\ref{sec:orb_sol}). The lightcurve displays the two stellar eclipses at phases 0 and 0.5 and the two broad attenuations bracketing the eclipses, as reported by \citet{hauck2018eclipsing}. Furthermore, the broad attenuations exhibit a remarkable level of repeatability, suggesting that whatever the cause, it remains consistently stable {throughout many orbital periods}.

\subsection{{Differential Spectroscopy}}

A total of 92 spectra in the optical range (380 -- 860\,nm), with  resolving power of {$R\sim 53000$}, were obtained with the 1-m telescopes of the Las Cumbres Observatory using the NRES instrument \citep[Network of Robotic Echelle Spectrographs,][]{siverd2018nres}. The employed strategy was to obtain 15-min (SNR $\sim 35$) exposures during the stellar eclipses and 30-min exposure ($\rm SNR \sim 50$) otherwise. 
We have also analyzed an older \Ha observation ($\rm SNR \sim 18$) from Bernard Heathcote, available in the BESS database \citep{neiner2011bess}.

\subsection{{Absolute Spectroscopy}}

The International Ultraviolet Explorer (IUE) has observed \VCAR two times with {$R \sim 300$} and {$\rm SNR \sim 10$}, one using the Short-Wavelength Prime spectrograph (SWP, in the 115 -- 198\,nm range)  and another with the Long-Wavelength Prime spectrograph (LWP, in the 185 -- 335\,nm range). The observations were conducted in the absence of any eclipse or attenuation, at phase 0.253.

Gaia Data Release 3 \citep[GDR3,][]{gaiamission, gaiddr3} has acquired optical (336 -- 1020\,nm) spectra of \VCAR 1670 times in the BP (blue) band and 2609 times in the RP (red) band, defined in \citet{gaia2021rpbp}.
These observations were taken at 48 different epochs, but only their time average values ({$R\sim 340$}, {$\rm SNR \sim 60$}) are available. Unfortunately, the specific epochs at which the spectra were obtained are unknown. As a result, the GDR3 spectrum is influenced by eclipses and attenuations in a manner that cannot be precisely quantified. Despite these limitations, the GDR3 spectrum remains a valuable source of information about the star. 

A total of 21 NIR spectra (945 -- 2465\,nm), with mid resolving power ({$R\sim 3500$}, {$\rm SNR \sim 300$}), were obtained with the TripleSpec {cross-dispersed} spectrograph \citep{Schlawin2014triplespec} attached to the 4.1\,m Southern Astrophysical Research (SOAR) telescope. {We adopted an ABBA dither pattern with an exposure time of 60~sec per frame, where A and B correspond to on-source positions spaced by {15\arcsec}, respectively. The SOAR data was processed using a modified version of the \texttt{Spextool} pipeline \citet{Cushing2004triplespec}. The data reduction steps include flat-field correction, wavelength calibration using a CuHeAr arc lamp, removal of emission sky features by subtracting A and B exposures, and extraction of the one-dimensional spectra. We used observations of the telluric standard star HIP\,54830 observed at similar airmass values to perform the telluric correction and flux calibration of the spectrum.}
%

\begin{table*}[t]
    \centering
    \caption{Summary of all observations. $N_{\rm obs}$ represents the number of observations, and $\overline{\mathrm{SNR}}$  denotes the mean SNR.}
    \begin{tabular}{cccccc} \hline
        Source & R & Observation period  & $N_{\rm obs}$ &$\overline{\mathrm{SNR}}$
        & Range  \\ \hline
        TESS & -- & Mar/2019 -- Apr/2023 & 5 Sectors & 1300 & 6000 -- 10000 \AA \\
        NRES & 53000 & Nov/2021 -- Jan/2024 & 92 & 35 -- 50 & 3800 -- 8600 \AA \\
        IUE(SWP) & 300 & Dec/1994 & 1 & 10 & 1150 -- 1980 \AA \\ 
        IUE(LWP) & 300 & Dec/1994 & 1 & 10 & 1850 -- 3350 \AA \\ 
        GDR3 & 340 & Jul/2014 -- May/2017 & 4279 & 60 & 3360 -- 10200 \AA \\ 
        SOAR & 3500 & May/2022 -- Jun/2025 & 21 & 300 & 9450 -- 24650 \AA \\
        OPD - B & -- & Dec/2022 -- Dec/2023 & 16 & 24 & 3600 -- 5600 \AA \\ 
        OPD - V & -- & Mar/2022 -- Dec/2023 & 96 & 25 & 4700 -- 7000 \AA \\ 
        OPD - R & -- & Mar/2022 -- Dec/2023 & 90 & 25 & 5500 -- 9000 \AA \\ 
        OPD - I & -- & Mar/2023 -- Dec/2023 & 16 & 25 & 7000 -- 9200 \AA \\ \hline \hline 
    \end{tabular}
    \label{tab:Spec_data}
\end{table*}

\subsection{{Polarization}}
Finally, multi-epoch broad-band linear polarization has been acquired in the $BVRI$ Johnson filters
using both the 0.6\,m and the 1.6\,m telescopes from the Pico dos Dias observatory (OPD). The data acquisition and reduction procedures are explained in \citet{carciofi2007}. {To determine the intrinsic polarization of \VCAR, it is first necessary to estimate the interstellar polarization (ISP) component present in the observed data. This was done using the field star method, which relies on measuring the polarization of nearby stars that are not expected to exhibit intrinsic polarization. For such stars, the observed polarization can be attributed entirely to interstellar effects. We observed {fours} stars selected among the brightest and nearest (both angularly and radially) stars to \VCAR. The procedure used to estimate the ISP is explained in Appendix.~\ref{Ap:Pol}.}

\section{Model-independent analysis} \label{sec:NoModel}


\begin{figure}
    \centering
    \includegraphics[width=\linewidth]{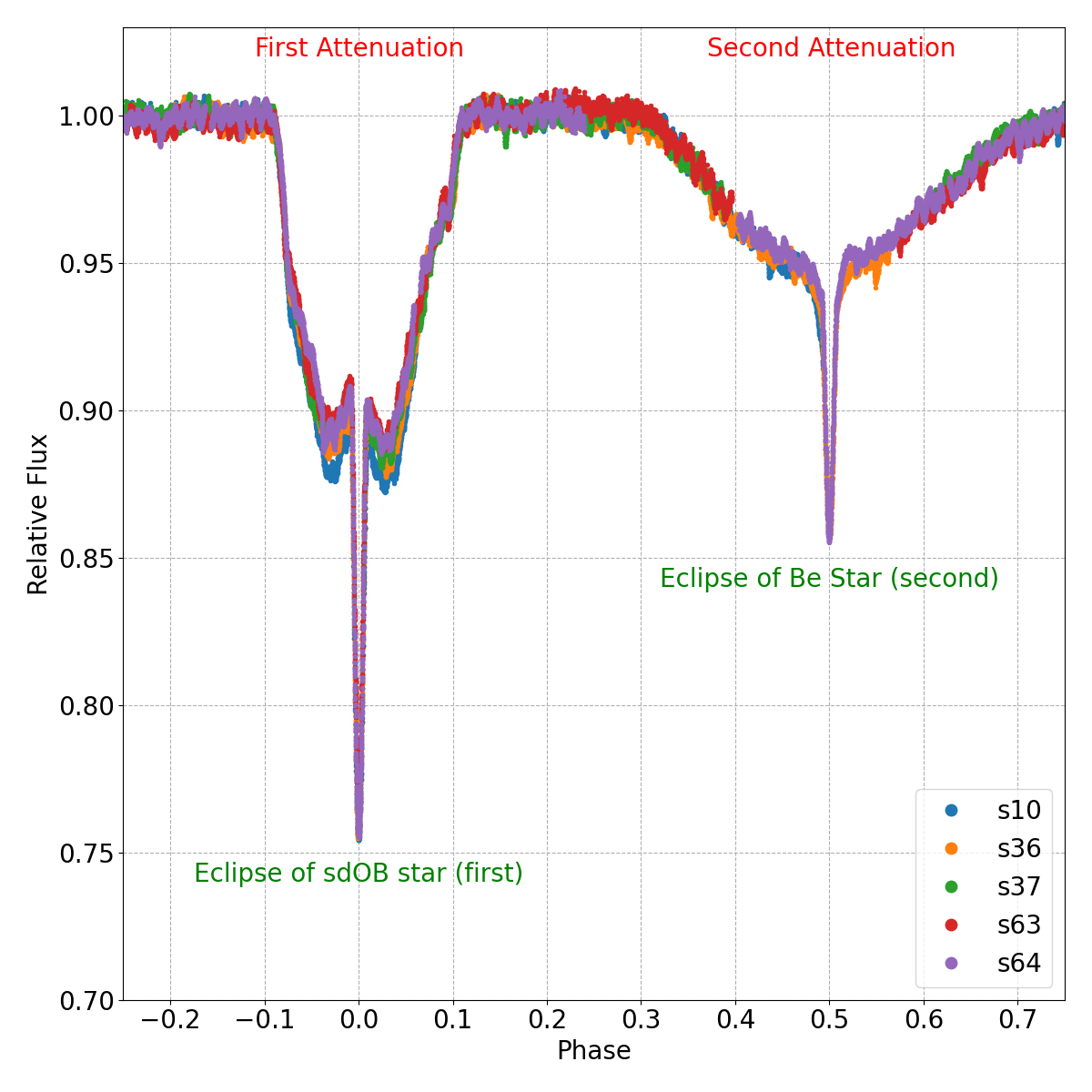}
    \includegraphics[width=\linewidth]{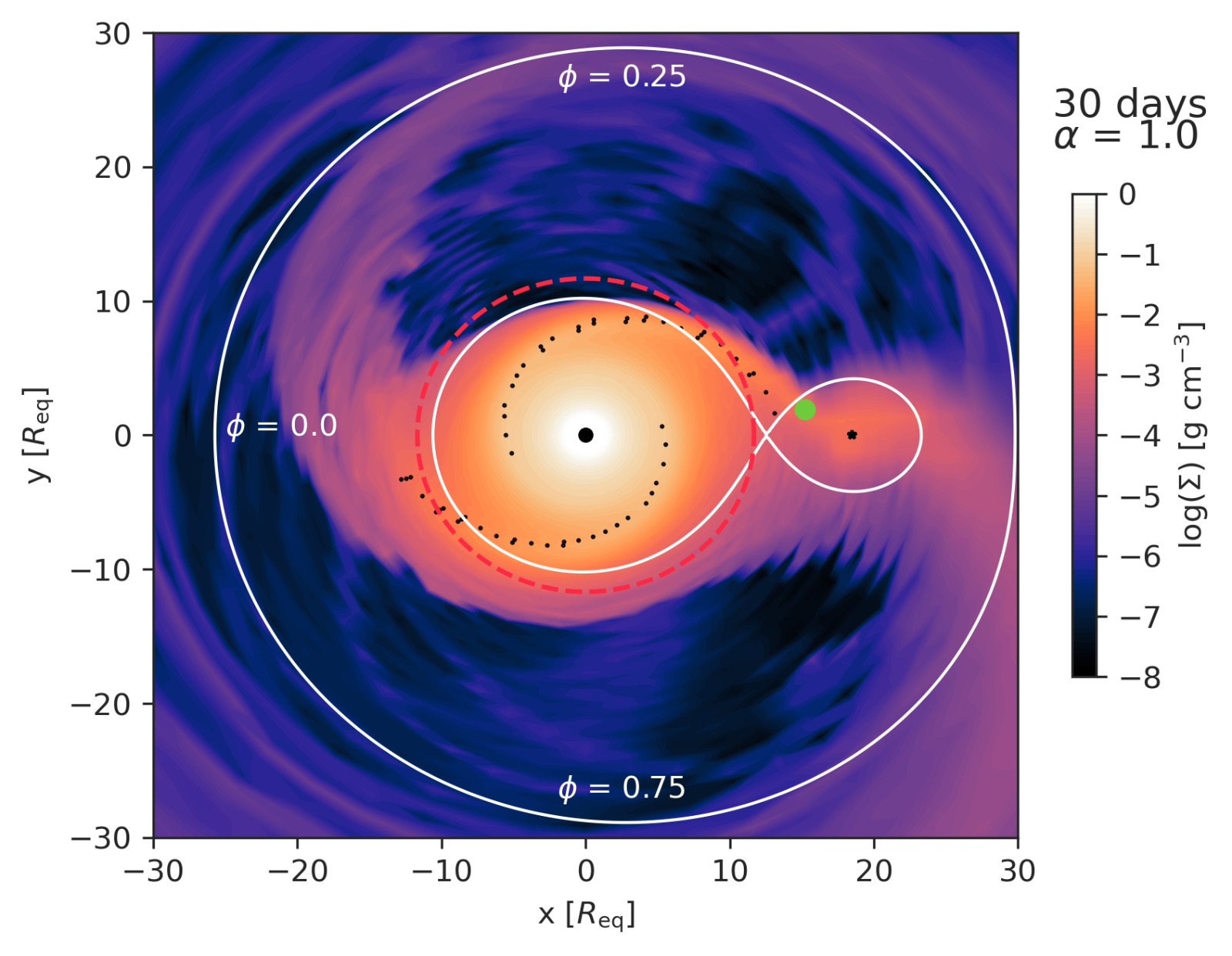}
    \caption{{Top: Phase-folded TESS light curve of \VCAR, each color representing a different sector (see legend). The main eclipses/attenuations are indicated. Bottom: SPH simulation for a binary system with $\alpha = 1$ \citep[viscosity parameter defined by][]{Shakura1973alpha}, $P = 30 \,\rm d$, and ratio of stellar masses of $M_{\rm Be}/M_{\rm sdOB} = 6$. $\phi$ indicates the corresponding observed orbital phase, with the system rotating counterclockwise. {The white solid line correspond to the calculated Roche Lobe}. More details about the simulation can be found in \citet{rubio2025}. Shown is the surface density, displaying the two-armed disk (dotted lines) and the transfer of mass from the Be disk to the companion. Considering the disk as Keplerian, we can map the physical positions of the double-peaked profile (red dashed circle) and oscilating emission (green dot) in \ion{Fe}{2}~9997~\AA\ based on their peak velocities (see Sect.~\ref{sec:helium_emission} for details). Since the latter varies over the orbital period, it is represented as a point, while the former remains stable and is therefore represented as a circle.
    }
    }
    \label{fig:SPH}
\end{figure}

This section provides {a qualitative analysis of the dataset, along with an orbital and dimensional solution}, aiming to establish a foundational basis for the model described in Sect.~\ref{sec:Modeling}. The underlying hypothesis driving the analysis is that \VCAR\ consists of an active Be star and an sdOB star, both possessing their own disks. Throughout this section and the two that follows, evidence will be provided to support this hypothesis.

{To assist the reader, we present in Fig.~\ref{fig:SPH} our main observable, the TESS light curve (see Sect.~\ref{sec:TESS}), alongside a recent smoothed particle hydrodynamic (SPH) model from \citet{rubio2025}, depicting a binary system with a Be star and a companion. We emphasize that this is not a model specifically of \VCAR; however, as we will demonstrate, it shares very similar parameters (see Tab.~\ref{tab:OrbSol}). For example, the period and stellar masses are sufficiently close to allow for a meaningful comparison of the observing phase in both plots.}

\begin{table}
    \caption{Orbital parameters were discussed in Sect.~\ref{sec:orb_sol}. The other parameters were explained in Sect.\ref{sec:Modeling}.}
    \begin{tabular}{cccc}\hline
        Parameter & State & \multicolumn{2}{c}{Solution} \\ \hline
        Component & & Be star & Companion\\
       $ K \, \rm [km \, s^{-1}]$ & Free & $12.2 (2.0)$ & $102.1 (2)$ \\
       $ M \, \rm [M_{\odot}]$ & Free & $4.45 (16)$  & $0.53 (11)$ \\
       $ a \, \rm [R_{\odot}]$ & Free & $7.8 (1.3)$  & $64.92 (13) $ \\
       $R_{\rm lobe} \, \rm [R_{\odot}]$ & Derived & $41.0 (7)$  & $15.8 (3) $ \\
       $R_{\rm p} [\rm R_{\odot}]$ & Free & 2.40(1) & 0.97(1) \\ 
       $R_{\rm eq} [\rm R_{\odot}]$ & Derived & 3.37(2) & 0.97(1) \\
       $T_{\rm p} [\rm kK]$ & Free & 18.0(5) & 21.0(5) \\
       $V_{\rm rot} [\rm km \, s^{-1}]$ & Derived & 452(8) & 23(1) \\
       $W$ & Fixed & 0.9 & -- \\
       $L [\rm L_{\odot}]$ & Derived & 347(40) & 163(18) \\ \hline
       Component & & \multicolumn{2}{c}{Be disk} \\
       $m$ & Fixed & \multicolumn{2}{c}{2.0} \\
       $\beta$ & Fixed & \multicolumn{2}{c}{1.5} \\
       $\rho_0 [\rm g~cm^{-3}]$ & Free & \multicolumn{2}{c}{$6.01(4) \times 10^{-11}$} \\
       $R_{\rm disk} [\rm R_{\odot}]$ & Fixed & \multicolumn{2}{c}{41} \\ \hline 
       Component &  & \multicolumn{2}{c}{System} \\
       $ e $ & Free & \multicolumn{2}{c}{$0.004 (2)$} \\
       $ P \, \rm [d]$ & Free & \multicolumn{2}{c}{$32.18534 (2)$} \\
       $ T_0 \, \rm [BJD]$ & Free & \multicolumn{2}{c}{$2459545.368(1)$} \\
       $ V_{\rm g} \, \rm [km \, s^{-1}]$ & Free & \multicolumn{2}{c}{$78.1 (2)$} \\
       $i [^{\circ}]$ & Free & \multicolumn{2}{c}{$88.6(1)$} \\
       $d [\rm pc]$ & Free & \multicolumn{2}{c}{$663(5)$} \\
       $E(B-V) [\rm mag]$ & Free & \multicolumn{2}{c}{$0.12(1)$} \\
       $P_{\rm max} [\%]$ & Fixed & \multicolumn{2}{c}{$0.61$} \\
       $\theta_{\rm IS} [^{\circ}]$ & Free & \multicolumn{2}{c}{$142.5(1.0)$} \\
       \hline \hline
    \end{tabular}
    \tablecomments{ $K$ is the orbital velocity of each star. $M$ represents the masses of each star and $a$ their orbital major axis, and $R_{\rm lobe}$ their Roche lobe. $V_{\rm g}$ is  the radial velocity of the system's center of mass, $e$  the eccentricity. The free parameters contain uncertainties estimated by the Levemberg-Marquadt method. It is important to stress that they do not reflect the real uncertainties, which should be much higher.}
    \label{tab:OrbSol}
\end{table}

\begin{figure*}
    \centering
    \includegraphics[width=\linewidth]{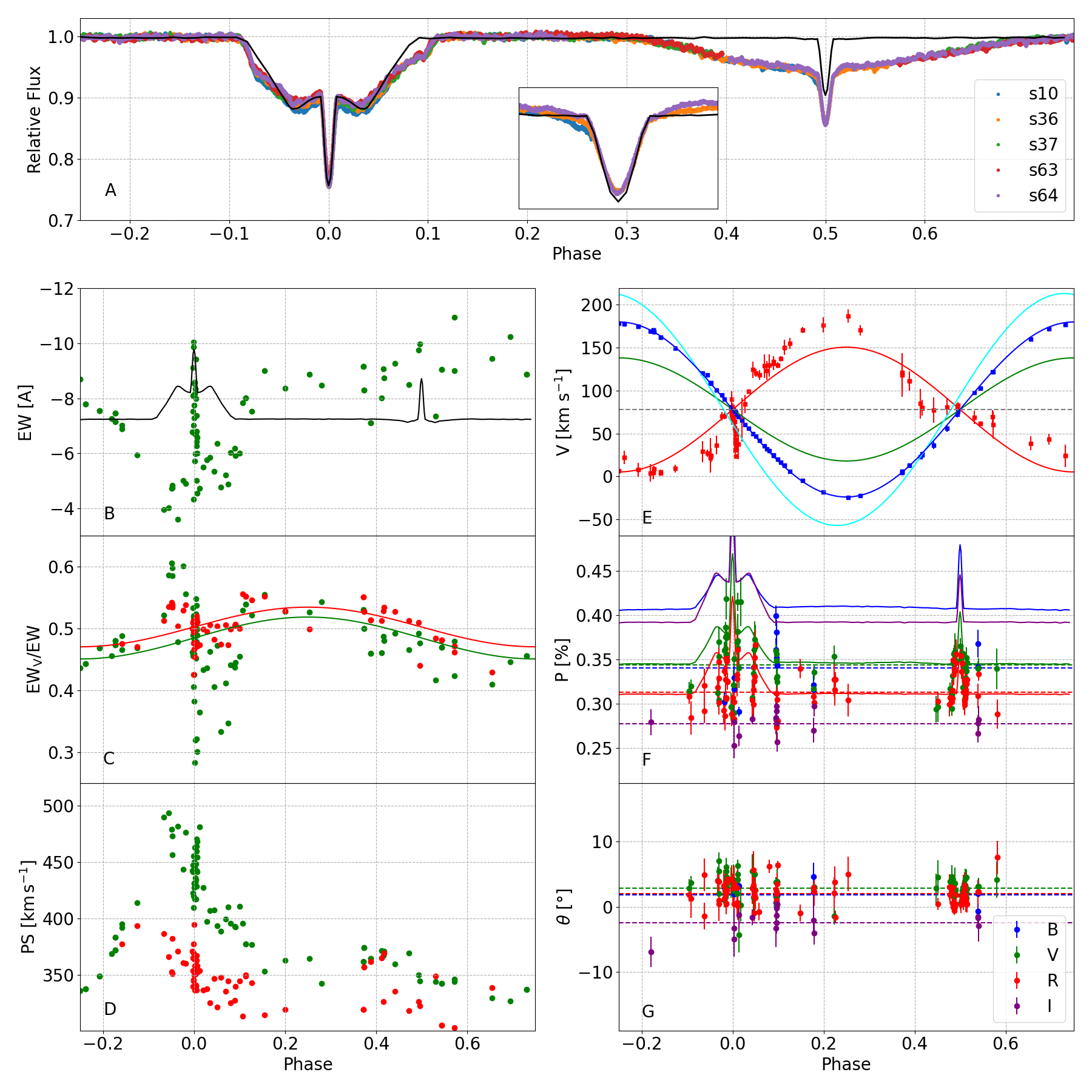}
    \caption{
    Phase-folded observations of \VCAR.
    Panel A: TESS light curve, each color representing a different sector (see legend); the inset zooms in the second stellar eclipse, where the model were shifted to match the observed level. 
    %
    Panels B, C and D illustrate the EW, line asymmetries and peak separation of \Ha, respectively.
    The red dots represent the standard quantities (V/R and PS, see text) and the green dots, the weighted quantities (Eqs.~\ref{eq:EW} and~\ref{eq:PSw}).
    The green and red solid lines in Panel C show the result of a sine fit to the data (see text).
    Panel E: Radial velocities, where the dots are the observations and the solid lines are the best fit obtained. Red, green, blue and cyan lines correspond to, respectively, the Be star (absorption lines), the \Ha wings (see Appendix~\ref{Ap:Ha_center}), the companion, and the travelling emission in \ion{Fe}{2}\,9997~\AA\ line. The amplitude of the former two were multiplied by 6, for better visualization. The dotted gray line represents the radial velocity of the center of mass.
    Panels F and G show the intrinsic polarization level and angle, respectively. The dots show the observations, with colors indicating each band, and the dashed lines represent the corresponding average values. 
    In panels A, B, F and G the solid lines represent our best model.
    }
    \label{fig:masterplot}
\end{figure*}

\subsection{TESS light curve}\label{sec:TESS}


The TESS light curve for \VCAR\ (Fig.~\ref{fig:masterplot}, panel A) shows two short, deep, and {narrow} stellar eclipses {at orbital phases 0 and 0.5 (hereafter, first and second stellar eclipses, respectively)}. The first stellar eclipse is about 50\% deeper than the second (relative to the adjacent brightness level) and {neither shows flat-bottom features, as would be expected from total eclipses between stars of different radii}.

Similar to what \citet{hauck2018eclipsing} observed, the TESS light curve also shows two broad and shallower dimmings, centered on the stellar eclipses, that do not resemble normal eclipses (hereafter, first and second attenuations, {following the phase convention for the stellar eclipses}). 
As in Algol-like eclipsing stars (Sect.~\ref{sec:intro}), the broad attenuations are likely linked to absorption by circumstellar material (see Sect.~\ref{sec:balmer}). 

The broad attenuations {(at phases ranging from -0.1--+0.1 and 0.33--0.67, respectively)} differ significantly in their characteristics. The first displays a steep variation in brightness and is marked by abrupt transitions. It is {distinctly} asymmetric, with the ingress\footnote{The terms ingress and egress are borrowed from eclipse terminology to describe the initial and final portions of the broad attenuations, respectively.} occurring more rapidly and appearing slightly shallower than the egress, that ends with a particularly sharp return to baseline near phase 0.1.
Another distinctive aspect of the first attenuation is a localized brightening near the first eclipse, which, ignoring the eclipse, imparts a W-like appearance to the light curve. In contrast, the second attenuation is nearly symmetric around the second stellar eclipse, shows gradual brightness changes with smooth departures from the baseline, and extends broadly in phase, from 0.33 to 0.67, using a 1\% deviation from baseline as a threshold across all TESS sectors.

\subsection{Optical and NIR Spectroscopy }
\label{sec:vis_ir_spec}
{Figure~\ref{fig:Dynamical} presents the main features seen in the optical and NIR spectra of \VCAR. The presence of clear emission and absorption lines that can be traced to each individual star is a remarkable feature.} This distinguishes \VCAR as one of the few known {SB2 Be systems in the optical/NIR range}. A notable example is the Be star HD\,55606, a SB2 comprising a rapidly rotating Be star and a hot, compact companion, which displays shell features and phase-dependent emission variability \citep{chojnowski2018hd55606}. 

\subsubsection{Balmer emission lines}\label{sec:balmer}

Emission in Balmer lines is a key indicator of disk presence in Be stars {\citep{rivinius2013classical}}.
The characteristic double-peaked emission lines are often symmetric, indicating an axisymmetric disk (see, e.g., the Be star $\beta$~CMi, \citealt{klement2015betacmi}, and the spectroscopic atlas of \citealt{hanuschik1996}). However, some Be stars exhibit more complex emission lines that deviate significantly from this simple pattern, with several possible underlying causes{, such as the one-armed} precessing wave, frequently observed in Be stars and usually resulting in large-amplitude V/R variations \citep{Okazaki1996m1waves,stefl2009m1wave}. 
Another cause is the tidal perturbation of the disk by a binary companion, which becomes more pronounced with decreasing binary separation \citep{panoglou2018discs}. In fact, all known short-period Be stars display complex profiles often with more than two {well-defined} peaks. A prime example is the Be+sdO star 59~Cyg with a period of about 28 days \citep{Peters201359cyg}.

\begin{figure*}
    \centering
    \includegraphics[width=0.3\linewidth]{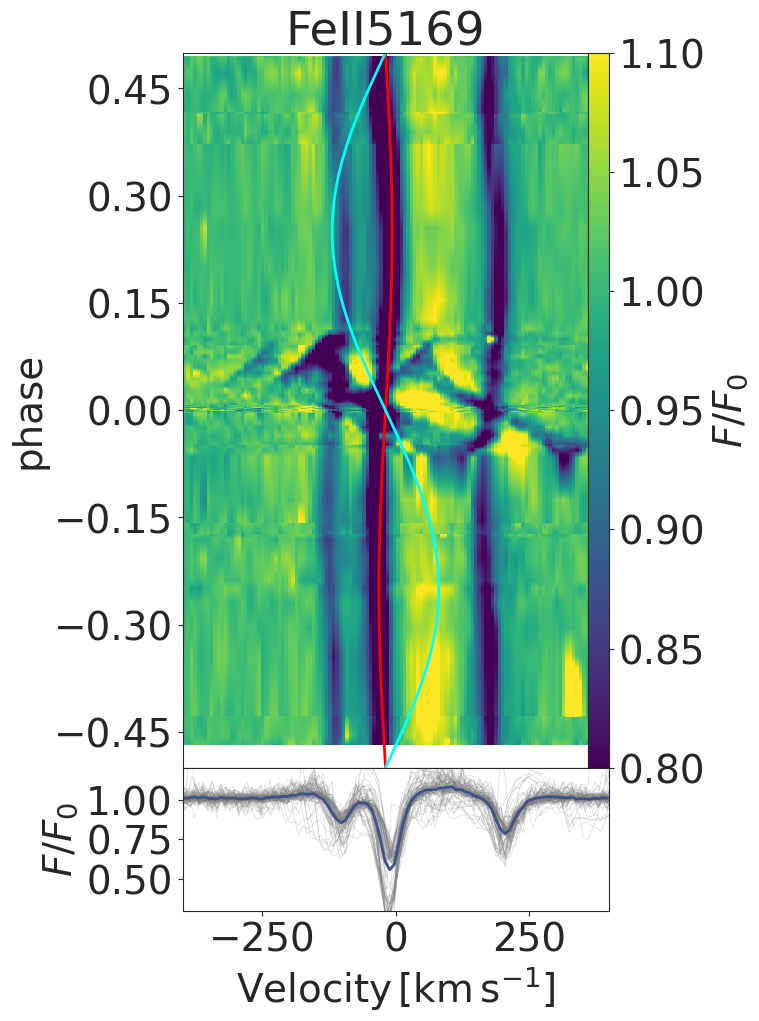}
    \includegraphics[width=0.3\linewidth]{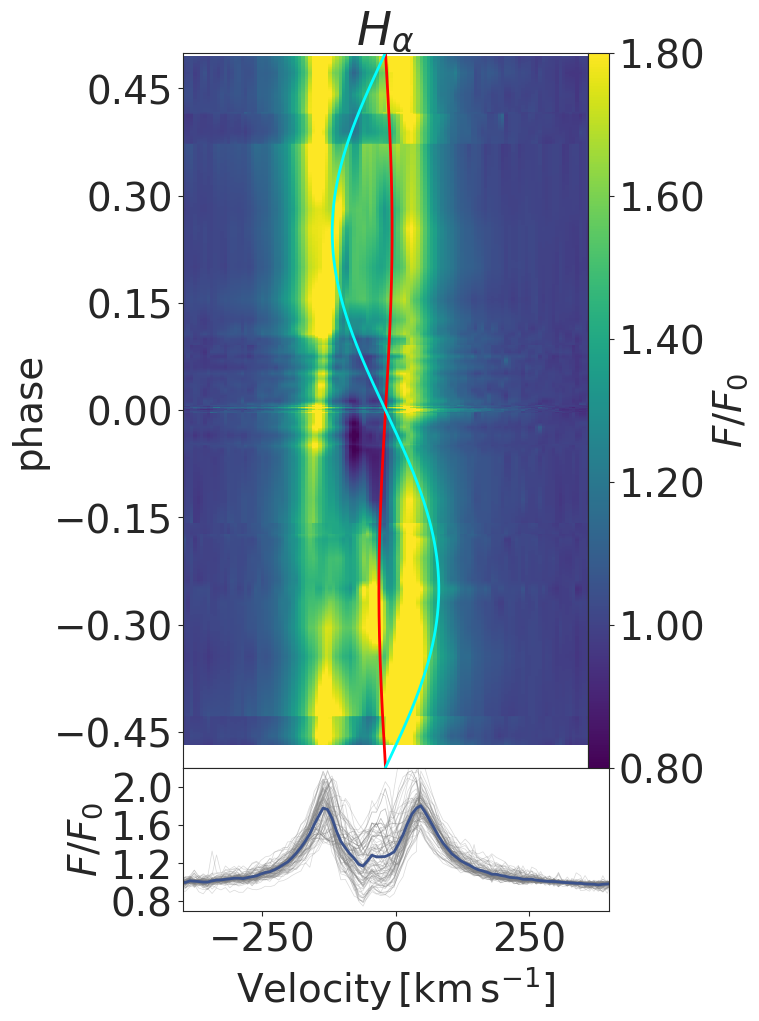}
    \includegraphics[width=0.3\linewidth]{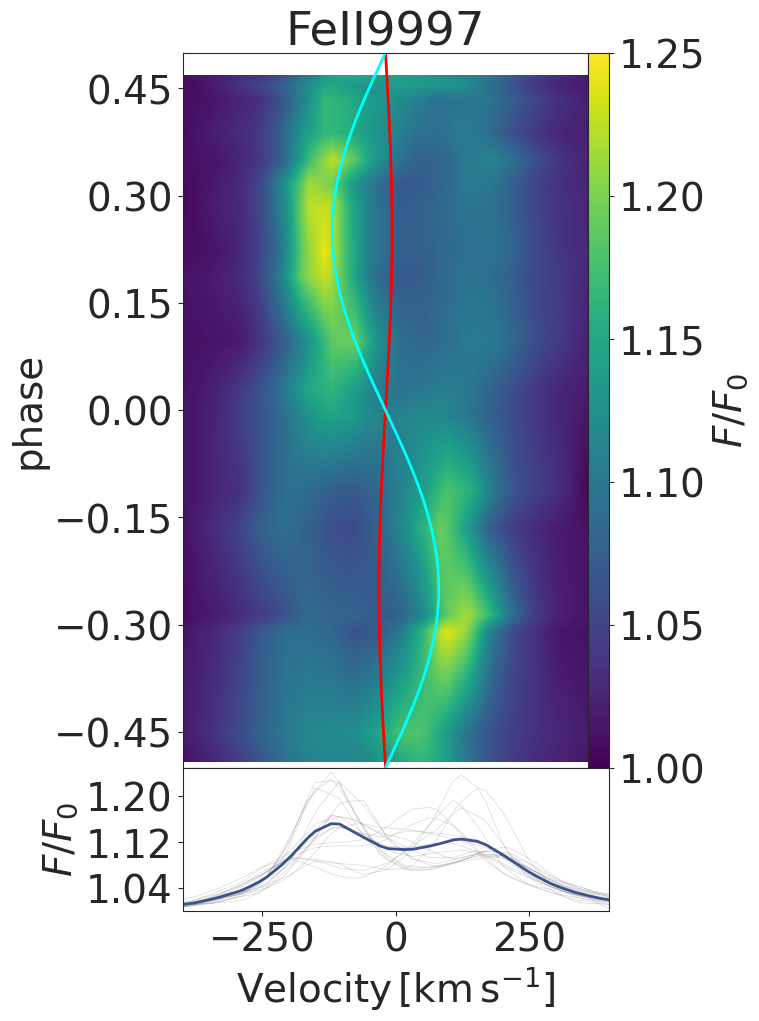}
    \includegraphics[width=0.3\linewidth]{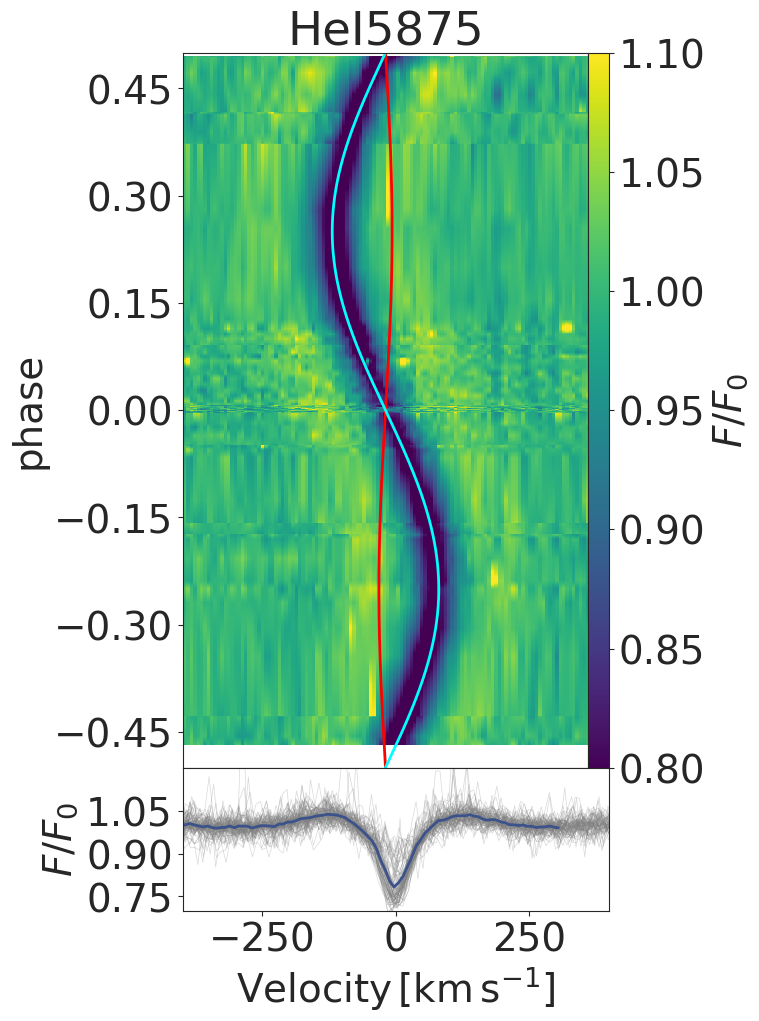}
    \includegraphics[width=0.3\linewidth]{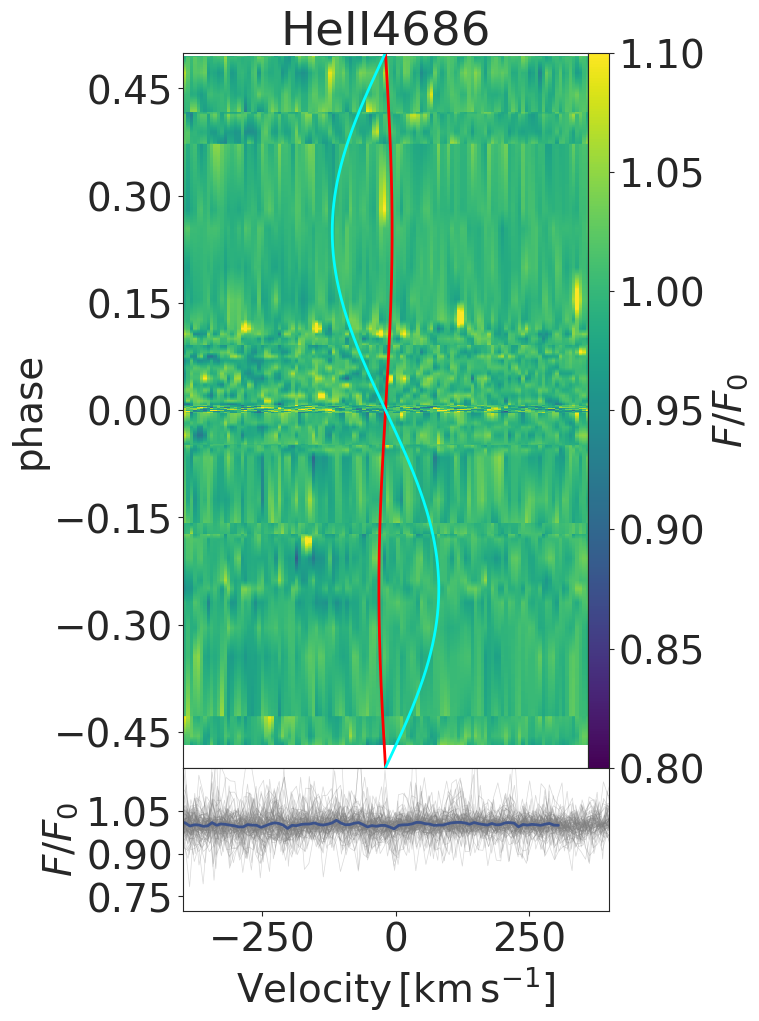}
    \includegraphics[width=0.3\linewidth]{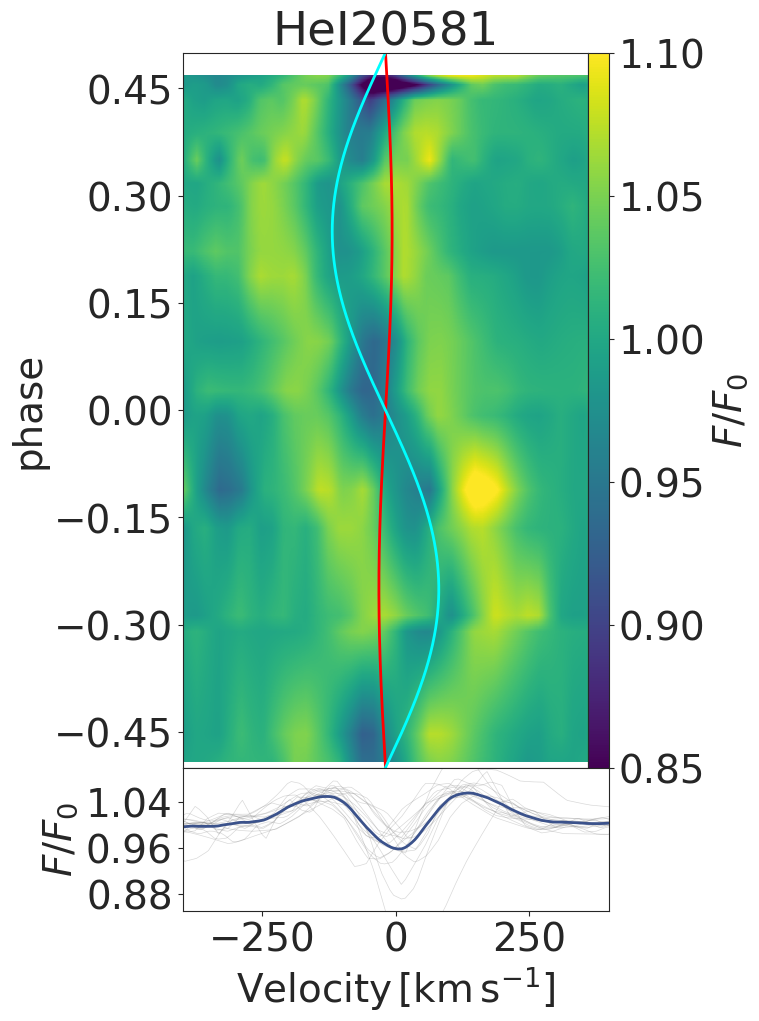}
    
    \caption{Dynamical spectral plots for \VCAR. The top panels present lines dominated by the Be star and/or its disk. The bottom panels present lines (or lack of) from the companion star and/or its disk. The red and cyan lines indicates the orbital velocity of the Be star and the companion, respectively. In this figure, we show selected single examples of each element for illustration. Appendix \ref{Ap:other_lines} contains more lines that show the same behavior.
    }
    \label{fig:Dynamical}
\end{figure*}

\begin{figure*}
    \includegraphics[width=0.49\linewidth]{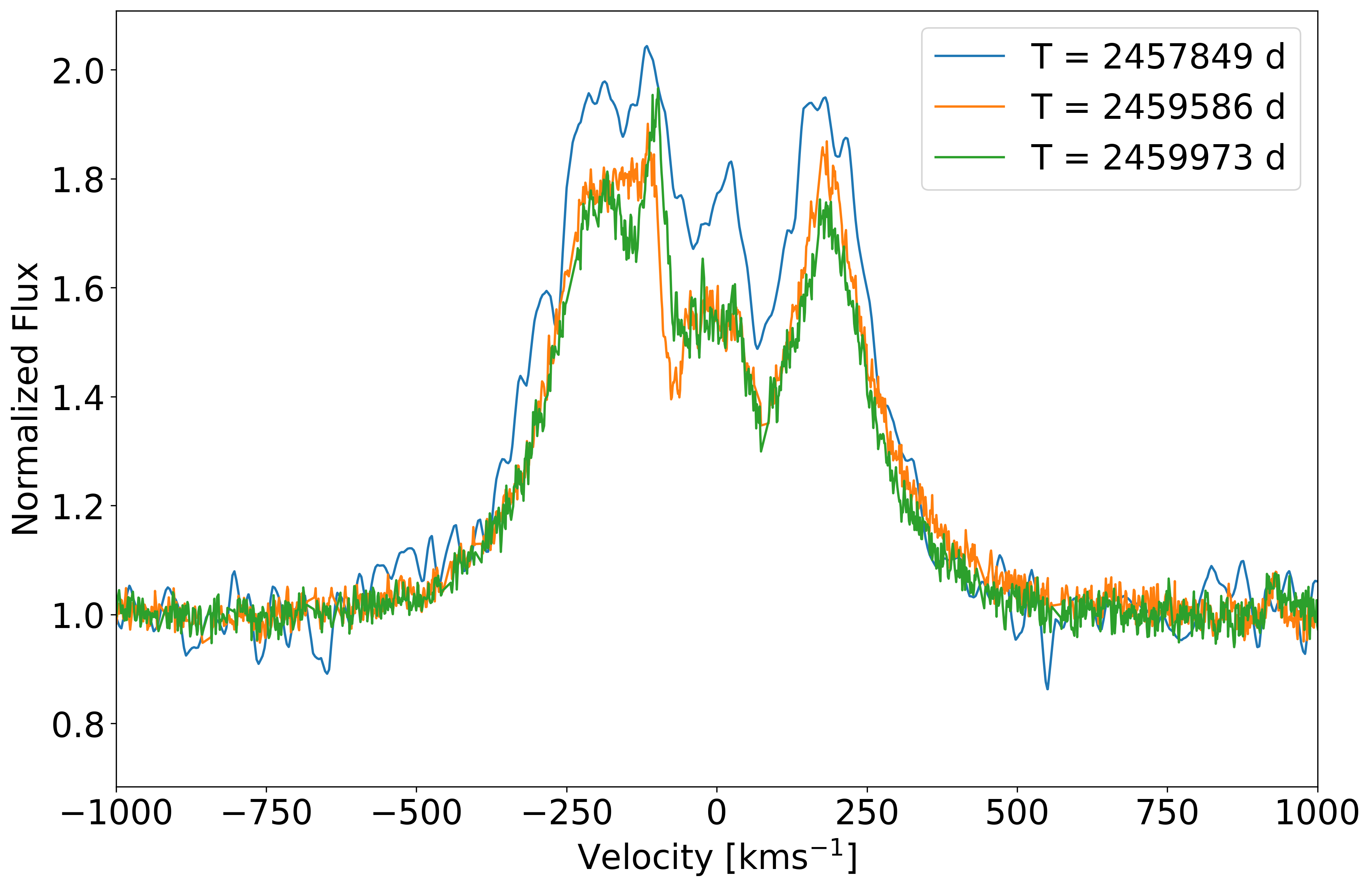}
    \includegraphics[width=0.49\linewidth]{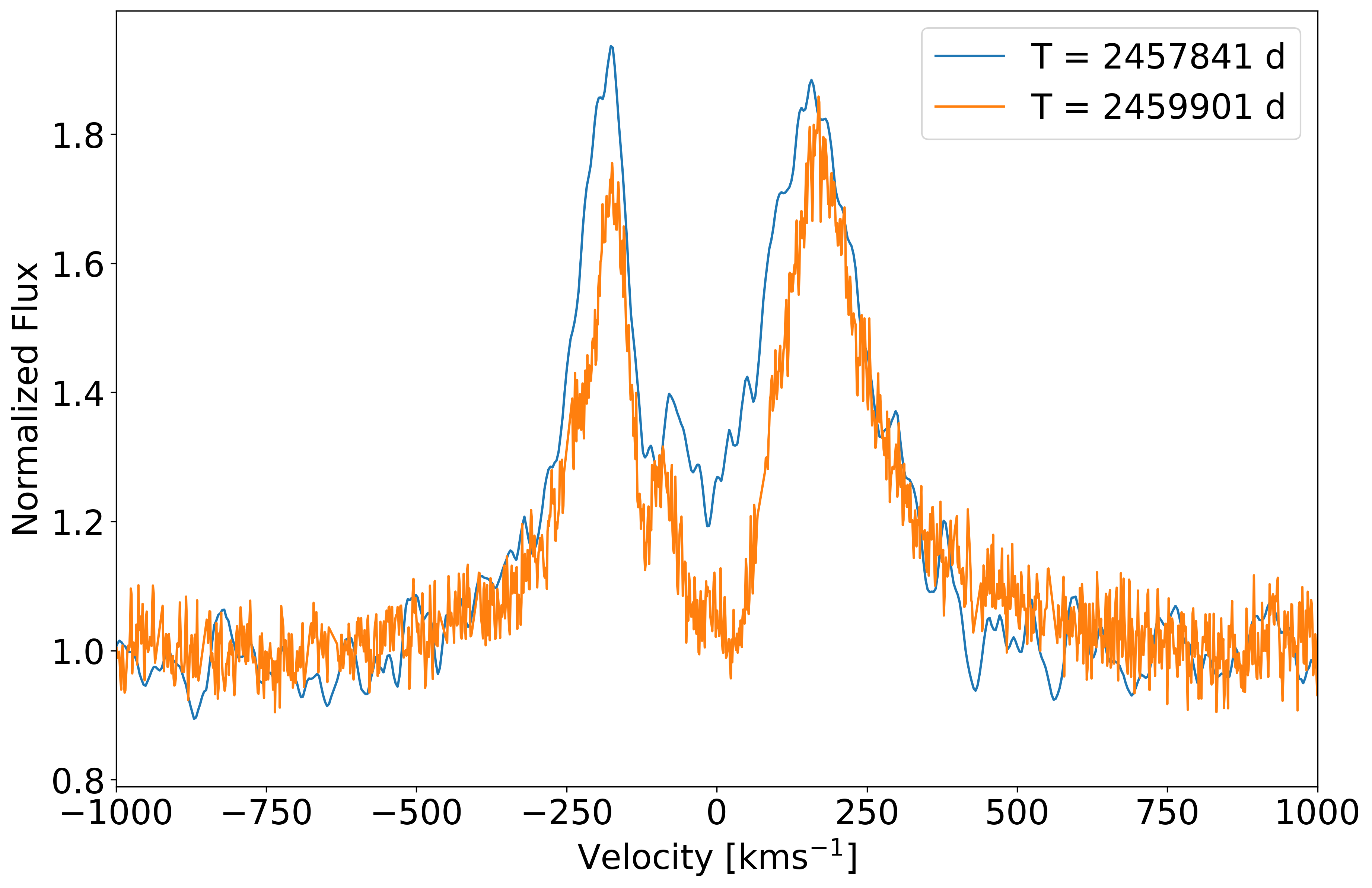}
    \caption{
    Variation of \Ha, corrected for the Be radial velocity, across many orbital periods at the same orbital phase, 0.3 and 0.043 for the left and right, respectively. The blue line was observed by Bernard Heathcote, and made available by BESS. The orange and green lines represent NRES observations.}
    \label{fig:Ha_time}
\end{figure*}

\VCAR's line profiles are complex and diverse, exhibiting a strong phase dependence. While they demonstrate a notable level of repeatability across different cycles, some cycle-to-cycle variations are also observed. This can be seen in the dynamical plots of Fig.~\ref{fig:Dynamical} and the selected profiles in Fig.~\ref{fig:Ha_time}. Dynamical plots for additional lines can be found in Fig.~\ref{fig:Dynamicalappendix}.


To quantify the temporal variability of the line emission, particularly its dependence on orbital phase, several quantities were measured, in addition to the standard equivalent width (EW) {of the profile}. 
To measure the individual peak heights relative to the adjacent continuum level, a Gaussian was fitted to the emission peaks (when they can be clearly identified), 
which also provided each peak position (in km\,s$^{-1}$). The latter are used to calculate the peak separation (PS).
Instead of the usual measurement of peak asymmetry (V/R), we adopt the V/(V+R) ratio. This {quantity} has the distinct advantage of being symmetric around 0.5, which represents the value of a symmetric profile. 
EW is plotted as green symbols in panel B of Fig.~\ref{fig:masterplot}, while the other two quantities, V/(V+R) and PS, are plotted as red symbols in panels C and D, respectively.

However, some profiles were too complex to allow for a meaningful definition of the V/(V+R) and PS; for instance, the three profiles on the left of Fig.~\ref{fig:Ha_time} show multiple peaks, rendering the peak definition subjective. To circumvent this, we adopted two quantities that can be applied to every line profile without subjectivity. 
The first quantity {(hereafter line asymmetry)} is defined as
\begin{equation}
    \frac{\rm EW_V}{\rm EW} \equiv \frac
    {\int_{0}^{\lambda_{0}}  (1 - \frac{F_{\lambda}}{F_{C}}) d\lambda}
    {\int_{0}^{\infty} (1 - \frac{F_{\lambda}}{F_{C}}) d\lambda},
    \label{eq:EW}
\end{equation}
where $F_{\lambda}$ and $F_{\rm C}$ are the observed line and continuum flux, respectively, for a selected wavelength, $\lambda$, and $\lambda_0$ represents the {\Ha rest wavelength Doppler-shifted to the radial velocity of the Be star (see sect~\ref{sec:orb_sol})}\footnote{In Appendix~\ref{Ap:Ha_center}, we explore an alternative approach by defining the center of the line as the midpoint between the \Ha wings.}. 
$\rm EW_{V}/EW$ is a measure of line asymmetry, which qualitatively behaves similarly to the V/(V+R) ratio (see green and red points on panel C of Fig.~\ref{fig:masterplot}).

The second quantity {is defined as the weighted peak separation ($ \rm PS_{w}$), given by}
\begin{equation}
    \rm PS_{\rm w} = \upsilon_{R} - \upsilon_{V} = 
    \frac{\int_{\upsilon_{0}}^{\infty} \upsilon (1 - \frac{F_{\upsilon}}{F_{C}}) d\upsilon}{\int_{\upsilon_{0}}^{\infty} (1 - \frac{F_{\upsilon}}{F_{C}}) d\upsilon} - 
    \frac{\int_{-\infty}^{\upsilon_{0}} \upsilon (1 - \frac{F_{\upsilon}}{F_{C}}) d\upsilon}{\int_{-\infty}^{\upsilon_{0}} (1 - \frac{F_{\upsilon}}{F_{C}}) d\upsilon},
    \label{eq:PSw}
\end{equation}
where $\upsilon$ is the velocity shift from the line center
${\upsilon = c (\lambda-\lambda_0)/\lambda_0}$.
$ \rm PS_{w}$ is displayed as green symbols in panel D of Fig.~\ref{fig:masterplot}.

The phase dependency of \VCAR's EW can be qualitatively understood by two competing mechanisms:
\begin{enumerate}
    \item \emph{Continuum suppression}: The first attenuation is assumed to be caused by absorption of light from the companion by the disk, primarily due to continuum opacity (mainly hydrogen), or by the Be star.
    As the continuum is suppressed, the EW is expected to increase in magnitude. Clearly, continuum suppression will be strongest during the first stellar eclipse.
    \item \emph{Shell absorption}: light from the companion will also undergo line absorption within the disk, resulting in a net decrease of the \Ha flux from the companion star reaching us. This should cause a reduction in the measured EW.
\end{enumerate}
From panel B of Fig.~\ref{fig:masterplot} we observe that the EW decreases in magnitude during the broad attenuation, suggesting that shell absorption is the dominant mechanism during these phases. However, during the first stellar eclipse, there is a sudden spike in EW, driven by continuum suppression.
Across the second broad attenuation, there seems to be a slight increase of the EW {magnitude} as a function of the phase, perhaps suggestive of a small-amplitude continuum suppression. Unfortunately, {optical spectra} are scarce during the second eclipse.
{The asymmetry of EW during the first attenuation suggests that the disk is likely asymmetric, leading} to a phase-dependent modulation of the EW, as the projected disk structure changes along the orbit.
This would inhomogeneously enhance continuum suppression and/or shell absorption, also introducing a phase dependency in the disk emission.

%
{We} note that part of the scatter in the EW curve arises from cycle-to-cycle variations that, while small, are not negligible. 
In the left panel Fig.~\ref{fig:Ha_time}, the three profiles {roughly correspond to the same orbital phase} ($\approx0.3$), where both the \Ha EW and the TESS light curve are nearly flat. The equivalent widths, from the oldest (green) to the most recent (orange) observation, are  {$-11.6(2)$\,\AA, $-9.2(1)$\,\AA\ and $-8.6(1)$\,\AA}, respectively. This suggests a gradual decrease in disk emission over the past six years, with little to no change in the overall line profile. 

The \Ha line asymmetry (panel C of Fig.~\ref{fig:masterplot}) displays a roughly sinusoidal behavior, if one ignores the phases around the first eclipse. The solid curves represent a {sinusoidal} fit to the data, excluding the data points between phases -0.01 and 0.01.
%
The peak of the blue-shifted emission is located approximately at phase $0.25$, the quadrature when the companion star is approaching us. It is interesting to note that, for the most part,  $\rm V/(V+R)$ and $\rm EW_{V}/EW$ {are in good} agreement with each other, with the exception of the phases around the first eclipse, during which $\rm EW_{V}/EW$ attains much larger amplitudes. 
{As seen in the right panel of Fig.~\ref{fig:Ha_time}, shortly after the stellar eclipse, the red peak is comparable in amplitude to the violet peak but is considerably broader. This explains the discrepancy between the two measurements.}
{The most likely explanation for this phenomenon, once again, involves shell absorption. Prior to the first eclipse, shell absorption is expected to occur on the red side of the \Ha line, diminishing its intensity and thereby enhancing $\rm EW_{V}/EW$. After the stellar eclipse, the shell absorption shifts to the blue side, leading to a reduction in that part of the profile.}

The sinusoidal nature of the {$\rm EW_{V}/EW$} variations (excluding the complexities introduced by the eclipses) suggests a causal link between the observed asymmetries and perturbations in the Be star's disk induced by the companion. Similar patterns have been observed in other stars. For instance, phase-locked V/R variations with comparable behavior were reported for $\nu$~Gem by \citet[][see their Fig.~4]{Miroshnichenko2023SB1}.

\begin{figure}
    \centering  \includegraphics[width=\linewidth]{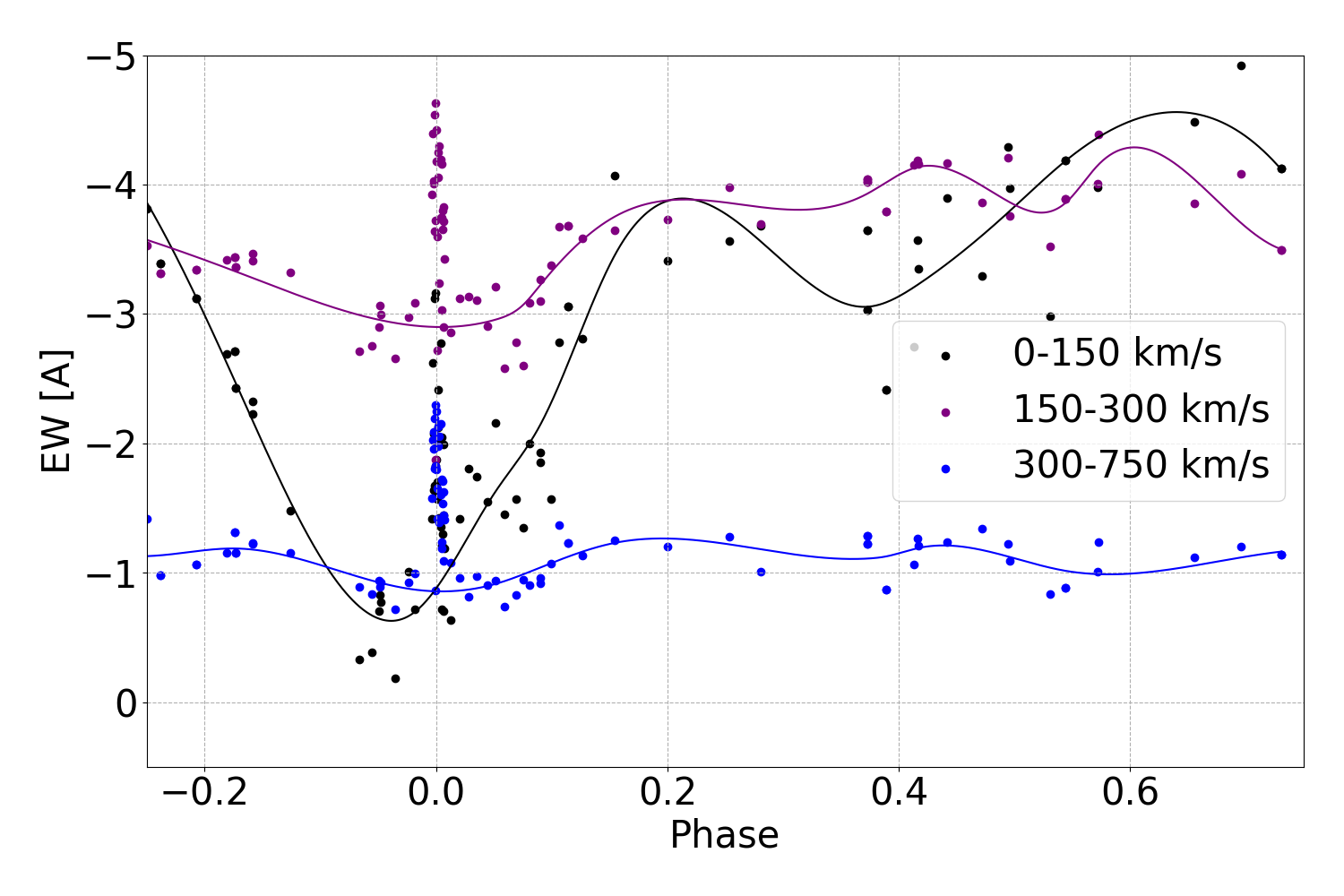}
    \caption{
    Partial EW of \Ha for different velocity bins relative to the Be radial velocity. Black: line center (-150 to 150\,$\rm km\,s^{-1}$); Pink: line peaks (-300 to -150\,$\rm km\,s^{-1}$ and 150 to 300\,$\rm km\,s^{-1}$); Blue: line wings (-750 to -300\,$\rm km\,s^{-1}$ and 300 to 750\,$\rm km\,s^{-1}$). The dots correspond to observations, while the lines are cubic splines to aid visualization of the variation outside the stellar eclipse.
    }
    \label{fig:EW_cut}
\end{figure}

Panel D of Fig.~\ref{fig:masterplot} shows a different pattern for the PS or $\rm PS_w$ of \Ha. Outside of the disk attenuations and stellar eclipses, both quantities have a similar behavior (as was the case for the line asymmetry in panel C), with average values of $332 (25) \,\rm km\,s^{-1}$ and $357 (17) \rm\,km\, s^{-1}$, respectively.
During the first attenuation, however, both PS and $\rm PS_w$ show a pronounced and highly asymmetric increase, which is anti-correlated with the absolute value of the EW. Notably, the amplitude of this increase differs significantly between the two quantities.
{The comparison between the standard definition (PS, red dots) and the new quantity ($\rm PS_w$, green dots)} reveals an important distinction: the standard PS traces variations in regions of the disk contributing most at intermediate velocities, where the peaks are located, while $\rm PS_w$ reflects changes across the entire \Ha line profile. The higher amplitude observed in $\rm PS_w$ suggests either a weakening of low-velocity emission or an enhancement at high velocities. However, the latter is inconsistent with the observed decrease in the absolute EW. Figure~\ref{fig:EW_cut}, which separates the relative EW {of \Ha} by velocity range, supports the first scenario: the low-velocity component ($\upsilon < 150\, \rm km\,s^{-1}$) dominates the change, likely due to increased dimming or added shell absorption. 
Finally, it is worth noting that, during the first eclipse, there is a sharp transition from {high values to lower values of PS and $\rm PS_w$}.

\subsubsection{Absorption lines}\label{sec:absorption_lines}

Most Be + sdOB binaries are SB1 systems, in which only the orbital motion of the Be star is detectable {in the optical range \citep{wang2021detection}}. This is primarily due to the large luminosity contrast between the components: the Be star typically dominates the optical spectrum, often outshining the companion by factors exceeding 30, as seen in systems like $\kappa$ Draconis \citep{Klement2022dynamical}. In SB1 systems, information about the companion might be obtained by other means, such as UV spectroscopy \citep{wang2021detection} or astrometry \citep{Klement2022dynamical}; in many cases, however, such information is not available. \VCAR\ belongs to a rare subset of SB2 Be stars, where 
the companion is bright enough to contribute significantly to the optical and NIR spectrum, allowing for the detection of its photospheric lines and thereby enabling a more complete determination of the orbital characteristics.
This is clearly illustrated in Fig.~\ref{fig:Dynamical}, where high-excitation lines, such as {those of} \ion{He}{1}, display a well-defined radial velocity curve in anti-phase with that of low-excitation lines like {those of} \ion{Fe}{2}. Furthermore, the radial velocity amplitudes of the high-excitation lines are significantly larger than those of the low-excitation lines, indicating that the hotter component is also the less massive one. As a final piece of the puzzle, we note that the Balmer emission lines do not exhibit high-amplitude radial velocity variations, showing behavior more consistent with that of the photospheric low-excitation lines (see, for instance, the two top-left panels of Fig.~\ref{fig:Dynamical})
Therefore, it is safe to conclude that the Be star is the most massive component and possesses the {lower} effective temperature. {Although the companion is hotter, the lack of absorption features in \ion{He}{2}~4686~\AA\ suggests that it is an early B-type star rather than an O-type star.}

\begin{figure*}
    \includegraphics[width=0.47\linewidth]{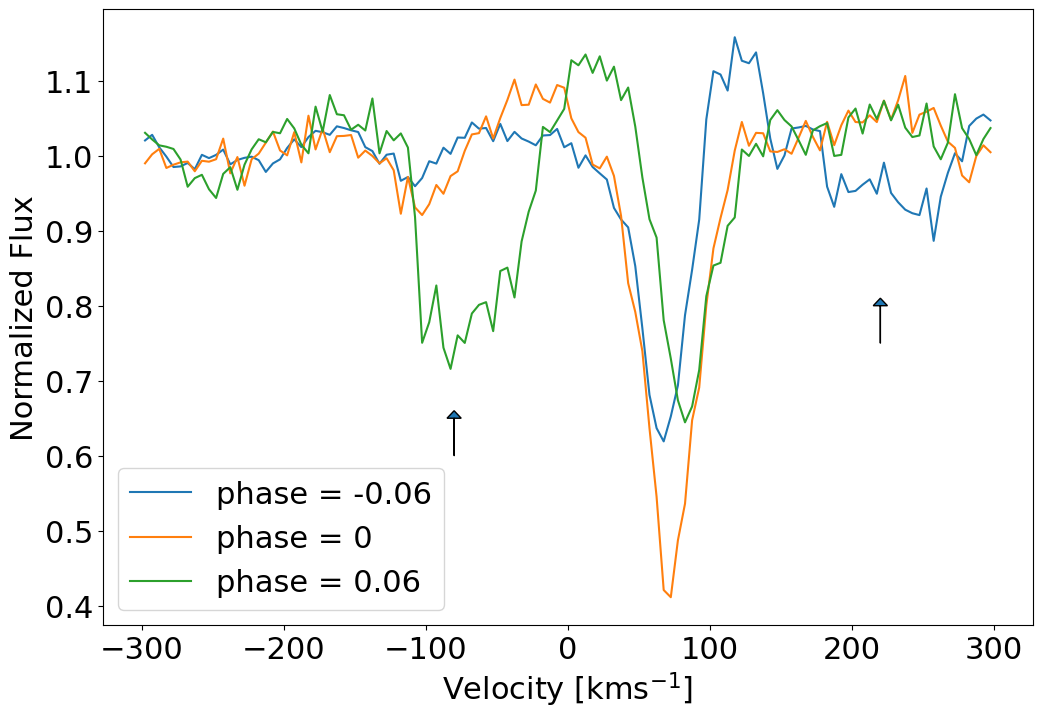}
    \includegraphics[width=0.49\linewidth]{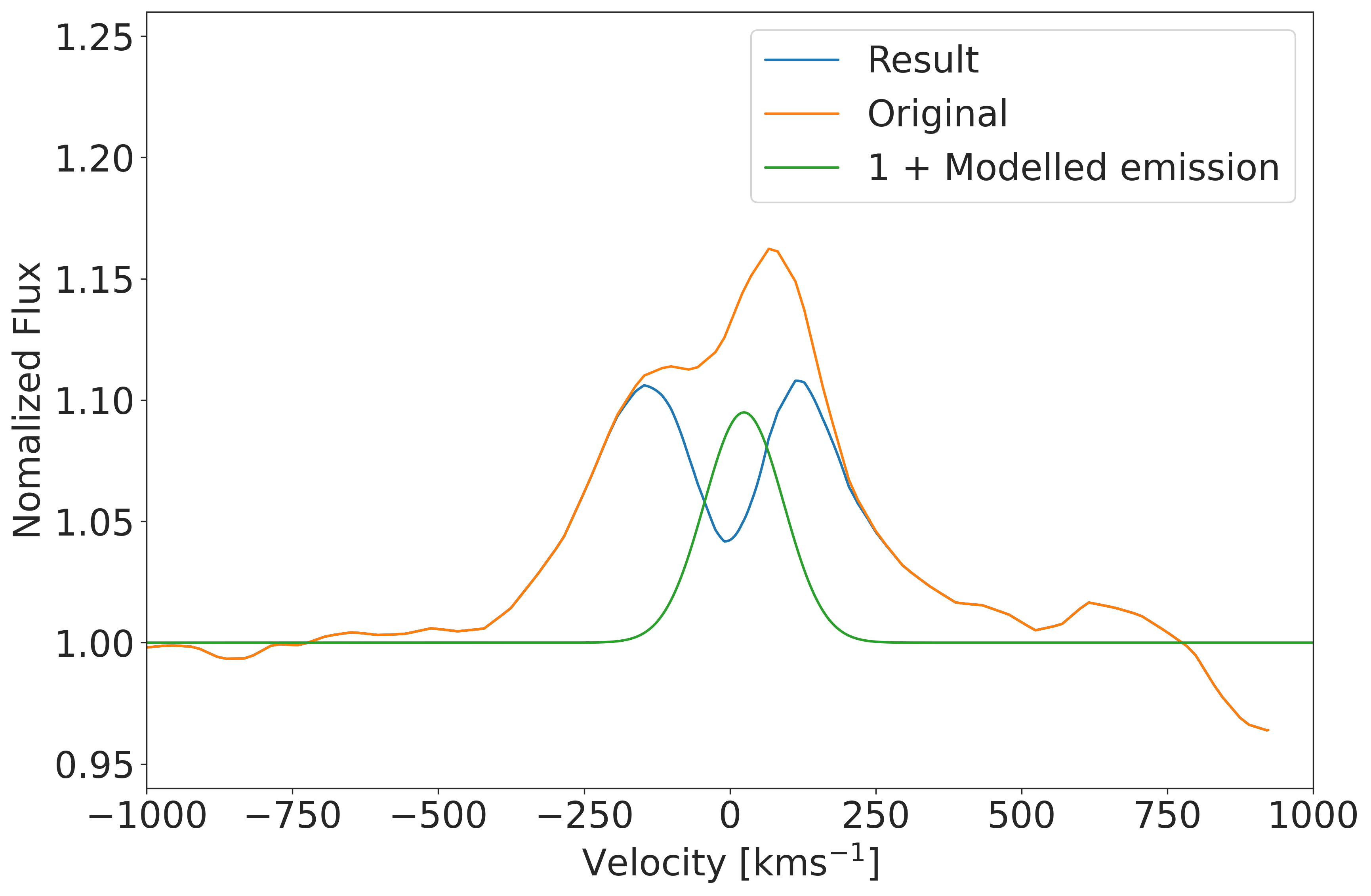}
    \caption{
    Left: Variation of \ion{Fe}{2}\,5316~\AA\ across different phases in the same orbital cycle. We highlighted the appearance and motion of the travelling features with black arrows. Right: The \ion{Fe}{2}~9997~\AA\ line was fitted with a three-Gaussian profile (see Sect.~\ref{sec:helium_emission}). Here, we show the observed spectrum at phase 0.508 in orange and the best-fit oscillating Gaussian emission component 
    in green. The blue line shows the subtraction of the latter from the former, revealing a nearly symmetric structure.
    }
    \label{fig:Iron_phase}
\end{figure*}


An unusual and intriguing feature of the system is the presence of traveling absorption components in the low-excitation lines (see the {\ion{Fe}{2}~5169~\AA} panel of Fig.~\ref{fig:Dynamical} and the three top panels of Fig.~\ref{fig:Dynamicalappendix}). These features can be interpreted as shell absorption lines forming when the companion star passes behind the Be disk. As its light is partially absorbed by cooler, dense material in the disk, a transient shell-like signature emerges, shifting from red to blue as the orbit progresses. The absence of such features during the second attenuation strongly suggests that no optically thick disk surrounds the companion. 

\subsubsection{Helium and iron emission lines}\label{sec:helium_emission}

The \ion{He}{1} lines exhibit weak double-peaked emission, with the features in the NIR lines being significantly stronger than those in the optical, suggesting the presence of a tenuous circumstellar structure co-moving with the companion’s orbit.
Perhaps the best example of a similar behavior is HD\,55606, that also displays weak emission structures co-moving with the lines showing large-amplitude radial variations \citep{chojnowski2018hd55606}.
A notable characteristic of both \VCAR and HD\,55606 is the approximately symmetrical double-peaked feature in each case, indicating that the structure around the companion is approximately symmetric.


An intriguing behavior is observed in the IR iron lines, which, unlike their optical counterparts, exhibit clear emission features. These emissions can be interpreted as the combination of two distinct components: a stable, symmetric double-peaked profile with a low velocity amplitude, consistent with orbital motion from the Be star and thus likely originating from its disk, and a superimposed single-peaked component. To separate these components, we fitted a triple-Gaussian profile:
\begin{align}
f_\upsilon(\phi) &= 1 
+ G_\upsilon(\upsilon_1, A_{12}, \sigma_{12}) \nonumber \\
&\quad + G_\upsilon(\upsilon_2, A_{12}, \sigma_{12}) \nonumber \\
&\quad + G_\upsilon\left(\upsilon_3 \sin\left(2\pi(\phi - \phi_0)\right), A_3, \sigma_3 \right) ,
\end{align}
where $G_\upsilon$ represents a standard Gaussian profile in velocity space,
\begin{equation}
G_\upsilon(\upsilon_0, A, \sigma) = A \cdot \exp\left( -\frac{(\upsilon - \upsilon_0)^2}{\sigma^2} \right)\,.
\end{equation}
The function $f_\upsilon$, at a given phase $\phi$, is composed of two fixed and symmetric Gaussians (with the same amplitude $A_{12}$ and width $\sigma_{12}$) located at $\upsilon_1$ and $\upsilon_2$, and a third Gaussian with an initial phase $\phi_0$, oscillating between $-\upsilon_3$ and $+\upsilon_3$ with fixed amplitude $A_3$ and width $\sigma_3$. To compare with the data, we first subtracted the Be star's radial motion (see Sect.~\ref{sec:orb_sol}) from the \ion{Fe}{2}~9997~\AA\ line, and then used a Levenberg–Marquardt algorithm to minimize the eight parameters ($\upsilon_1$, $\upsilon_2$, $\upsilon_3$, $A_{12}$, $A_3$, $\sigma_{12}$, $\sigma_3$, $\phi_0$) across all observed phases simultaneously. The best-fit solution consists of a double-peaked profile with $A_{12} = 0.097(1)$ and a peak separation of $|\upsilon_2 - \upsilon_1| = 316 (15)~\rm{km~s^{-1}}$. The oscillating Gaussian (see cyan line in panel E of Fig.~\ref{fig:masterplot} and green line in the right panel of Fig.~\ref{fig:Iron_phase}) has a similar amplitude, $A_3 = 0.095(1)$, and a phase slightly ahead of the companion star, $\phi_0 = -0.020(1)$, corresponding to approximately {$7.2(4)^\circ$}. Its orbital velocity amplitude is $\upsilon_3 = 138(5)~\rm km~s^{-1}$.

\subsection{Orbital Solution}\label{sec:orb_sol}

To trace the orbital motion, we selected the three most prominent absorption lines associated with each star: \ion{Fe}{2}\,5169~\AA, \ion{Fe}{2}\,5316~\AA, and \ion{Fe}{2}\,5362~\AA\ for the Be star, and \ion{He}{1}\,4713~\AA, \ion{He}{1}\,5875~\AA, and \ion{He}{1}\,6678~\AA\ for the companion. Our initial approach involved fitting Gaussian profiles to these lines to derive their radial velocities. However, due to the presence of traveling absorption features at certain orbital phases (see Fig.~\ref{fig:Iron_phase}), this method proved insufficiently robust. As an alternative, we adopted a template-matching strategy: starting from an initial orbital solution, we constructed a Doppler-shifted average spectrum, which was then used as a reference template to refine the radial velocity measurements. For each observed spectrum and for all the selected line sets, the radial velocity was determined as the value that minimized the difference between the observed profile and the template—effectively performing a least-squares match in velocity space. This new set of radial velocities was then used to compute an updated orbital solution, and the process was repeated iteratively until no further refinements were required. The measured radial velocities for each star and their uncertainties are shown as the symbols in panel E of Fig.~\ref{fig:masterplot}. We acknowledge that they are still affected to some extent by the traveling features, but their impact is considerably reduced with this method.

The TESS data provide a precise means of determining the orbital period and the ephemeris of the first eclipse. We applied a simple Gaussian fitting routine to identify the best-fit parameters across all TESS sectors in which the first stellar eclipse was observed. The resulting period, $P = 32.18534(2)$\,d, and ephemeris, $T_0[\mathrm{HJD}] = 2459545.368(1)$, are consistent with the values reported by \citet{hauck2018eclipsing}, who found $P = 32.1854(1)$\,d and $T_0[\mathrm{HJD}] = 2452786.438(2)$ (where $T_0 + 210P = 2459545.372$). Assuming both values of $T_0$ are accurate, the optimal period that reconciles them is $P = 32.18538(1)$\,d. 
The radial velocities, period and ephemerid were used as input to the code of \citet{Milson2020pythonbinaries} to calculate the orbital parameters, displayed in Table~\ref{tab:OrbSol}. The results for the masses and semi-major axis depend on the inclination of the orbital plane, $i$. The occurrence of eclipses suggests that $i$ is very close to 90$^{\circ}$. Therefore, it is reasonable to assume $\sin{(i)} = 1$. The actual uncertainty in the orbital velocity of the Be star, as obtained by the code, was $\pm 0.2~\rm km~s^{-1}$. However, it is not accounting for the possible systematic errors that might be introduced in the radial velocities of the Be star by the traveling features discussed in Sect.~\ref{sec:helium_emission}.  We argue that the perturbations may have a significant impact on the radial velocities; thus, we have manually increased the uncertainty of the orbital solution by a factor of ten to reflect this concern. Clearly, the errors in the derived quantities shown in Table~\ref{tab:OrbSol} are correspondingly increased.

The orbital solution offers a straightforward means of estimating the possible sizes of the circumstellar structures surrounding each component. {Independently, the light curve constrains the extent of the attenuating material. This allows for a direct comparison between the two approaches.} Assuming the attenuated star as a point source, the size of the disk of star $j$, $R_d^j$, responsible for an attenuation from phases $\phi_0$ to $\phi_1$ can be estimated by
\begin{equation}
    2 R_d^j = (K_1 + K_2) P \sin{i} \int_{\phi_0}^{\phi_1} \cos \left(2\pi \phi \right) d\phi\,. 
\end{equation}
The first attenuation lasts about 1/5 of the orbital period, from phases $-$0.09 to 0.11, yielding a size of $\sim 42.6 \, \rm R_{\odot}$. 
The second attenuation lasts for almost 2/5 of the orbital period, from phases 0.33 to 0.67, resulting in a size of the attenuating region of $\sim 63.7 \, \rm R_{\odot}$. On the other hand, using Eq.~2 from \citet{Eggleton1983Roche} to estimate the size of the Roche lobe, we find that the Roche lobe radii of the Be star and its companion are approximately $41.0~\rm R_{\odot}$ and $15.8~\rm R_{\odot}$, respectively.
{In the case of the Be star, the disk extent derived from the light curve and inferred from the Roche lobe geometry} are consistent, suggesting that the disk---likely filling {the Roche Lobe of the Be star}---is responsible for the first attenuation. In contrast, the maximum possible size of the companion’s disk, as limited by its Roche lobe, is far too small to account for the second attenuation (provided that the Be star is the one being attenuated). Combined with the absence of absorption signatures attributable to the companion's disk in the spectra (Sect.~\ref{sec:absorption_lines}), this strongly suggests that any circumstellar structure around the companion is of low density and does not contribute significantly to the light curve. Nevertheless, emission-line spectroscopy confirms the presence of such a structure, indicating that it exists but has limited optical depth.

\subsection{Polarization}


{Panels F and G of Fig.~\ref{fig:masterplot} show the orbital modulation of the intrinsic polarization level and angle. Within the uncertainties, the latter remains constant, with mean values for each band of $\theta_{B} = 1.9 (1.2) ^{\circ}$, $\theta_{V} = 2.7 (1.6) ^{\circ}$, $\theta_{R} = 1.9 (1.8) ^{\circ}$ and $\theta_{I} = -2.3 (1.7) ^{\circ}$, all consistent within $\sim2\sigma$. This result indicates that the plane of the disk (the source of polarization) is predominantly coplanar with the orbital plane.}{The behavior of the intrinsic polarization level, however, is more complex and is better understood with the help of models (see Sect.~\ref{sec:mod_sol}.} 

\section{A Three-Component Model}
\label{sec:Modeling}

In the previous section, it was outlined that \VCAR is an eclipsing Be binary, where a disk is present around the Be star and a low-density structure surrounds the {companion}. Additionally, the disk of the Be star likely extends to the companion, feeding this circumsecondary structure.

{These elements are qualitatively similar to what is found in SPH simulations {of binary Be stars}. 
In the model shown in Fig.~\ref{fig:SPH}, the disk around the Be star shows clear indications of the perturbations induced by the companion in the form of two strong spiral density arms. The main arm \citep{Cyr2020} extends up to the Roche lobe of the companion, a structure referred to as the ``bridge'' by \citet{rubio2025}. Material flows from the Be star's disk to the companion through this bridge, feeding a circumsecondary structure. The properties of this structure are discussed in detail by \citeauthor{rubio2025}; in general terms, it can be either rotationally dominated (e.g., for models with low viscosity) or best described by a radial outflow (as in the case of high-viscosity models, see their Fig.~11).}

{Fully describing a complex system such as the one depicted in Fig.~\ref{fig:SPH} requires accounting for intricate geometries and kinematics, disk thermodynamics, and the non-spherical shape of the stars involved. Additionally, in the case of an edge-on configuration, the modeling must also consider both eclipses and attenuation by and within the circumstellar structure.}
{There is no existing tool capable of fully addressing all these requirements}.
{Thus, we} developed a simplified {model} that does not account for the material around the companion and adopts a much simpler prescription for the Be star disk. This strategy serves as a proof of concept, allowing us to extract most of the system's key physical parameters. Support for this simplified model includes
\begin{itemize}
    \item Based on the strength of the \ion{He}{1} {emission} lines {and the lack of \ion{He}{1} traveling shell features} during the second attenuation, it is reasonable to assume that the circumsecondary structure is rather tenuous. Circumstantial evidence supporting this interpretation comes from the analysis of the duration of the secondary attenuation, which suggests it is unlikely to be caused by the circumsecondary material.
    \item Recent high-resolution SPH simulations of binary Be stars \citep{rubio2025}, particularly models with parameters closely matching those of \VCAR, indicate that the circumsecondary structure is physically much smaller than the {Be star}’s disk. Moreover, the simulations show that the density in the circumsecondary region is at least an order of magnitude lower than that in the {Be star} disk. 
\end{itemize}

To build the three-component model outlined above, we combine the radiative transfer code HDUST with a newly-developed ray-tracing (RT) routine. Each code contributes to part of the solution, as outlined below.

HDUST \citep{carciofi2006,carciofi2008} is a Monte Carlo radiative transfer code that has been extensively used in modeling Be star disks. Given a prescription for the central star (e.g., mass, rotation rate, luminosity, etc.) and for the circumstellar disk (density, geometry, etc.), HDUST calculates the electron temperature and hydrogen level populations in the circumstellar gas and generates the emergent spectrum, including the SED, hydrogen line profiles, polarization, and images. However, the current version of HDUST is limited to simulating only a single star, which poses a challenge when modeling systems like \VCAR, where the presence of another stellar component might cause relevant effects.

The RT routine was developed specifically to allow for the inclusion of an additional source of radiation---the companion star. It takes the HDUST simulation as input, incorporating the properties of the Be star and its disk. It is assumed that the companion’s orbit lies in the equatorial plane of the {Be} star ({i.e.}, the orbit is not tilted). The HDUST model is essential for providing the disk opacities, which are then used to calculate the attenuation of the {companion stellar} light as it passes through the disk. It should be noted that the RT code is capable of addressing only the continuum absorption by the disk, without accounting for line emission/absorption, continuum emission, or electron scattering. The latter processes are accounted for by HDUST. 

To integrate these codes, we developed a modeling framework that couples HDUST with the RT routine. The observed {simulated} spectrum at a given epoch, $F_\lambda(t)$, is given by the following expression:
\begin{equation}\label{eq:all_flux}
    F_\lambda(t) = F_\lambda^1 A_\lambda^{\rm s1}(t) + (F_\lambda^{\rm e} + F_\lambda^{\rm s}) A_\lambda^{\rm d} (t) + F_\lambda^2 A_\lambda^{\rm s2}(t)\,.
\end{equation}
Here, the first term corresponds to the Be star spectrum, $F_\lambda^1$, attenuated by the disk 
or eclipsed by the companion, $A_\lambda^{\rm s1}(t)$.
The second term represents the contribution of the light scattered off and emitted by the disk ($F_\lambda^{\rm s}$ and $F_\lambda^{\rm e}$, respectively), 
{partially eclipsed by the stars,} $A_\lambda^{\rm d}(t)$.
Finally, the third term accounts for the companion's spectrum, $F_\lambda^2$, attenuated by the 
{disk or eclipsed by the Be star}, $A_\lambda^{\rm s2}(t)$.
All attenuation terms are time-dependent, as they vary with the positions of the stars relative to the observer.
To represent the spectrum of each star, we use the database from \citet{Coelho2014spectra} for the Be star and \citet{Lanz2007Tlusty} for the companion star.
To compare with the TESS data, a lightcurve, {LC($t$)}, is constructed as a function of phase by convolving the model flux with the TESS bandpass:
\begin{equation}
    \mathrm{LC}(t) = F_\lambda(t) \ast T_{\rm TESS}\,.
\end{equation}

The polarization arises from light scattering off the Be star disk. To account for the depolarizing effect of the companion, the model polarization, {$P_\lambda(t)$}, is calculated as
\begin{equation}
    P_\lambda(t) = P_\lambda^{1} \frac{F_\lambda^1 + F_\lambda^{\rm e} + F_\lambda^{\rm s}}{F_\lambda(t)} A_\lambda^{\rm d}(t)\,,
\end{equation}
where $P_\lambda^{1}$ is the polarization calculated by HDUST for the Be star and its disk.

\begin{figure*}
    \centering
    \includegraphics[width=\linewidth]{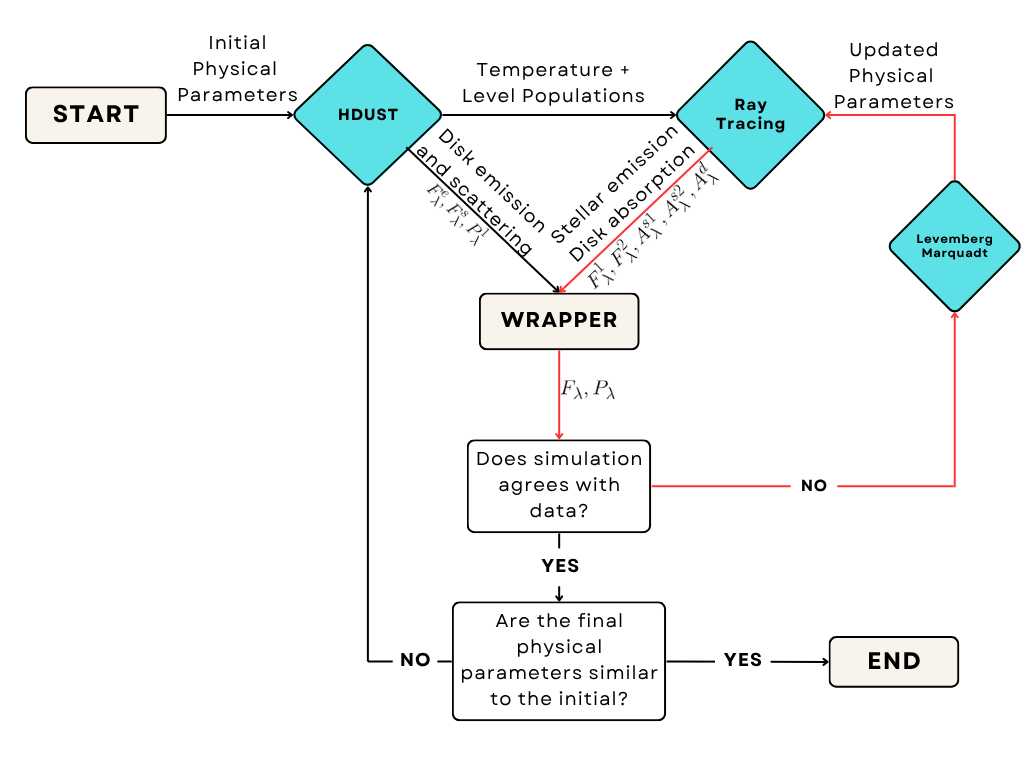}
    \caption{Flowchart illustrating the modeling approach described in Section~\ref{sec:Modeling}. Red lines indicate the inner loop, while the entire diagram represents the outer loop. See text for details.}
    \label{fig:Schematic}
\end{figure*}

\subsection{Model components}\label{sec:vdd_model}

The viscous decretion disk (VDD) model, originally proposed by \citet{lee1991VDD}, has inspired numerous implementations {for Be stars} over the years. {At its core, it describes a disk in which matter is ejected by the central star, and shear viscosity transports angular momentum outward, causing the disk to grow.} Hydrodynamical simulations, such as those by \citet{okazaki2002} and \citet{haubois2012buildup}, consistently demonstrate that during phases of disk build-up and dissipation, the disk develops a complex radial density structure that deviates significantly from a simple steady-state profile. In the idealized case of a disk that is both isothermal and fed at a constant rate for a sufficiently long time, the density profile asymptotically approaches a radial power law with index $-3.5$. Although such a steady-state configuration is never fully realized—even in theoretical models—a common approximation in the literature is to represent the radial density profile, {$\rho(r)$}, as a power law with a variable slope:
\begin{equation}
    \rho(r) = \frac{\rho_0}{H(r)} \left( \frac{r}{R_{\rm eq}} \right )^{-m}\,,
\end{equation}
where $m$ is the disk density slope, $\rho_0$ is the numerical density at the base of the disk and $R_{\rm eq}$ is the equatorial radius of the Be star. The scale height, {$H(r)$}, is also usually approximated as a power-law:
\begin{equation}
    H(r) = H_0 \left( \frac{r}{R_{\rm eq}} \right )^{\beta}\,,
\end{equation}
where $\beta$ is the flaring parameter and $H_0$ is the scale height at the base of the disk. For a disk in steady-state and isothermal \citep{bjorkman2005structure}, $m=3.5$, $\beta = 1.5 $ and ${H_0} = c_{\rm s} {R_{\rm eq}/V_{\rm orb}}$, where $V_{\rm orb}$ is the orbital velocity at the stellar equator, $c_{\rm s}$ is the sound speed, calculated with a temperature of 72 \% of the effective temperature \citep{carciofi2008}.
In our simulations, we adopted two values for the slope $m$: the ideal value of 3.5, and a smaller value of 3.0. The latter was motivated by the simulations of \citet{panoglou2016SPH}, who found that in short-period binary systems, such as \VCAR, the disk's density profile tends to exhibit a shallower slope. This effect, referred to by {these} authors as the ``mass accumulation effect,'' arises from the gravitational influence of the companion, which inhibits the outward transport of material in the disk. The only free parameter in this implementation is $\rho_0$. {As discussed in Sec.~\ref{sec:orb_sol}, the TESS light curve is consistent with a Be disk truncated at the Roche lobe of the star. Accordingly, we fixed the outer radius of the disk, $R_{\rm disk}$, at $41\rm \, R_{\odot}$}. 

The Be star was modeled as an oblate object, characterized by four input parameters: mass ($M$), rotation parameter ($W$), polar temperature ($T_{\rm p}$), and polar radius ($R_{\rm p}$). From these, we derived the rotational velocity ($V_{\rm rot}$), stellar luminosity ($L$), and equatorial radius ($R_{\rm eq}$), {as well as the equatorial temperature ($T_{\rm eq}$), based on the von Zeipel law \citep{vonZeipel1924} and adopting an exponent of 0.2 \citep{Espinosa2011gravity}}. The parameters $T_{\rm p}$ and $R_{\rm p}$ were treated as free parameters, while $M$ was fixed at 4.45~$\rm M_{\odot}$, based on the orbital solution (Sec.~\ref{sec:orb_sol}). The $W$ was fixed at 0.9, following the result of \citet{Huang2010wparameter}, who found that the fastest rotating diskless B star in the 4–5~$\rm M_{\odot}$ range had $W = 0.91$, suggesting a threshold above which all B stars form a disk.
The companion was modeled using the same parameters, but treated as a spherical object (i.e., the rotation velocity is very small) with a fixed mass of $M = 0.53~\rm M_{\odot}$.

In addition to the intrinsic parameters described above, we also adopted two free parameters essential for defining the model's SED: the distance and the reddening, $E(B - V)$. 
In Appendix~\ref{Ap:Pol}, we describe the measurement of the interstellar polarization, which is characterized by a maximum polarization level of $P_{\rm max} = 0.61(4)~\%$ and a polarization angle of $\theta_{\rm IS} = 138.4(1.5)^{\circ}$. However, due to the specific intrinsic polarization angle of \VCAR, even small variations in $\theta_{\rm IS}$ (of just a few degrees) can lead to significant differences in the derived intrinsic polarization. Therefore, we treated the polarization angle as a free parameter within $3\sigma$ of the measured value, while keeping $P_{\rm max}$ fixed.

\subsection{Fitting Procedure}\label{sec:fitting_procedure}

Figure~\ref{fig:Schematic} provides {an overview} of the adopted modeling procedure. We begin with an initial set of model parameters, {based on the stellar masses}. The parameters explored in the model are listed in Table~\ref{tab:OrbSol}. Given a set of parameters, the outer loop in Fig.~\ref{fig:Schematic} is initiated by first calculating the disk temperature, level populations, and emergent flux with HDUST and feeding the output to the RT code. The outputs from HDUST and the RT code are combined (wrapper box in Figure~\ref{fig:Schematic}) to produce the model lightcurve, polarization, and spectrum.

An interactive minimization loop is started (inner loop in Fig.~\ref{fig:Schematic}) in which the following sets of observables {are} compared to the data: UV spectrum, optical spectrum, IR spectrum, TESS light-curve, polarization and the EW of \Ha. {To properly account for the second eclipse, we {have} to manually remove the second attenuation using a gaussian fit, as the latter was not reproduced by the model}. Instead of using the standard $\chi^2$ as the merit function, we adopt a modified definition that offers the distinct advantage of normalizing the $\chi^2$ contribution from each observable. This approach accounts for the heterogeneous number of data points among the different datasets, ensuring a more balanced evaluation of the fit quality:
\begin{equation}
    \chi^2 = \left ( \sum^{\rm obs} \frac{\chi^2_{\rm obs}}{N_{\rm obs}}\right )
    \cdot \left( \sum^{\rm obs} N_{\rm obs} \right )\,,
\end{equation}
where $\chi^2_{\rm obs}$ and $N_{\rm obs}$ represent the $\chi^2$ and number of data points of each observable. In the inner loop, the Levenberg–Marquardt method is used to iteratively adjust the input parameters of the RT code in order to optimize the $\chi^2$. Once convergence is achieved, the resulting parameter set is passed to HDUST, triggering a new iteration of the outer loop. This outer loop continues until overall convergence is reached. This hybrid scheme, comprising nested inner and outer loops, was adopted because HDUST is significantly more computationally intensive than the RT code. By restricting the computationally cheaper RT code to the inner loop, the optimization process becomes substantially more efficient. {Given the combined convergence of the inner and outer loop, this approach is fully self-consistent within the current possibilities of both codes.}

{Absorption lines were not included in this procedure due to limitations in the current codes. The Be star’s spectral lines are heavily affected by disk absorption, and HDUST is currently unable to reproduce shell lines for non-hydrogen atoms. For the companion, accurate modeling would require tuning its chemical composition, as it is expected to be helium-enhanced, an effect not accounted for in the atmosphere models of \citet{Lanz2007Tlusty}. However, once the model converged, we estimated $V_{\rm rot}$ of the companion by applying the rotational broadening function from \citet{pyastronomy} to match the average absorption profile of the \ion{He}{1}~5875~\AA\ line.}

\section{Results}\label{sec:mod_sol}

\begin{figure*}
    \centering
    \includegraphics[width=\linewidth]{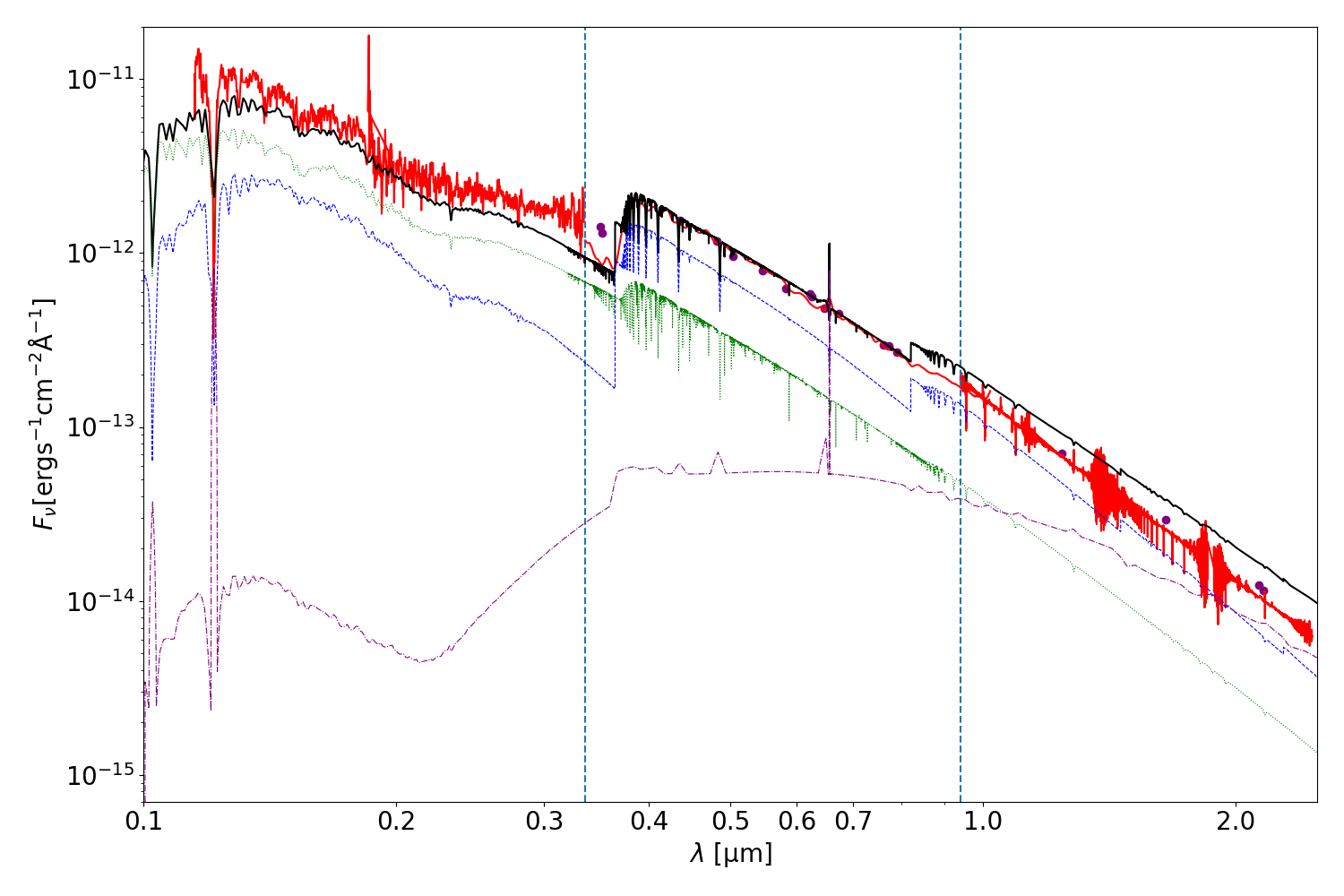}
    \caption{SED of \VCAR. The observations (UV from IUE, optical from GDR3 and IR from SOAR) were merged and presented in red to facilitate the visualization. Blue vertical dotted line separates the different sources. 
    The black line represents the solution obtained. 
    The other lines represent each component of the model: Be star in blue; companion in green; the sum of disk emission and scattering in purple.}
    \label{fig:Spec}
\end{figure*}

The best-fitting parameters are listed in Table~\ref{tab:OrbSol}. {As mentioned in Sect.~\ref{sec:vdd_model}, we tested two different values of $m$. However, since the final parameters and modeled observables were nearly identical, we focus our analysis on the ideal case of $m = 3.5$, for the sake of simplicity.}

We find that the Be star has a polar radius of $2.40(1)~\rm R_{\odot}$, consistent with the mass estimated in Sect.~\ref{sec:orb_sol} \citep{Andersen1991Binaries}. Given a polar temperature of $18.0(5)~\rm kK$, the resulting luminosity is $347(40)~\rm L_{\odot}$.
The companion star, in contrast, is significantly smaller, $0.97(1)~\rm R_{\odot}$, and hotter, $21.0(5)~\rm kK$. Surprisingly, despite its compact size, its luminosity is {still more than half} that of the Be star, consistent with the fact that \VCAR is an SB2. 
The duration and depth of the eclipses place tight constraints on the radii of both stars, providing a robust validation of the model's stellar sizes.
The rotation rate of the Be star was not included as a free parameter in our analysis. However, the best fit to the companion suggests it is a slow rotator, with $V_{\rm rot} = 23(1)\rm \,km\,s^{-1}$.
The full set of data fitted placed quite stringent values on the inclination angle of the system, found to be $88.6 (1)^{\circ}$.

{The modeled disk is composed primarily of ionized hydrogen, with a mean temperature of approximately $10~\mathrm{kK}$. Still, it is dense enough, with $\rho_0 = 6.01(4) \times 10^{-11}~\rm g~cm^{-3}$ to be radially very optically thick, reaching optical depths up to ${\tau \approx 10^{4}}$ for the TESS bandpass. {As can be seen in Fig.~8 of \citet{Vieira2017VDD}, the obtained $\rho_0$ value is higher than what is typical for similar stars.} Figure~\ref{fig:animation} offers a complete view of the disk optical depth structure in the line of sight. While Be stars exhibit intrinsic polarization (a feature not included in our model) due to their oblate shape, the scattering by free electrons is the dominant contribution to the observed polarization.}

The fit to the TESS light curve, \Ha EW and polarization is shown in Fig.~\ref{fig:masterplot}.
The model performs remarkably well in reproducing the main features of the first eclipse and attenuation. The depths of the eclipse offer tight constraints on the effective temperature of both stars. 
In particular, the model {nicely} reproduces the rise (decrease) in brightness before (after) the stellar eclipse (the W-shaped curve discussed in Sect.~\ref{sec:TESS}). This is a consequence of: the flaring of the disk, which makes the inner disk thinner, and a non-perfect edge-on orientation. When both factors are considered, the observed light from the companion star, as it nears phase zero, passes through progressively lower density regions ({See} Appendix~\ref{Ap:animation} for further insight).

The model successfully reproduces the duration of the first eclipse. As discussed previously in Sect.~\ref{sec:orb_sol}, this fit indicates that the disk size was well estimated by the Roche radius of the Be star. 
Another key feature of the model is that it reproduces well the depth of the attenuation, providing a strong constraint on the disk density: the overall column density along the line of sight is accurately captured by our simulations.
The only aspect of the first attenuation not reproduced by the model is the duration and shape of the egress. The modeled attenuation is symmetric about the center of the first eclipse, as expected from an axisymmetric disk. The failure to fit the observed egress suggests that the disk around \VCAR deviates from axial symmetry.

The second attenuation cannot be reproduced by our model, as the only potential source of eclipsing or attenuation during these phases, the companion star, is far too small to cause any significant variations in the light curve. On the other hand, the depth and shape of the second eclipse are well reproduced (see inset of panel A of Fig.~\ref{fig:masterplot}).
%

{The polarization level in the V and R bands, where most observations were obtained, is also well reproduced by the model.} Polarization depends on the system's geometry and the total number of scatterers \citep{Brown1977}. Given that the inclination angle is well constrained, the overall polarization level provides a strong diagnostic of the disk density. When combined with the depth {in the light curve} of the first attenuation, these two independent observables offer complementary constraints on the same physical quantity.
{The model, however, predicts polarization levels in the B and I bands that are higher than observed. The former should be taken with caution, as the data are highly scattered. The latter is more puzzling, with the I-band level exceeding those in the V and R bands, an unexpected result in standard Be star models. As will be discussed later in the context of the SED, similar discrepancies occur in the NIR region and may share a common origin.}

{The main modulations in the polarization level observed throughout the orbital cycle can be understood using Eq.~\ref{eq:all_flux}, which accounts for all flux components: the direct flux from the Be star and its companion, the disk emission, and the scattered flux. The latter is particularly important as the only significant source of polarization. Thus, any modulation in polarization is likely caused by changes in either the polarized (scattered) flux or the unpolarized flux (the other components).
For example, a clear increase in polarization is observed during the second eclipse, which can be interpreted as the eclipse of an unpolarized source (direct flux from the Be star). The first attenuation is more complex: an overall increase in polarization is also observed, which could similarly be attributed to the attenuation of another unpolarized source (companion star). However, the data are more scattered in this phase, and no well-defined trend is evident.
Notably, during the first eclipse, the polarization level is actually lower than during the surrounding attenuation. This counterintuitive result could only be explained if, alongside the eclipse of the unpolarized source (the companion), which would typically lead to an increase in polarization, the scattered flux is also being attenuated.
Except for this latter case, the model qualitatively reproduces the main observed modulations. It is important to emphasize, however, that the model accounts only for scattered flux originating from the Be star. The companion, though, may also contribute significantly to the polarized flux, which could be attenuated during the first eclipse.}

%

%
Figure~\ref{fig:Spec} shows the fit to the SED data, from the UV to the NIR{, during the flat portion of the light-curve}. The UV and optical ranges are mostly dominated by stellar emission (attenuated by the disk)---see green and blue lines in Fig.~\ref{fig:Spec}---and the fit to the data in these spectral regions is reasonably good. The UV region is completely dominated by the emission from the companion star, which is, on average, 3.5 times brighter than the Be star. This dominance is typically how stripped companions are detected in Be binary systems. Remarkably—and quite unusually—the companion still contributes as much as 31\% of the total flux in the optical range of the SED.
The NIR portion of the SED, however, is not well reproduced by our model. Specifically, the simulations predict an IR excess that is {around 30\%} higher than what is observed. Within the framework of our current model, reconciling this discrepancy would require reducing the disk density, as the IR emission is primarily driven by free-free and bound-free processes in the disk material. However, lowering the disk density would directly impact other key observables that are well matched by the model, most notably, the depth of the first attenuation and the overall polarization level, both of which are sensitive to the amount of material in the disk. This suggests that the IR mismatch may arise from limitations in the underlying assumptions of the model and places the IR SED as a challenge to it.

The SED is a key constraint on the values for the distance and extinction and both can be contrasted to the values measured by Gaia. {According to \citet{bailerjones2021}, using data from the {GDR3}, the photogeometric distance is $d = 0.80^{+0.20}_{-0.13} \rm \, kpc$ and the geometric distance of $d = 2.1^{+1.4}_{-1.2} \rm \, kpc$}. Although the inferred distances differ significantly, the large uncertainties associated with the geometric estimate render the values mutually consistent, as well as compatible with our modeling results. While this agreement is encouraging, it must be interpreted with caution. As discussed by \citet{Lindegren2021gaia}, reliable parallax solutions typically require a Re-normalized Unit Weight Error (RUWE) below 1.4. In unresolved binary systems, the orbital motion and associated photocenter shifts introduce systematics that degrade the astrometric solution, leading to inflated RUWE values and unreliable parallaxes. In such cases, elevated RUWE values can even serve as indicators of binarity rather than simply being poor-quality data \citep{kervella2022gaia}. For \VCAR, the RUWE value is 14.4, strongly suggesting that the Gaia astrometry is affected by the binary nature of the system.

GDR3 also lists the extinction of \VCAR as ${A_G = 0.145 (3) \, \rm mag}$ in the G band \citep[defined in][]{gaia2021rpbp}. However, the measured extinction for a single system is highly dependent on the object's characteristics, which Gaia cannot properly account for \citep[e.g., unresolved binaries, ][]{kervella2022gaia}. We then employed a more robust method that considers the mean extinction in the field. We selected stars within 10\% of the measured parallax from GDR3, a maximum of $0.5^{\circ}$ of spatial separation and a maximum 10\% parallax uncertainty. The result for the mean reddening was $A_G = 0.20 (2) \, \rm mag$. Following \citet{wang2019gaia}, this represents a reddening of $E(B-V) = 0.08 (1)\, \rm mag$. The model value is within $2\sigma$ from Gaia's reddening, which suggests our analysis was able to properly constrain this quantity.

Among all observables considered, the observed EW of \Ha posed the greatest challenge for the model. The base level, corresponding to the flat part of the light-curves (phases around 0.25 and 0.75) is not well reproduced: the observed values are about $\rm EW = - 8.3$~\AA, however, the model predicts $\rm EW = - 7.0$~\AA. These values suggest that the density of the model should be increased, which contradicts the IR SED. However, the greatest challenge is posed by the temporal variation of the EW during the first attenuation: the data indicate that the EW decreases (in absolute value) sharply while the model predicts an \emph{increase} during the same phase.
{On the other hand}, the behavior during the eclipse itself is qualitatively reproduced: both the observations and the model show a rise in EW, likely driven by the selective attenuation of the stellar continuum of the companion. This agreement lends some confidence to the treatment of the radiative transfer effects during the eclipse, but the pre- and post-eclipse mismatch highlights the need for further refinement to the model.

\section{Discussion}\label{sec:final_discussion}

The scenario presented in the previous section, in which the model was directly confronted with a wide range of observational diagnostics, reveals a mixed picture: one of notable successes alongside meaningful discrepancies. Among the main achievements, we highlight its ability to correctly reproduce the broad first attenuation, accurately fit both stellar eclipses, and yield a polarization level {and behavior during the first attenuation and second eclipse that are} consistent with observations. Additionally, the model provides a good match to the ultraviolet and optical portions of the SED and captures the baseline value of the \Ha equivalent width reasonably well. {These results suggest that we were able to robustly reproduce the observations governed by the stars and the inner disk structure}. On the other hand, several mismatches emerge. The model overpredicts the NIR flux, fails to reproduce the expected polarization during the first {eclipse}, does not account for the second attenuation event, and predicts temporal variations of the \Ha equivalent width that deviate significantly from the observed behavior. These issues point to limitations in some of the simplifying assumptions adopted in the modeling framework and will be further explored in the following discussion.

One important conclusion that can be drawn from the successes of our model is the broad validation of the proposed scenario. We have demonstrated, with strong observational and modeling support, that \VCAR is a classical Be star surrounded by a disk, and orbited by a hot subdwarf (sdB) companion. This makes it only the second known Be+sdB system, following the recent identification of the first such system by \citet{Klement2022dynamical}. The fact that \VCAR is an SB2 system has allowed us to determine the orbital parameters and stellar masses with high precision. Furthermore, the added advantage of being the first known Be eclipsing binary enabled us to place tight constraints on the radii of both stars. These results provide critical empirical input for future models of binary evolution, and open a promising pathway for testing theories of mass transfer and angular momentum evolution in post-interaction systems{, once additional eclipsing Be systems are identified}.

Nonetheless, the limitations of our model cannot be overlooked, and are likely rooted in structural simplifications. First, recent studies \citep{rubio2025, panoglou2016SPH, panoglou2018discs, suffak2024} have shown that the disks of Be stars in binary systems are often strongly elongated and exhibit spiral arm structures, as seen in Fig.~\ref{fig:SPH}. This finding is consistent with our interpretation of the observed asymmetry in the first attenuation event, which suggests that the disk around the {Be star} is not axisymmetric. \citet{rubio2025} also proposed that mass transfer from the Be disk to the companion occurs primarily through a dominant spiral arm, forming a bridge-like structure that feeds the circumsecondary material. The authors also discuss the properties of the circumsecondary structure, which, depending on the model parameters, may be either outflowing or rotationally supported (see Fig.~11 in \citet{rubio2025}).

Our model and data provide evidence for some of the structures discussed in \citet{rubio2025}. We find clear evidence of a structure around the companion, supported by weak emission peaks around \ion{He}{1} lines, present in the optical lines but much stronger in the IR (Fig.~\ref{fig:Dynamical}). The lack of shell features during the second attenuation indicates that this structure is rather tenuous. The triple-Gaussian model presented in Sect.~\ref{sec:helium_emission} indicates that the \ion{Fe}{2}~9997~\AA\ line consists of a roughly double-peaked structure (see red dashed circle in Fig.~\ref{fig:SPH}), matching the radial motion of the Be star, with a PS of $316 (15)~\rm{km~s^{-1}}$, plus a traveling peak with a radial velocity of $138 (5)~\rm km~s^{-1}$, larger than that of the companion star ($102.1 (2) ~\rm km~s^{-1}$), and a phase shift of $-0.020(1)$. This suggests that the feature is generated at a radius closer to the star than the companion and leads it by about {$7.2(4)^\circ$} (see green dot in Fig.~\ref{fig:SPH}). Furthermore, the radial velocity of this structure is slightly smaller than the peak separation of the double-peaked component, suggesting that the latter originates at a slightly smaller radius. Taken together, these facts strongly suggest that the extra emission in the \ion{Fe}{2} lines likely originates from the bridge. Both detections---the bridge and the circumsecondary structure---pave the way for future studies exploring the detailed structure of Be disks in binary systems.

Another important limitation of the current model is the treatment of radiative transfer: only the Be star is considered as a source of radiation. This assumption becomes problematic in the case of \VCAR, where the companion is not only comparable in optical brightness, but also emits a strong UV flux. The presence of such a UV-bright source can significantly alter the ionization and thermal structure of the disk region facing the companion. This, in turn, may affect not only the \Ha emission but also the observed polarization. Future modeling efforts will need to account for radiative feedback from both stars in order to properly capture these effects and improve the accuracy of the simulations.

{In this paper, we were unable to model the absorption lines for two main reasons: the lack of disk models that include non-hydrogen atoms, and the absence of stellar atmosphere models that account for the anomalous composition of the companion, which is expected to be helium-enriched. Future modeling efforts that address these limitations will provide deeper insights into the structure surrounding the companion and yield more accurate estimates of its temperature and evolutionary state. A similar limitation applies to the iron lines from the Be star (see Sect.~\ref{sec:absorption_lines}), which exhibit traveling features in the optical spectrum as a result of the companion’s light being partially absorbed by the disk. These features, evident in Fig.~\ref{fig:Iron_phase}, can be used to map the local radial velocities of the disk regions responsible for absorbing the companion’s radiation. A more in-depth analysis of this effect could, in principle, constrain the density, kinematics and temperature of the disk.
}

{The {\ion{Fe}{2}~9997~\AA\ line is} also of interest, as it shows a traveling emission feature associated with the companion (see Sect.~\ref{sec:helium_emission}), but with higher velocity amplitude. By combining disk models that include iron with external illumination and SPH simulations, we can investigate the precise location of this localized emission. The velocity amplitude and orbital behavior suggest that it originates in the outer disk of the Be star, in the direction of the companion.}

{In a recent study, \citet{deamorim2025} analyzed the photometric behavior of \VCAR {the same model as employed here}. The authors identified an interesting pattern: their model successfully reproduced, within uncertainties, the $\rm{B} - \rm{V}$ color and V-band magnitude during both the flat portion of the light curve and the first stellar eclipse (see their Fig.~2). However, their model also predicted that the system becomes bluer during the disk attenuation, whereas the observations show that it becomes redder. A bluer color during disk attenuation is, in fact, the expected outcome from the model, as bound-free absorption is stronger at longer wavelengths, causing greater attenuation of redder light. This discrepancy poses yet another challenge to the disk assumptions adopted in our model.}

\subsection{{V658~Car in the Be + sdOB context}} \label{sec:survey}

\begin{figure}[t]
    \centering
    \includegraphics[width=\linewidth]{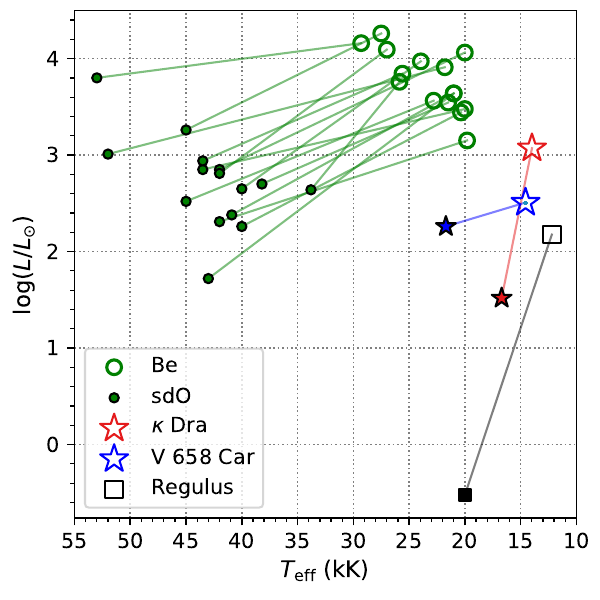}
    \caption{The luminosity and temperature of 15 known Be+sdO binaries \citep{Mourard2015phiper,Gies1998phiper,wang2021detection,klement2024interfsdo,Peters2016hr2142,chojnowski2018hd55606,Peters2008fycma,Klement2022dynamical,Peters201359cyg,Wang201760cyg} are plotted as open (Be) and filled (sdO) circles, with a line connecting the two components of each binary system. The Regulus system is plotted in a similar fashion with black squares, and $\kappa$ Draconis is plotted with red stars. This figure was updated from \citet{Klement2022dynamical} to contain the solution (Tab.~\ref{tab:OrbSol}) of \VCAR in blue.
    }
    \label{fig:HRknown}
\end{figure}

The first subdwarf companion of a Be star, $\phi$ Per, was identified decades ago by \citet{phiper1981}. However, as recently as ten years ago, only two such systems, $\phi$ Per and 59 Cyg \citep{Peters201359cyg}, had been confirmed. These binaries are understood to be in a post-mass-transfer evolutionary phase, where the stripped companion is visible, but the rapidly rotating Be star has not yet evolved significantly away from the main sequence \citep{vanbever1997number, shao2014formation}. Over the past decade, thanks largely to ultraviolet spectroscopy, a growing number of Be + sdOB systems have been identified. Notably, \citet{wang2021detection} alone reported nine new early-type Be + sdOB binaries. To date, only two such systems have been found with a lower-mass Be primary: $\kappa$ Draconis, consisting of a B6e star and a hot subdwarf companion \citep{Klement2022dynamical}, and \VCAR. A few related analogs have also been studied, such as the Regulus system---comprising a rapidly rotating B8 star and a “pre-white dwarf” companion \citep{gies2020spectroscopic, rappaport2009past}, and several Be/X-ray binaries. These latter systems represent a more evolved phase, in which the companion is a neutron star formed via a supernova explosion \citep{Rappaport1982xray}. Figure~\ref{fig:HRknown} summarizes this evolutionary landscape, highlighting the unique position occupied by \VCAR.

Compared to other Be + sdOB systems, \VCAR holds particular significance due to the presence of eclipses. In this paper, we emphasize how crucial these eclipses are for understanding the interaction between the two stars and the circumstellar disk, a key factor in identifying and characterizing additional Be binaries. Remarkably, \VCAR remains the only known eclipsing Be + sdOB system. Given its high photometric precision and sky coverage, the TESS mission may be a powerful tool for discovering more such systems. 

\section{Conclusions}

We present the most extensive observational dataset and modeling effort to date for \VCAR, the only known eclipsing binary system with a Be star and one of the few known SB2 systems.
By applying a three-component model, comprising an oblate central star, a symmetric disk, and a spherical companion, we were able to constrain most of the fundamental parameters of the system. 

The main robust conclusions drawn from this study are as follows:

\begin{itemize}
  \item \VCAR is the first confirmed eclipsing Be + sdOB binary system and, as such, has the one of the most accurately determined parameters among this class of objects.

  \item The system is one of the few known Be stars with a stripped companion that is visible in the optical range.

  \item The system is only the second known case of a late-type Be star with an sdB companion, offering valuable insight into the diversity of evolutionary pathways in interacting binaries.

  \item The three-component model, comprising an oblate central star, a symmetric disk, and a spherical companion, was able to reproduce most observed features, including the stellar eclipses, the first attenuation, and the polarization level {and behavior during the first attenuation and second eclipse}.

  \item The depth and duration of the eclipses provide strong constraints on the stellar radii and effective temperatures, validating the orbital solution and the system’s fundamental parameters.

  \item The Be star's disk appears to fill its Roche lobe and is dense enough to account for the depth of the first attenuation.

  \item The orbital inclination was determined with high precision ($88.6(1)^\circ$), supported by multiple independent diagnostics.

  \item Recent smoothed particle hydrodynamic models of Be star disks in binary systems describe mass transfer from the disk to the companion as occurring through a bridge-like structure, which gives rise to a circumsecondary structure. We provide strong spectroscopic evidence for the presence of both components.
\end{itemize}

Despite the model's broad success in reproducing key observables, important discrepancies remain. Some observables were not well fitted, and a few exhibited qualitative disagreement with the model's predictions. The main limitations are as follows:

\begin{itemize}
  \item {Equivalent width (EW) variability:} The model fails to capture the observed phase-dependent variations of the \Ha {EW}. Notably, it predicts variations during the first attenuation that are at odds with the observations.

  \item {Conflicting density constraints:} There is significant tension between the density inferred from the polarization and the fitting of the first attenuation and that required to match the {near-infrared portion of the SED}. The latter suggests a more tenuous disk than the former predicts.

  \item {Second attenuation:} The model does not reproduce the second attenuation in the TESS light curve, whose origin remains unexplained.

  \item {Disk asymmetries and shell absorption:} Observational evidence points to significant disk asymmetries and transient shell absorption features, particularly in the Fe and H lines. These effects are not accounted for in the model, which assumes an axisymmetric disk.

  \item {Circumsecondary structure:} The tenuous structure around the companion is not explicitly included in the model, despite evidence of its spectroscopic signature. While its contribution to continuum attenuation appears negligible, it should be incorporated in future modeling efforts.
\end{itemize}

Looking ahead, two key advancements are envisioned to improve the modeling framework. First, the inclusion of the companion's radiative impact on the Be disk, particularly heating by photoionization, will allow for a more physically realistic treatment of the disk’s thermal and ionization structure. Second, the adoption of an asymmetric disk geometry, guided by high-resolution SPH simulations, will enable the modeling of complex disk features such as spiral arms, the bridge, and the circumsecondary structure present in the system. Given its unique orientation and wealth of observational diagnostics, \VCAR stands out as a benchmark system for the study of Be star disks and their interactions with compact companions.

\section{Acknowledgments}

We are grateful for the valuable discussions and insights from many colleagues, particularly André Figueiredo, Ilfa Gabitova, Jorick Vink, and Prof. Carol Jones' group.

THA acknowledges support from the `Fundação de Amparo à Pesquisa do Estado de São Paulo' (FAPESP, grant 2021/01891-2) and the `Coordenação de Aperfeiçoamento de Pessoal de Nível Superior' (CAPES, grant 88887.834998/2023-00).
ACC acknowledges support from `Conselho Nacional de Pesquisa' (CNPq, grant 314545/2023-9) and FAPESP (grants 2018/04055-8 and 2019/13354-1).
JLB acknowledges support from the European Union (ERC, MAGNIFY, Project 101126182).
ACS acknowledges support from CAPES (grant 88887.673991/2022-00).
{FN acknowledges support from CNPq (grant 303093/2025-0)}.
PQ acknowledges support from CAPES (grant 8887.820796/2023-00).
ACR acknowledges support from the Max Planck Institute for Astrophysics in Garching, Germany.

Views and opinions expressed are, however, those of the authors only and do not necessarily reflect those of the European Union or the European Research Council. Neither the European Union nor the granting authority can be held responsible for them.

This work required a large amount of computing time and was only made possible with the access to the following computing facilities:
\begin{itemize}
    \item This work made use of the computing facilities of the Centro de Processamento de Dados do IAG/USP (CPD-IAG), whose purchase was made possible by the Brazilian agency FAPESP (grants 2019/25950-8,  2017/24954-4 and 2009/54006-4).
    \item Santos Dumont from the National Laboratory for Scientific Computing (LNCC/MCTI, Brazil).
\end{itemize}

We also want to acknowledge the observing facilities that obtained data for this work and made it possible:
\begin{itemize}
    \item Southern Astrophysical Research (SOAR) telescope, which is a joint project of the Minist\'{e}rio da Ci\^{e}ncia, Tecnologia e Inova\c{c}\~{o}es (MCTI/LNA) do Brasil, the US National Science Foundation’s NOIRLab, the University of North Carolina at Chapel Hill (UNC), and Michigan State University (MSU).
    
    \item Pico dos Dias Observatory (OPD), managed by the National Astrophysics Lab (LNA) part of the Ministry of Science, Technology and Innovation (MCTI) of Brazil.
    
    \item Network of Robotic Echelle Spectrographs (NRES) from the Las Cumbres Observatory (LCO) global telescope network.
    
    \item Bernard Heathcote \Ha observations obtained via the Be star spectrum (BeSS) database, operated at LESIA, Observatoire de Meudon, France: \url{http://basebe.obspm.fr}.
    
    \item European Space Agency (ESA) space mission Gaia. Gaia data are being processed by the Gaia Data Processing and Analysis Consortium (DPAC). Funding for the DPAC is provided by national institutions, in particular the institutions participating in the Gaia MultiLateral Agreement (MLA).

    \item International Ultraviolet Explorer (IUE) satellite, a joint project of the National Aeronautics and Space Administration (NASA), ESA, and the United Kingdom.
\end{itemize}
\vspace{5mm}


\appendix

\section{\texorpdfstring{$\mathrm{H}_\alpha$}{Hα}\ Line center determination}\label{Ap:Ha_center}
\restartappendixnumbering

A commonly-used procedure to measure the radial velocity of the Be star is through the the \Ha wings. We manually select a portion (from the continuum to approximately half of the peak) of the blue side of the \Ha wing and reflect it horizontally over a center $\upsilon_0$ to match the shape on the red side of the profile. The value of $\upsilon_0$ for each spectrum is chosen as the one that minimizes 
\begin{equation}
    \chi^2 = \sum_{k} \left( F_k - F_{\text{\rm inverted}, k} \right)^2 , 
\end{equation}
where $F_k$ represents the red side of the normalized flux value at point $k$, and $F_{\text{\rm inverted}, k}$ corresponds to the inverted blue segment. The comparison is carried out within a restricted velocity range centered on a reference value---in this case, the velocity of the center of mass---in order to avoid undesired fluctuations associated with noise, telluric absorption, or local asymmetries in the profile. This method is a variation of the one introduced by \citet{Shafter1986bisector}, known as the bisector method.

\begin{figure*}
    \centering
    \includegraphics[width=0.45\linewidth]{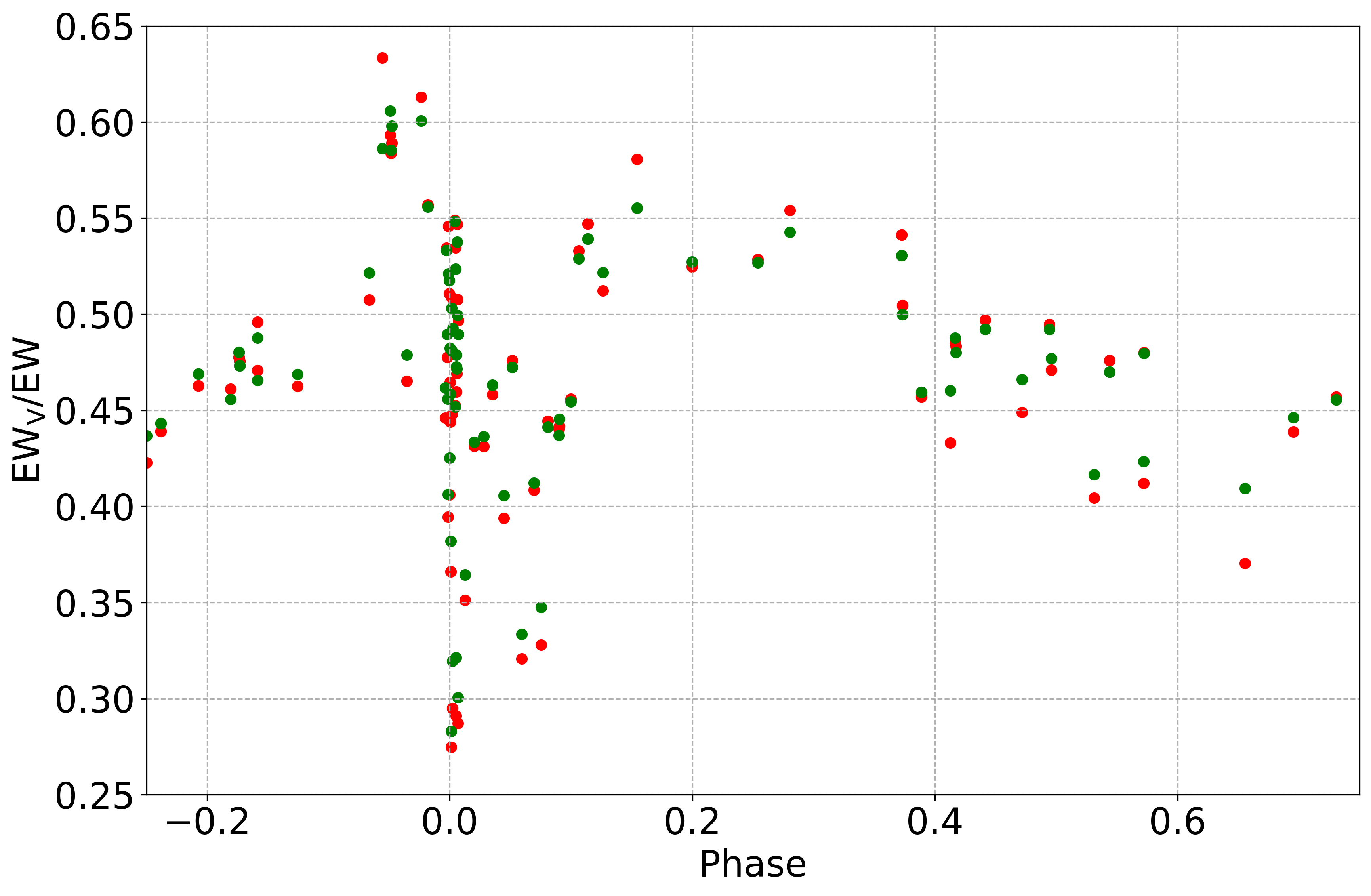}
    \includegraphics[width=0.45\linewidth]{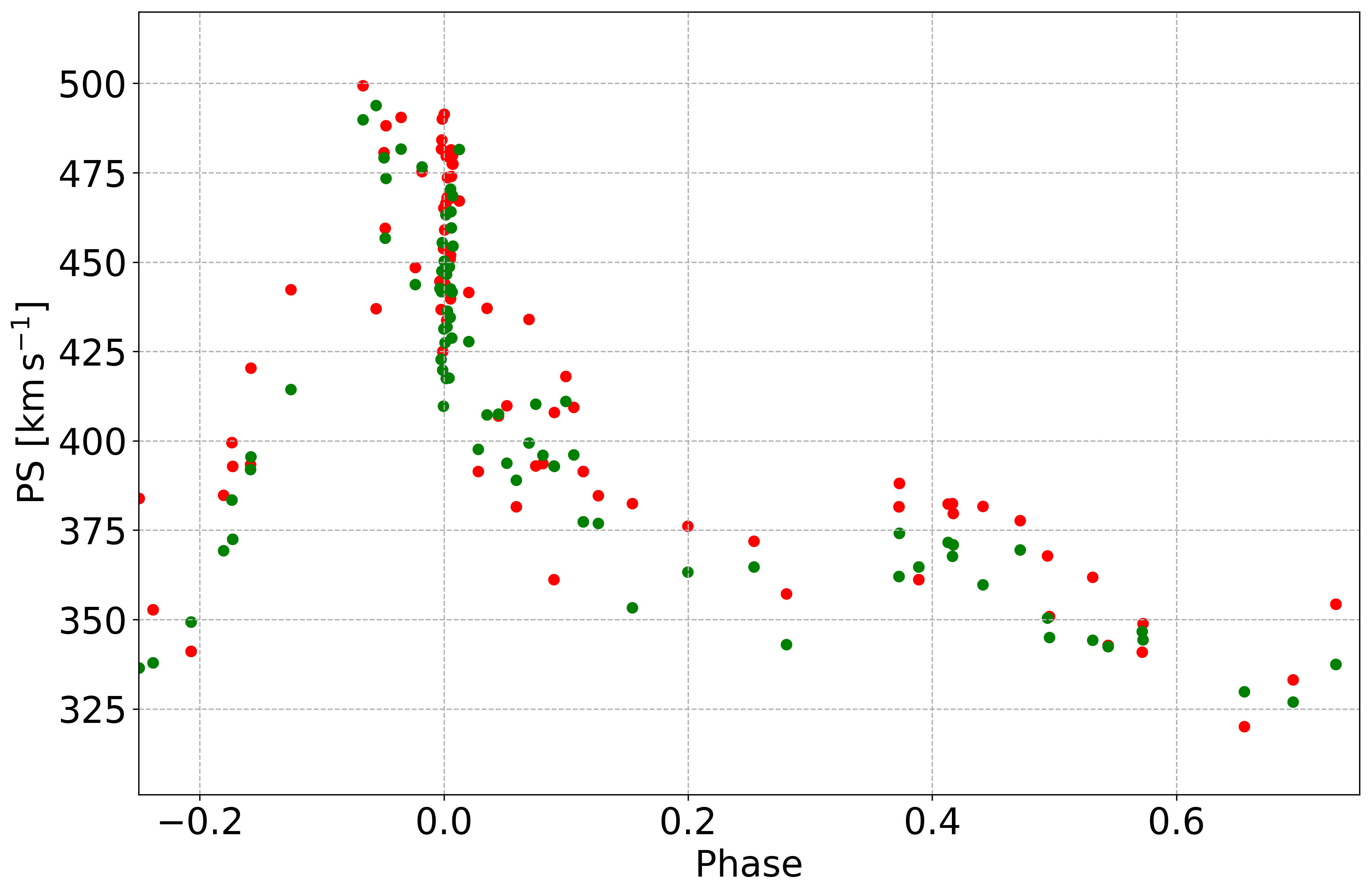}
    \includegraphics[width=0.45\linewidth]{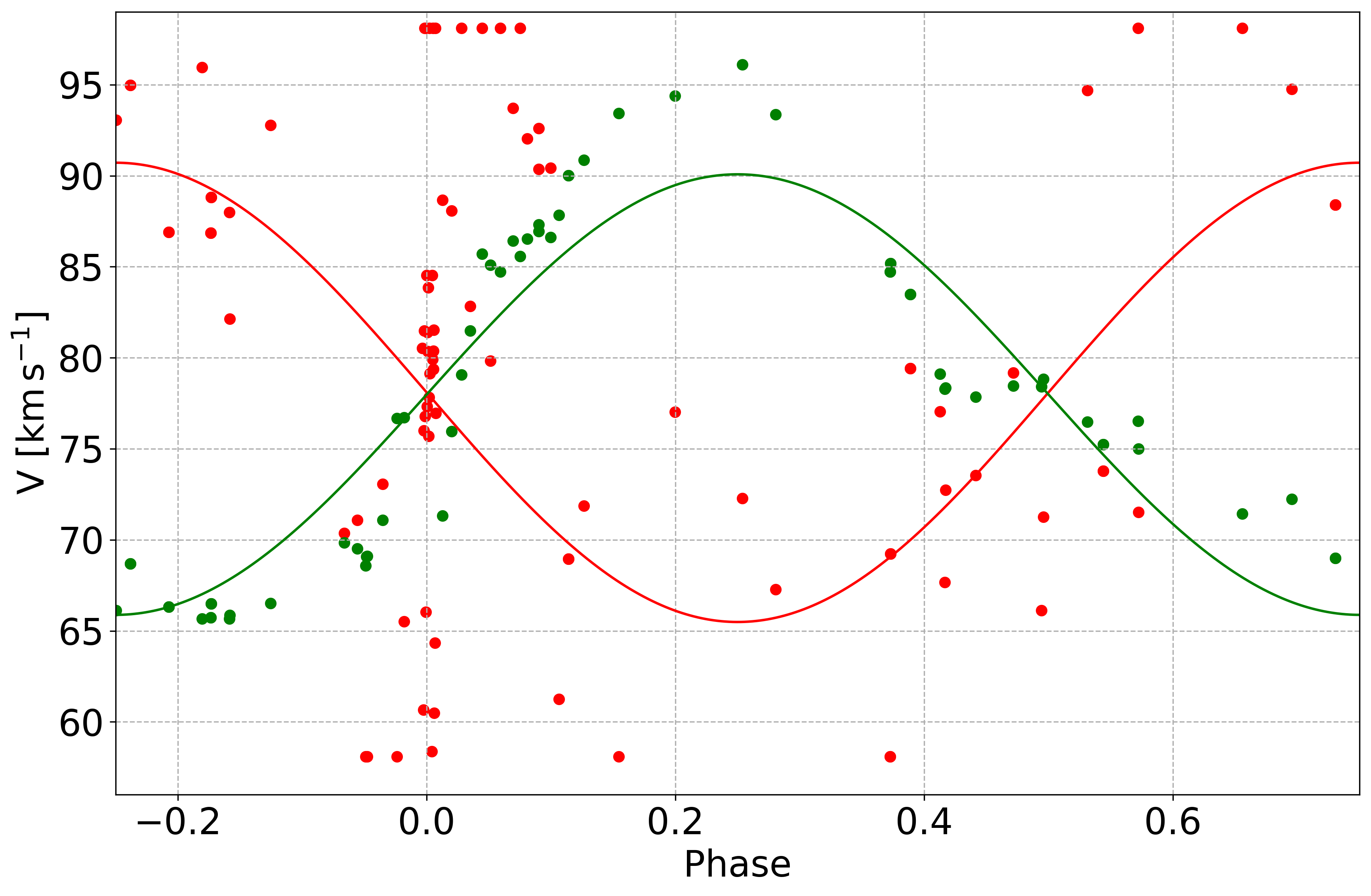}
    \caption{Reproduction of Fig.~\ref{fig:masterplot}, where Top-left is panel C, top-right is panel D and bottom is panel E. Here, however, the green points represent measurements using the Be star’s motion as the line center, while the red points use the \Ha wings. The green and red lines are the respective best-fits.}
    \label{fig:Compare_center}
\end{figure*}

As shown in panel E of Fig.~\ref{fig:masterplot}, this produces an unexpected result. While the orbital amplitude of the \Ha wings (green), $K=12.6(2.2) ~\rm km~s^{-1}$, is similar to that of the Be star (red), $K=12.2(2.0) ~\rm km~s^{-1}$, their orbital motion is in anti-phase. This result is counter-intuitive, as the \Ha wings are generally used to trace the motion of the Be star. At present, we are unable to explain the reason for this anti-phase behavior. To ensure that we are not producing any bias by choosing the center of the line as the Be orbital motion, we reproduce here the Panels B and C, using both line centers. Fig.~\ref{fig:Compare_center} shows that the results are almost identical. 

\section{Determination of the Interstellar Polarization} \label{Ap:Pol}
\restartappendixnumbering

To estimate the interstellar polarization (ISP) component present in the observed data of \VCAR, we have used the field star method. It relies on measuring the polarization of nearby stars that are not expected to exhibit intrinsic polarization. For such stars, the observed polarization can be attributed entirely to interstellar effects. We observed {four} stars selected among the brightest in the field of \VCAR (see Tab.~\ref{tab:serk}). In addition to brightness, two further criteria were applied in selecting suitable field stars:

\begin{enumerate}
\item The stars’ distances must bracket that of \VCAR, ensuring that the interstellar medium (ISM) is sampled both in the foreground and background relative to the target. For this analysis, we considered the photogeometric distance from \citet{bailerjones2021}.

\item When available, {the spectral type of the selected star} should not indicate the presence of circumstellar material, so as to minimize the risk of intrinsic polarization.
\end{enumerate}

\begin{table}
    \caption{Parameters of \VCAR's field stars. Both the distance and angular separation were calculated from GDR3 data.}
    \begin{tabular}{ccccccc}\hline
        \multirow{2}{*}{Object}
        & Spectral
        & Angular
        & Distance
        & $P_{\rm max}$
        & $\left \langle \theta  \right \rangle$
        & \multirow{2}{*}{$N$} \\ 
        & Type & Separation [$^{\circ}$] & [kpc] & [\%] & [$^{\circ}$] \\
        \hline
        Gaia ...$^{1}$ & --  & 0.076 & 2.28 (35) & $1.19 (2)$ & 126.4 (4) & 2 \\
        HD 303097 & G5 & 0.070 & 0.546 (4) & $0.561 (7)$ & 137.1 (3) & 3 \\
        HD 303181 & A2 & 0.065 & 0.448 (4) & $0.647 (8)$ & 140.1 (1) & 4 \\
        TYC 8613-2371-1 & -- & 0.065 & 2.17 (7) & $0.66 (2)$ & 118.3 (6) & 4 \\
         \hline \hline
    \end{tabular}
    \label{tab:serk}
    \justify
    \tablecomments{
        1 - Gaia DR2 5350619134384304384. \newline
        $N$ represents the number of filters used for fitting the Serkowski function. 
        }
\end{table}

Linear polarization in the ISM follows a spectral dependence described by the Serkowski Law \citep{Serkowski}
\begin{equation}
    \label{serkowski}
    \frac{P_{\rm IS}(\lambda)}{P_{\rm max}} = \exp \left[-K \ln^2 \left( \frac{\lambda_{\rm max}}{\lambda}\right)\right]\,,
\end{equation}
where $P_{\rm max}$ is the maximum polarization  at the wavelength $\lambda_{\rm max}$ and $K$ is a dimensionless constant that characterizes the inverse width of the polarization curve peaking around $\lambda_{\rm max}$ \citep{Cotton2}. If $K$ is treated as a third free parameter, it can be shown that $\lambda_{\rm max}$ and $K$ are linearly related following the relationship \citep{martin}
\begin{equation}
    \label{wilking}
    K=c_1 \lambda_{\rm max} + c_2\,.
\end{equation}
The values of $c_1=1.66 (9)$ and $c_2=0.01 (5)$ were defined by \citet{whittet}, with $\lambda_{\rm max}$ in $\mu$m. In addition to $P_{\rm max}$ and $\lambda_{\rm max}$, the {interstellar} position angle $\theta_{\rm IS}$ is also needed to fully characterize the ISP.

\begin{figure}
    \centering
    \includegraphics[width=0.49\linewidth]{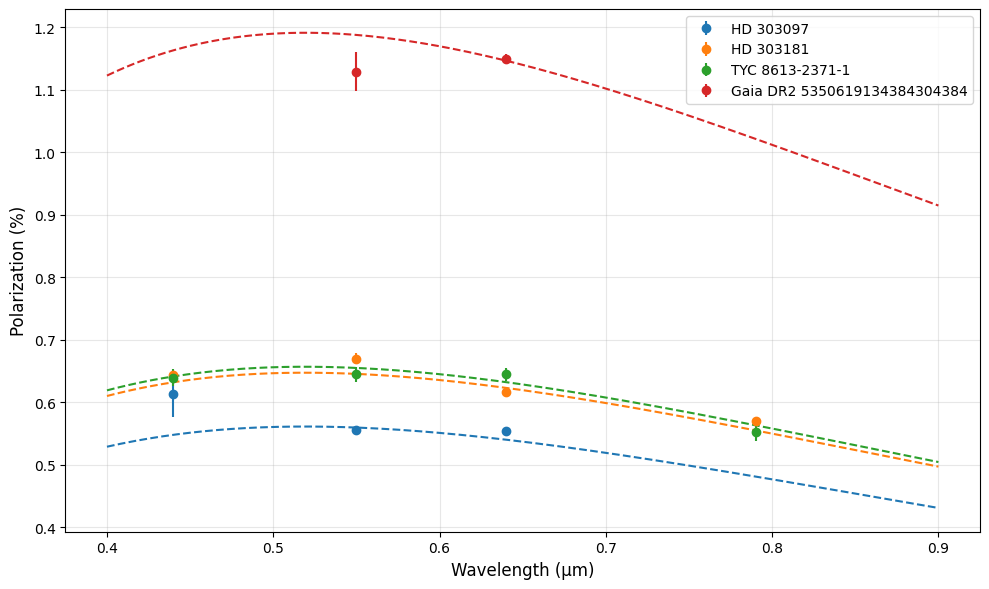}
    \includegraphics[width=0.475\linewidth]{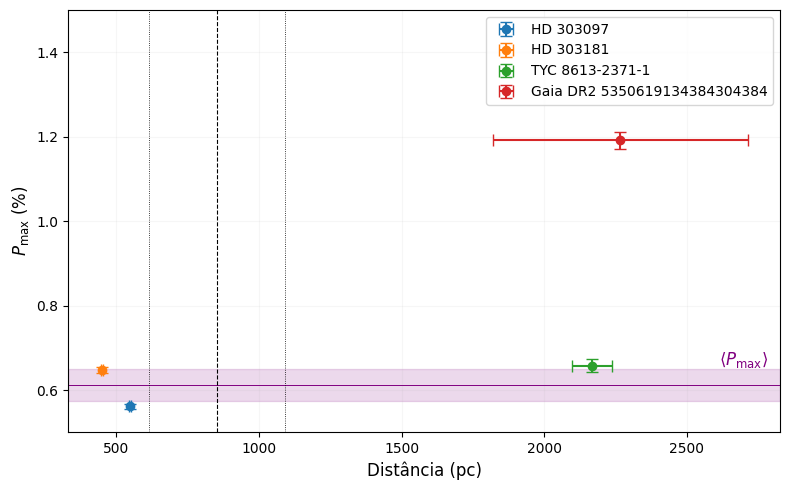}
    \includegraphics[width=0.5\linewidth]{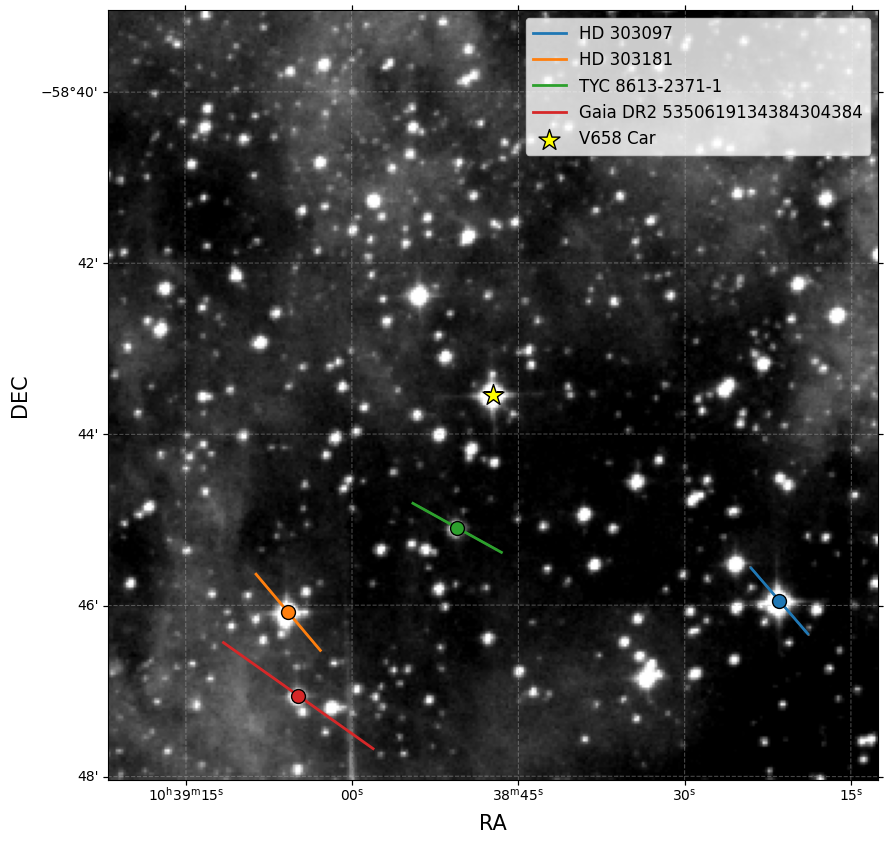}
    \caption{
    Polarization of stars in the same field as \VCAR.
    Top-left: Polarization level for each field star. The dots corresponds to the observation at each band while the dashed lines represent the fits to the Serkowski function. 
    Top-right: $P_{\rm max}$ for each field star. The diagonal dashed line indicate the best fit for the interstellar polarization. The vertical lines indicate the distance for the measured GDR3 parallax of \VCAR (dashed) and the 3 $\sigma$ region (dotted). The purple line is the weighted average $ \left< P_{\rm max}\right>$ and the light purple shades represents its weighted standard deviation.
    Bottom: Polarization map in Equatorial coordinates for the field of \VCAR. The vectors indicate the direction of the average polarization in all observed filters. The length of the vector is proportional to $P_{\rm max}$.}
    \label{fig:IPpos}
\end{figure}

Fig.~\ref{fig:IPpos} (right) shows the polarization as a function of wavelength for the four field stars. We modeled the polarization curves using Equation~\ref{serkowski}, fitting the parameters $P_{\rm max}$ and $\lambda_{\rm max}$ through a non-linear least-squares minimization using the Levenberg-Marquardt algorithm. 
To ensure consistency and minimize overfitting, we fixed $\lambda_{\rm max}$ to a common value for all four stars, allowing only $P_{\rm max}$ to vary individually. This approach is justified by the assumption that interstellar dust along similar lines of sight produces similar polarization peak wavelengths. The corresponding $P_{\rm max}$ are shown in Table~\ref{tab:serk} and $\lambda_{\rm max}=0.519 (16)~\mu$m.
Figure~\ref{fig:IPpos} (left) displays the positions of the stars in the sky along with their respective polarization pseudovectors (their length represents $P_{\rm max}$ and their direction $\left \langle \theta \right \rangle$, the average of the polarization angles $\theta$ for all observed filters).

To estimate $P_{\rm max}$ and $\theta_{\rm IS}$ for \VCAR we employed the Inverse Distance Weighting (IDW) method that uses, as a weight, the geometric distance ($r_{\rm GD}$) of each field star to \VCAR. This ensures that more distant stars contribute less to the final value, since nearby stars are more likely to share the same interstellar environment as \VCAR. We calculated $r_{\rm GD}$ using the stars' parallaxes and Equatorial coordinates obtained from Gaia DR3. The IDW method applies a weighting scheme where the influence of each star decreases with distance raised to a power $p$. We adopted $p = 2$, a commonly used value that balances local sensitivity with smoothness in the interpolation. We then obtained $P_{\rm max}=0.61(4)~\%$ and $\theta_{\rm IS}=138.4 (1.5)^{\circ}$. With the ISP towards \VCAR thus determined, Eq.~\ref{serkowski} was used to calculate the ISP for each filter used in the observations. The intrinsic polarization of \VCAR was then obtained by subtracting the ISP from the observed polarization in each filter.


\section{Supplementary optical and NIR lines}\label{Ap:other_lines}
\restartappendixnumbering

\begin{figure*}
    \centering
    \includegraphics[width=0.3\linewidth]{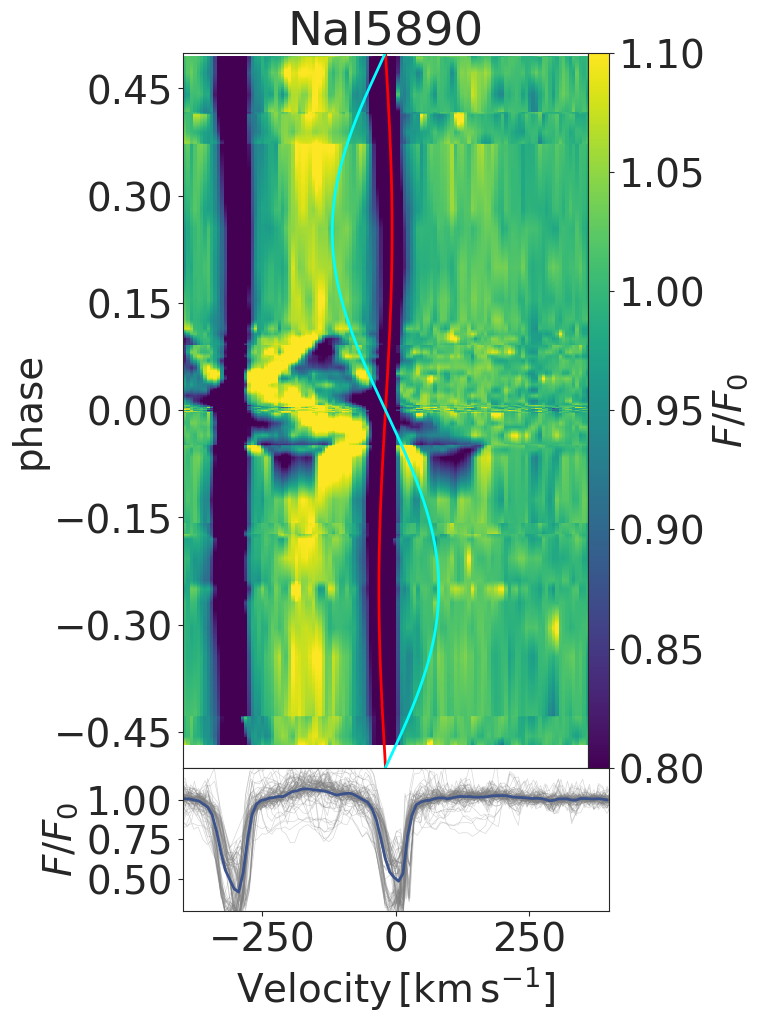}
    \includegraphics[width=0.3\linewidth]{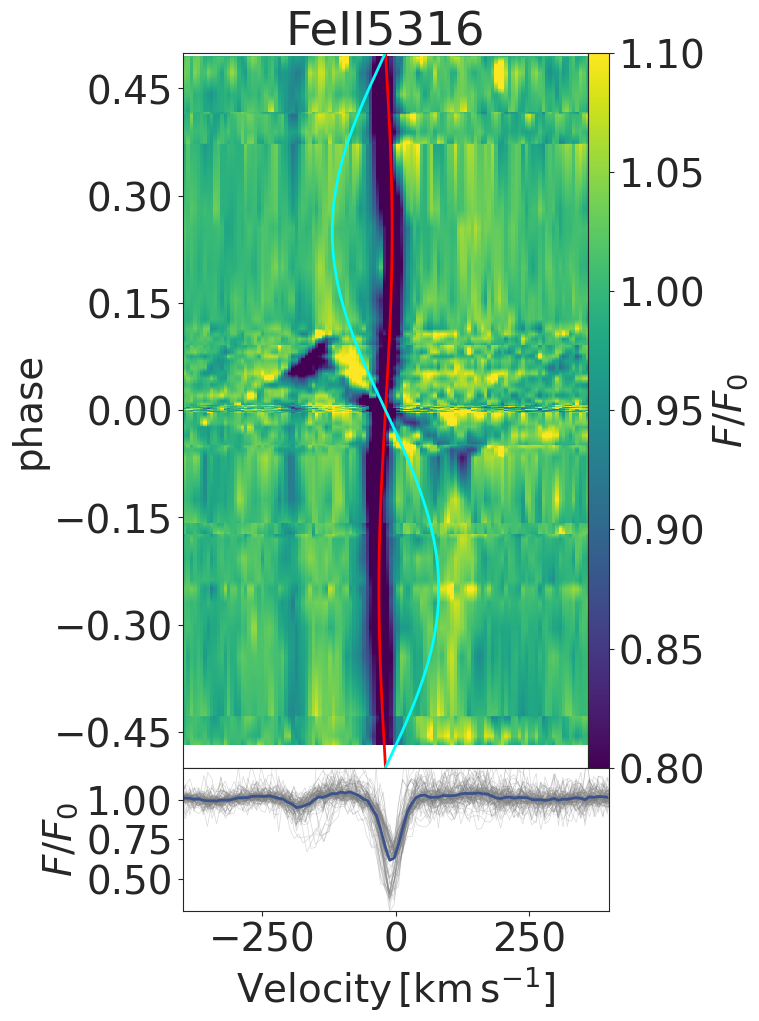}
    \includegraphics[width=0.3\linewidth]{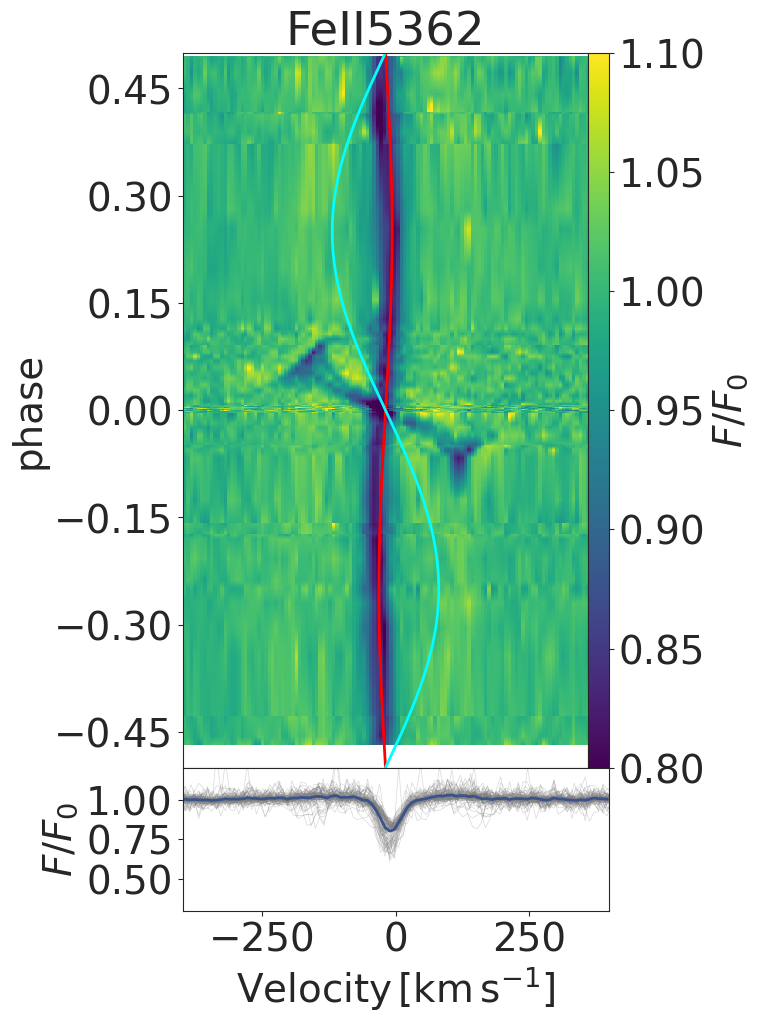}
    \includegraphics[width=0.3\linewidth]{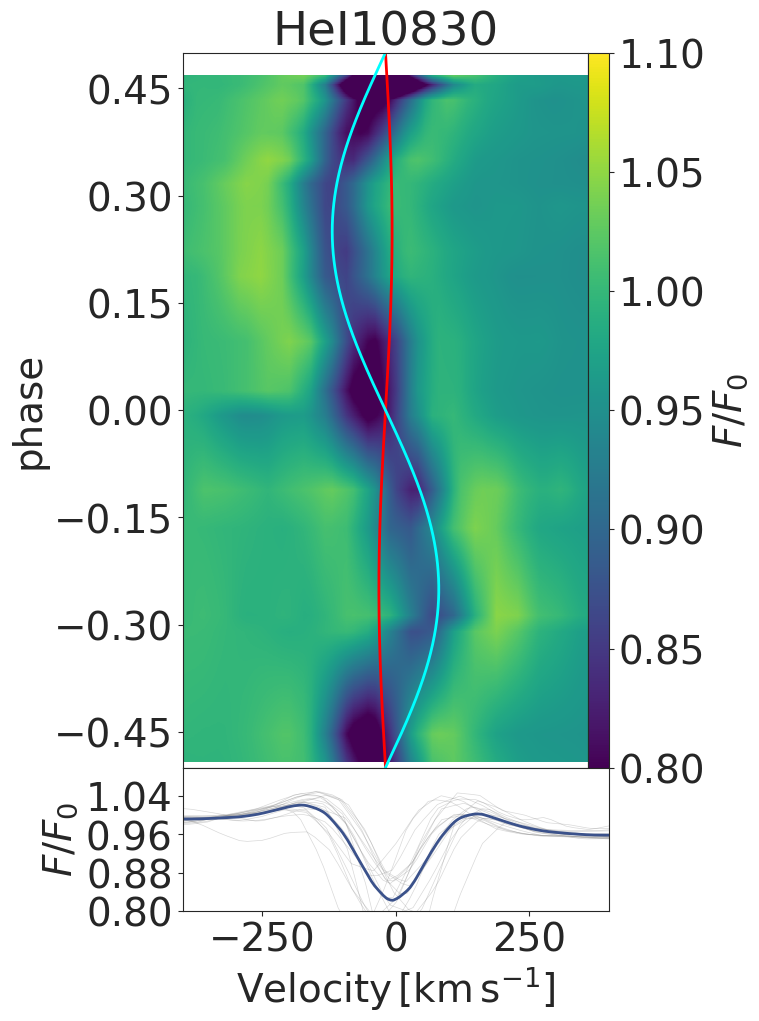}
    \includegraphics[width=0.3\linewidth]{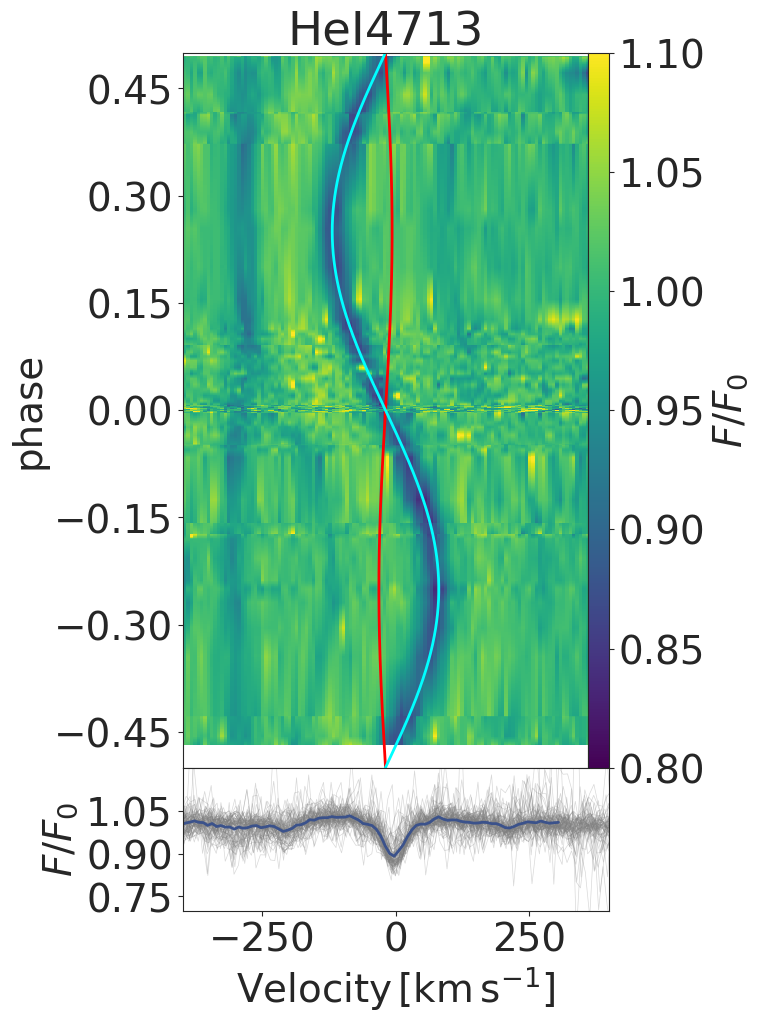}
    \includegraphics[width=0.3\linewidth]{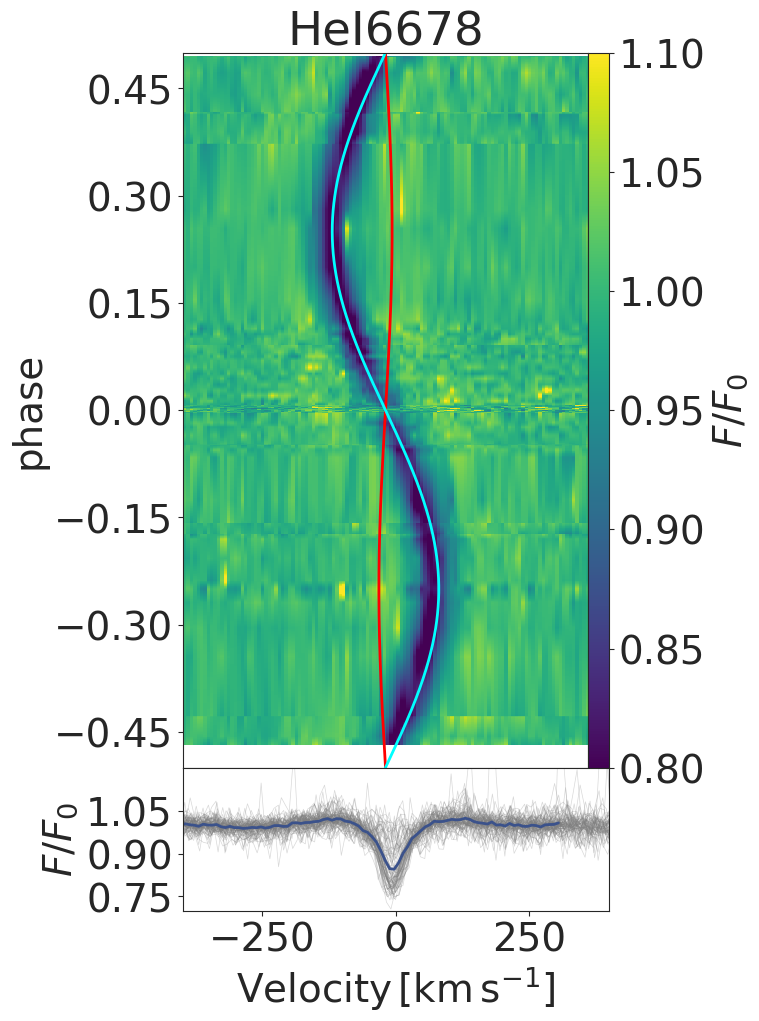}
    \includegraphics[width=0.3\linewidth]{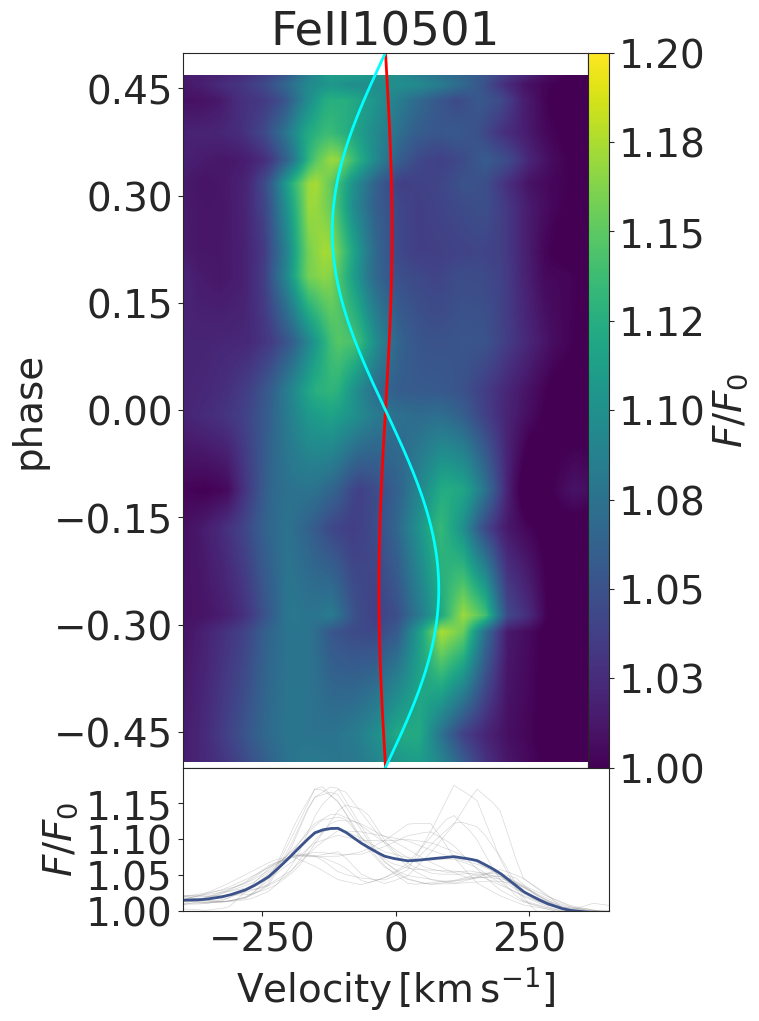}
    \includegraphics[width=0.3\linewidth]{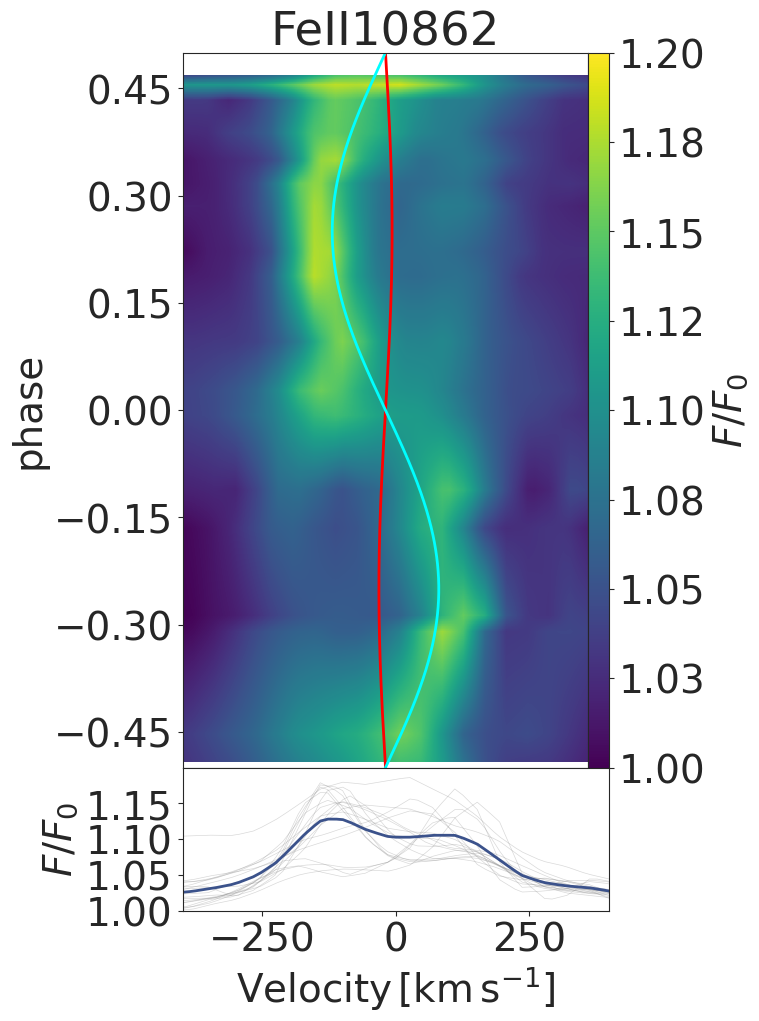}
    \includegraphics[width=0.3\linewidth]{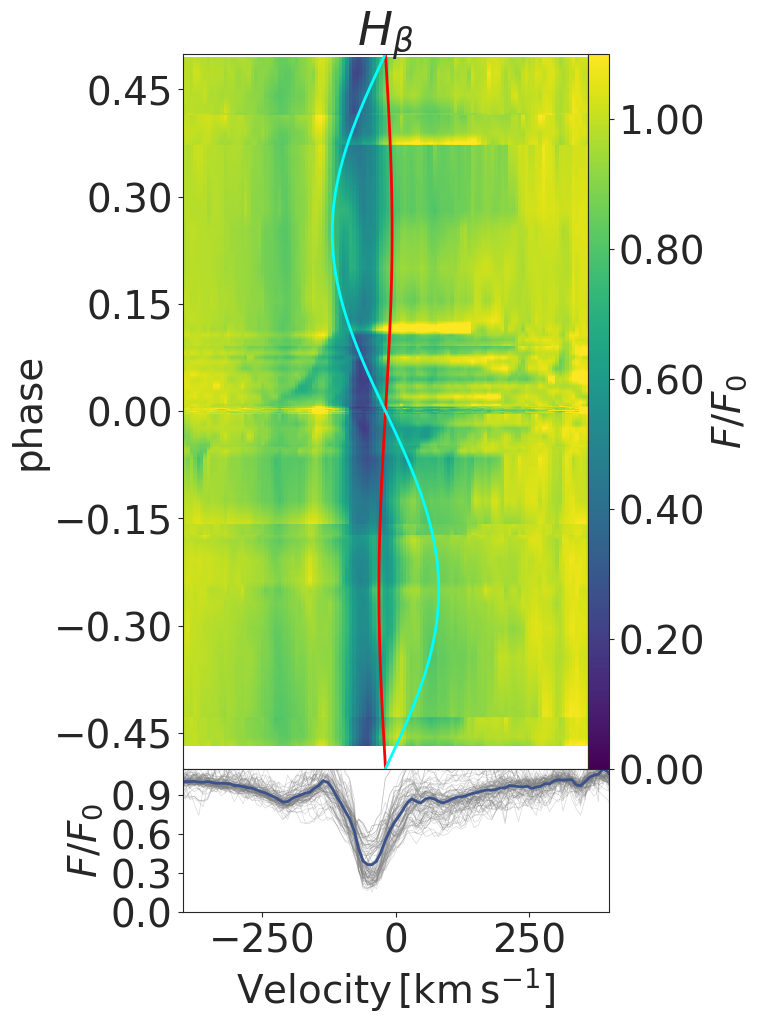}
    
    \caption{Same as Fig.~\ref{fig:Dynamical} for additional lines, as labeled.}
    \label{fig:Dynamicalappendix}
\end{figure*}

In Fig.~\ref{fig:Dynamical}, we present one example for each observed behavior. Here (Fig.~\ref{fig:Dynamicalappendix}), for completeness, we show more lines that undergo the same effects.

\section{Animation for the best-fit model}\label{Ap:animation}

\begin{figure*} \label{fig:animation}
    \begin{interactive}{animation}{Binaries.mp4}
        \includegraphics[width=0.9\linewidth, trim={0.3cm 0 0 0}, clip]{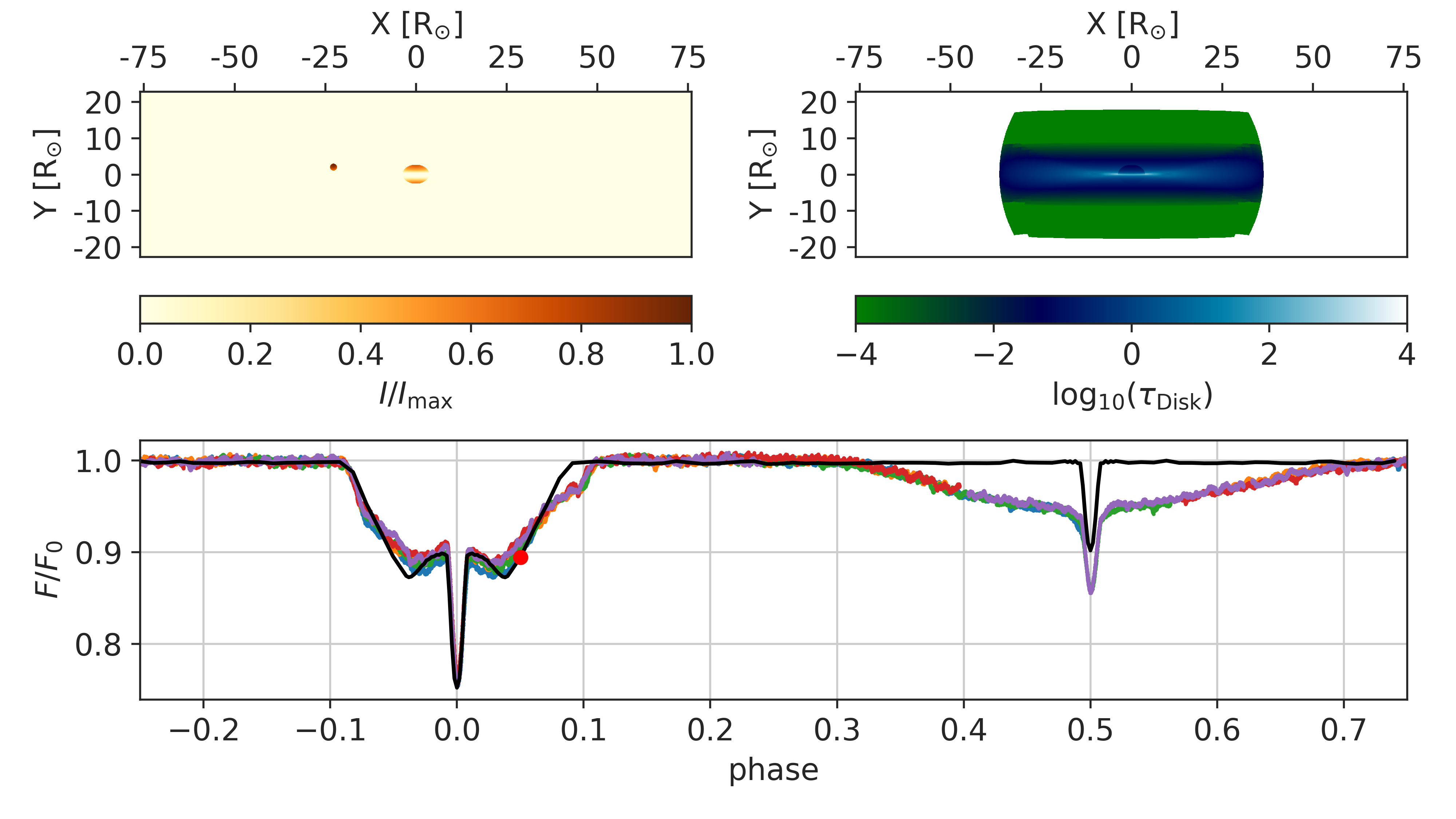}
    \end{interactive}
    \caption{
    Schematic representation of the model at each phase. Top left: relative intensity in the line of sight. Top right: Disk optical depth in the line of sight. Bottom: The colored lines represent the TESS observations. The black line correspond to the modelled light curve. The red dot represents the current phase being plotted. (An \underline{\href{https://drive.google.com/file/d/1igh7zU_BdYKuj_2FsEoINBKFXyg-zYho/view?usp=sharing}{animation}} of this figure is available on-line.)
    }
\end{figure*}

In order to facilitate the visualization of our best-fit model (as described in Tab.~\ref{tab:OrbSol}) at each phase, we have created the video present in Fig.~\ref{fig:animation}. {The top-left panel shows the relative intensity at an inclination of $88.6^{\circ}$. Notably, the companion star (hotter, and therefore appearing brownish) becomes dimmer (appearing more yellow) in the vicinity of the first stellar eclipse, due to attenuation by the disk. The top-right panel shows the disk optical depth in the line of sight, calculated from the far side of the envelope.}

\bibliography{bibliography}{}

\begin{thebibliography}{}
\expandafter\ifx\csname natexlab\endcsname\relax\def\natexlab#1{#1}\fi
\providecommand{\url}[1]{\href{#1}{#1}}
\providecommand{\dodoi}[1]{doi:~\href{http://doi.org/#1}{\nolinkurl{#1}}}
\providecommand{\doeprint}[1]{\href{http://ascl.net/#1}{\nolinkurl{http://ascl.net/#1}}}
\providecommand{\doarXiv}[1]{\href{https://arxiv.org/abs/#1}{\nolinkurl{https://arxiv.org/abs/#1}}}

\bibitem[{{Andersen}(1991)}]{Andersen1991Binaries}
{Andersen}, J. 1991, \aapr, 3, 91, \dodoi{10.1007/BF00873538}

\bibitem[{{Baade}(1982)}]{Baade1982pulsation}
{Baade}, D. 1982, \aap, 105, 65

\bibitem[{{Baade} {et~al.}(2023){Baade}, {Labadie-Bartz}, {Rivinius}, \& {Carciofi}}]{baade2023gammaCas}
{Baade}, D., {Labadie-Bartz}, J., {Rivinius}, T., \& {Carciofi}, A.~C. 2023, \aap, 678, A47, \dodoi{10.1051/0004-6361/202244149}

\bibitem[{{Bailer-Jones} {et~al.}(2021){Bailer-Jones}, {Rybizki}, {Fouesneau}, {Demleitner}, \& {Andrae}}]{bailerjones2021}
{Bailer-Jones}, C.~A.~L., {Rybizki}, J., {Fouesneau}, M., {Demleitner}, M., \& {Andrae}, R. 2021, \aj, 161, 147, \dodoi{10.3847/1538-3881/abd806}

\bibitem[{{Bak{\i}{\c{s}}} {et~al.}(2025){Bak{\i}{\c{s}}}, {Hilal Y{\i}ld{\i}z}, {Bak{\i}{\c{s}}}, \& {Y{\"u}cel}}]{Bakis2025RSsgr}
{Bak{\i}{\c{s}}}, H., {Hilal Y{\i}ld{\i}z}, {\"O}., {Bak{\i}{\c{s}}}, V., \& {Y{\"u}cel}, G. 2025, arXiv e-prints, arXiv:2506.01516.
\newblock \doarXiv{2506.01516}

\bibitem[{{Bjorkman} \& {Carciofi}(2005)}]{bjorkman2005structure}
{Bjorkman}, J.~E., \& {Carciofi}, A.~C. 2005, in Astronomical Society of the Pacific Conference Series, Vol. 337, The Nature and Evolution of Disks Around Hot Stars, ed. R.~{Ignace} \& K.~G. {Gayley}, 75

\bibitem[{{Brown} \& {McLean}(1977)}]{Brown1977}
{Brown}, J.~C., \& {McLean}, I.~S. 1977, \aap, 57, 141

\bibitem[{{Carciofi} \& {Bjorkman}(2006)}]{carciofi2006}
{Carciofi}, A.~C., \& {Bjorkman}, J.~E. 2006, \apj, 639, 1081, \dodoi{10.1086/499483}

\bibitem[{{Carciofi} \& {Bjorkman}(2008)}]{carciofi2008}
---. 2008, \apj, 684, 1374, \dodoi{10.1086/589875}

\bibitem[{{Carciofi} {et~al.}(2007){Carciofi}, {Magalh{\~a}es}, {Leister}, {Bjorkman}, \& {Levenhagen}}]{carciofi2007}
{Carciofi}, A.~C., {Magalh{\~a}es}, A.~M., {Leister}, N.~V., {Bjorkman}, J.~E., \& {Levenhagen}, R.~S. 2007, \apjl, 671, L49, \dodoi{10.1086/524772}

\bibitem[{{Chojnowski} {et~al.}(2018){Chojnowski}, {Labadie-Bartz}, {Rivinius}, {Gies}, {Panoglou}, {Borges Fernandes}, {Wisniewski}, {Whelan}, {Mennickent}, {McMillan}, {Dembicky}, {Gray}, {Rudyk}, {Stringfellow}, {Lester}, {Hasselquist}, {Zharikov}, {Levenhagen}, {Souza}, {Leister}, {Stassun}, {Siverd}, \& {Majewski}}]{chojnowski2018hd55606}
{Chojnowski}, S.~D., {Labadie-Bartz}, J., {Rivinius}, T., {et~al.} 2018, \apj, 865, 76, \dodoi{10.3847/1538-4357/aad964}

\bibitem[{{Coelho}(2014)}]{Coelho2014spectra}
{Coelho}, P.~R.~T. 2014, \mnras, 440, 1027, \dodoi{10.1093/mnras/stu365}

\bibitem[{{Cotton} {et~al.}(2019){Cotton}, {Marshall}, {Frisch}, {Kedziora-Chudzer}, {Bailey}, {Bott}, {Wright}, {Wyatt}, \& {Kennedy}}]{Cotton2}
{Cotton}, D.~V., {Marshall}, J.~P., {Frisch}, P.~C., {et~al.} 2019, \mnras, 483, 3636, \dodoi{10.1093/mnras/sty3318}

\bibitem[{{Cushing} {et~al.}(2004){Cushing}, {Vacca}, \& {Rayner}}]{Cushing2004triplespec}
{Cushing}, M.~C., {Vacca}, W.~D., \& {Rayner}, J.~T. 2004, \pasp, 116, 362, \dodoi{10.1086/382907}

\bibitem[{{Cyr} {et~al.}(2020){Cyr}, {Jones}, {Carciofi}, {Steckel}, {Tycner}, \& {Okazaki}}]{Cyr2020}
{Cyr}, I.~H., {Jones}, C.~E., {Carciofi}, A.~C., {et~al.} 2020, \mnras, 497, 3525, \dodoi{10.1093/mnras/staa2176}

\bibitem[{{Cyr} {et~al.}(2017){Cyr}, {Jones}, {Panoglou}, {Carciofi}, \& {Okazaki}}]{cyr2017}
{Cyr}, I.~H., {Jones}, C.~E., {Panoglou}, D., {Carciofi}, A.~C., \& {Okazaki}, A.~T. 2017, \mnras, 471, 596, \dodoi{10.1093/mnras/stx1427}

\bibitem[{{Czesla} {et~al.}(2019){Czesla}, {Schr{\"o}ter}, {Schneider}, {Huber}, {Pfeifer}, {Andreasen}, \& {Zechmeister}}]{pyastronomy}
{Czesla}, S., {Schr{\"o}ter}, S., {Schneider}, C.~P., {et~al.} 2019, {PyA: Python astronomy-related packages}.
\newblock \doeprint{1906.010}

\bibitem[{{Dallas} {et~al.}(2022){Dallas}, {Oey}, \& {Castro}}]{Dallas2022binaries}
{Dallas}, M.~M., {Oey}, M.~S., \& {Castro}, N. 2022, \apj, 936, 112, \dodoi{10.3847/1538-4357/ac8988}

\bibitem[{{Davidge}(2022)}]{Davidge2022AV367Cygni}
{Davidge}, T.~J. 2022, \aj, 164, 149, \dodoi{10.3847/1538-3881/ac8b00}

\bibitem[{{de Amorim} {et~al.}(2025){de Amorim}, {Carciofi}, {Zanardo}, {Colesanti}, {Jacques}, {Kulh}, {Mattei}, {Domingues}, {Rocca}, {Cacella}, {Silva}, {Napoleão}, \& {Labadie-Bartz}}]{deamorim2025}
{de Amorim}, T.~H., {Carciofi}, A.~C., {Zanardo}, A., {et~al.} 2025, \gal, submitted

\bibitem[{{Dodd} {et~al.}(2024){Dodd}, {Oudmaijer}, {Radley}, {Vioque}, \& {Frost}}]{Dodd2024Gaia}
{Dodd}, J.~M., {Oudmaijer}, R.~D., {Radley}, I.~C., {Vioque}, M., \& {Frost}, A.~J. 2024, \mnras, 527, 3076, \dodoi{10.1093/mnras/stad3105}

\bibitem[{{Eggleton}(1983)}]{Eggleton1983Roche}
{Eggleton}, P.~P. 1983, \apj, 268, 368, \dodoi{10.1086/160960}

\bibitem[{{Espinosa Lara} \& {Rieutord}(2011)}]{Espinosa2011gravity}
{Espinosa Lara}, F., \& {Rieutord}, M. 2011, \aap, 533, A43, \dodoi{10.1051/0004-6361/201117252}

\bibitem[{Gabitova {et~al.}(2025)Gabitova, Zharikov, Miroshnichenko, Carciofi, Khokhlov, Agishev, \& Prendergast}]{Gabitova2025}
Gabitova, I.~A., Zharikov, S.~V., Miroshnichenko, A.~S., {et~al.} 2025, \gal, 13, 80, \dodoi{10.3390/galaxies13040080}

\bibitem[{{Gaia Collaboration} {et~al.}(2016){Gaia Collaboration}, {Prusti}, {de Bruijne}, {Brown}, {Vallenari}, {Babusiaux}, {Bailer-Jones}, {Bastian}, {Biermann}, {Evans}, {Eyer}, {Jansen}, {Jordi}, {Klioner}, {Lammers}, {Lindegren}, {Luri}, {Mignard}, {Milligan}, {Panem}, {Poinsignon}, {Pourbaix}, {Randich}, {Sarri}, {Sartoretti}, {Siddiqui}, {Soubiran}, {Valette}, {van Leeuwen}, {Walton}, {Aerts}, {Arenou}, {Cropper}, {Drimmel}, {H{\o}g}, {Katz}, {Lattanzi}, {O'Mullane}, {Grebel}, {Holland}, {Huc}, {Passot}, {Bramante}, {Cacciari}, {Casta{\~n}eda}, {Chaoul}, {Cheek}, {De Angeli}, {Fabricius}, {Guerra}, {Hern{\'a}ndez}, {Jean-Antoine-Piccolo}, {Masana}, {Messineo}, {Mowlavi}, {Nienartowicz}, {Ord{\'o}{\~n}ez-Blanco}, {Panuzzo}, {Portell}, {Richards}, {Riello}, {Seabroke}, {Tanga}, {Th{\'e}venin}, {Torra}, {Els}, {Gracia-Abril}, {Comoretto}, {Garcia-Reinaldos}, {Lock}, {Mercier}, {Altmann}, {Andrae}, {Astraatmadja}, {Bellas-Velidis}, {Benson}, {Berthier}, {Blomme}, {Busso}, {Carry}, {Cellino}, {Clementini},
  {Cowell}, {Creevey}, {Cuypers}, {Davidson}, {De Ridder}, {de Torres}, {Delchambre}, {Dell'Oro}, {Ducourant}, {Fr{\'e}mat}, {Garc{\'\i}a-Torres}, {Gosset}, {Halbwachs}, {Hambly}, {Harrison}, {Hauser}, {Hestroffer}, {Hodgkin}, {Huckle}, {Hutton}, {Jasniewicz}, {Jordan}, {Kontizas}, {Korn}, {Lanzafame}, {Manteiga}, {Moitinho}, {Muinonen}, {Osinde}, {Pancino}, {Pauwels}, {Petit}, {Recio-Blanco}, {Robin}, {Sarro}, {Siopis}, {Smith}, {Smith}, {Sozzetti}, {Thuillot}, {van Reeven}, {Viala}, {Abbas}, {Abreu Aramburu}, {Accart}, {Aguado}, {Allan}, {Allasia}, {Altavilla}, {{\'A}lvarez}, {Alves}, {Anderson}, {Andrei}, {Anglada Varela}, {Antiche}, {Antoja}, {Ant{\'o}n}, {Arcay}, {Atzei}, {Ayache}, {Bach}, {Baker}, {Balaguer-N{\'u}{\~n}ez}, {Barache}, {Barata}, {Barbier}, {Barblan}, {Baroni}, {Barrado y Navascu{\'e}s}, {Barros}, {Barstow}, {Becciani}, {Bellazzini}, {Bellei}, {Bello Garc{\'\i}a}, {Belokurov}, {Bendjoya}, {Berihuete}, {Bianchi}, {Bienaym{\'e}}, {Billebaud}, {Blagorodnova}, {Blanco-Cuaresma}, {Boch},
  {Bombrun}, {Borrachero}, {Bouquillon}, {Bourda}, {Bouy}, {Bragaglia}, {Breddels}, {Brouillet}, {Br{\"u}semeister}, {Bucciarelli}, {Budnik}, {Burgess}, {Burgon}, {Burlacu}, {Busonero}, {Buzzi}, {Caffau}, {Cambras}, {Campbell}, {Cancelliere}, {Cantat-Gaudin}, {Carlucci}, {Carrasco}, {Castellani}, {Charlot}, {Charnas}, {Charvet}, {Chassat}, {Chiavassa}, {Clotet}, {Cocozza}, {Collins}, {Collins}, {Costigan}, {Crifo}, {Cross}, {Crosta}, {Crowley}, {Dafonte}, {Damerdji}, {Dapergolas}, {David}, {David}, {De Cat}, {de Felice}, {de Laverny}, {De Luise}, {De March}, {de Martino}, {de Souza}, {Debosscher}, {del Pozo}, {Delbo}, {Delgado}, {Delgado}, {di Marco}, {Di Matteo}, {Diakite}, {Distefano}, {Dolding}, {Dos Anjos}, {Drazinos}, {Dur{\'a}n}, {Dzigan}, {Ecale}, {Edvardsson}, {Enke}, {Erdmann}, {Escolar}, {Espina}, {Evans}, {Eynard Bontemps}, {Fabre}, {Fabrizio}, {Faigler}, {Falc{\~a}o}, {Farr{\`a}s Casas}, {Faye}, {Federici}, {Fedorets}, {Fern{\'a}ndez-Hern{\'a}ndez}, {Fernique}, {Fienga}, {Figueras}, {Filippi},
  {Findeisen}, {Fonti}, {Fouesneau}, {Fraile}, {Fraser}, {Fuchs}, {Furnell}, {Gai}, {Galleti}, {Galluccio}, {Garabato}, {Garc{\'\i}a-Sedano}, {Gar{\'e}}, {Garofalo}, {Garralda}, {Gavras}, {Gerssen}, {Geyer}, {Gilmore}, {Girona}, {Giuffrida}, {Gomes}, {Gonz{\'a}lez-Marcos}, {Gonz{\'a}lez-N{\'u}{\~n}ez}, {Gonz{\'a}lez-Vidal}, {Granvik}, {Guerrier}, {Guillout}, {Guiraud}, {G{\'u}rpide}, {Guti{\'e}rrez-S{\'a}nchez}, {Guy}, {Haigron}, {Hatzidimitriou}, {Haywood}, {Heiter}, {Helmi}, {Hobbs}, {Hofmann}, {Holl}, {Holland}, {Hunt}, {Hypki}, {Icardi}, {Irwin}, {Jevardat de Fombelle}, {Jofr{\'e}}, {Jonker}, {Jorissen}, {Julbe}, {Karampelas}, {Kochoska}, {Kohley}, {Kolenberg}, {Kontizas}, {Koposov}, {Kordopatis}, {Koubsky}, {Kowalczyk}, {Krone-Martins}, {Kudryashova}, {Kull}, {Bachchan}, {Lacoste-Seris}, {Lanza}, {Lavigne}, {Le Poncin-Lafitte}, {Lebreton}, {Lebzelter}, {Leccia}, {Leclerc}, {Lecoeur-Taibi}, {Lemaitre}, {Lenhardt}, {Leroux}, {Liao}, {Licata}, {Lindstr{\o}m}, {Lister}, {Livanou}, {Lobel}, {L{\"o}ffler},
  {L{\'o}pez}, {Lopez-Lozano}, {Lorenz}, {Loureiro}, {MacDonald}, {Magalh{\~a}es Fernandes}, {Managau}, {Mann}, {Mantelet}, {Marchal}, {Marchant}, {Marconi}, {Marie}, {Marinoni}, {Marrese}, {Marschalk{\'o}}, {Marshall}, {Mart{\'\i}n-Fleitas}, {Martino}, {Mary}, {Matijevi{\v{c}}}, {Mazeh}, {McMillan}, {Messina}, {Mestre}, {Michalik}, {Millar}, {Miranda}, {Molina}, {Molinaro}, {Molinaro}, {Moln{\'a}r}, {Moniez}, {Montegriffo}, {Monteiro}, {Mor}, {Mora}, {Morbidelli}, {Morel}, {Morgenthaler}, {Morley}, {Morris}, {Mulone}, {Muraveva}, {Musella}, {Narbonne}, {Nelemans}, {Nicastro}, {Noval}, {Ord{\'e}novic}, {Ordieres-Mer{\'e}}, {Osborne}, {Pagani}, {Pagano}, {Pailler}, {Palacin}, {Palaversa}, {Parsons}, {Paulsen}, {Pecoraro}, {Pedrosa}, {Pentik{\"a}inen}, {Pereira}, {Pichon}, {Piersimoni}, {Pineau}, {Plachy}, {Plum}, {Poujoulet}, {Pr{\v{s}}a}, {Pulone}, {Ragaini}, {Rago}, {Rambaux}, {Ramos-Lerate}, {Ranalli}, {Rauw}, {Read}, {Regibo}, {Renk}, {Reyl{\'e}}, {Ribeiro}, {Rimoldini}, {Ripepi}, {Riva}, {Rixon},
  {Roelens}, {Romero-G{\'o}mez}, {Rowell}, {Royer}, {Rudolph}, {Ruiz-Dern}, {Sadowski}, {Sagrist{\`a} Sell{\'e}s}, {Sahlmann}, {Salgado}, {Salguero}, {Sarasso}, {Savietto}, {Schnorhk}, {Schultheis}, {Sciacca}, {Segol}, {Segovia}, {Segransan}, {Serpell}, {Shih}, {Smareglia}, {Smart}, {Smith}, {Solano}, {Solitro}, {Sordo}, {Soria Nieto}, {Souchay}, {Spagna}, {Spoto}, {Stampa}, {Steele}, {Steidelm{\"u}ller}, {Stephenson}, {Stoev}, {Suess}, {S{\"u}veges}, {Surdej}, {Szabados}, {Szegedi-Elek}, {Tapiador}, {Taris}, {Tauran}, {Taylor}, {Teixeira}, {Terrett}, {Tingley}, {Trager}, {Turon}, {Ulla}, {Utrilla}, {Valentini}, {van Elteren}, {Van Hemelryck}, {van Leeuwen}, {Varadi}, {Vecchiato}, {Veljanoski}, {Via}, {Vicente}, {Vogt}, {Voss}, {Votruba}, {Voutsinas}, {Walmsley}, {Weiler}, {Weingrill}, {Werner}, {Wevers}, {Whitehead}, {Wyrzykowski}, {Yoldas}, {{\v{Z}}erjal}, {Zucker}, {Zurbach}, {Zwitter}, {Alecu}, {Allen}, {Allende Prieto}, {Amorim}, {Anglada-Escud{\'e}}, {Arsenijevic}, {Azaz}, {Balm}, {Beck}, {Bernstein},
  {Bigot}, {Bijaoui}, {Blasco}, {Bonfigli}, {Bono}, {Boudreault}, {Bressan}, {Brown}, {Brunet}, {Bunclark}, {Buonanno}, {Butkevich}, {Carret}, {Carrion}, {Chemin}, {Ch{\'e}reau}, {Corcione}, {Darmigny}, {de Boer}, {de Teodoro}, {de Zeeuw}, {Delle Luche}, {Domingues}, {Dubath}, {Fodor}, {Fr{\'e}zouls}, {Fries}, {Fustes}, {Fyfe}, {Gallardo}, {Gallegos}, {Gardiol}, {Gebran}, {Gomboc}, {G{\'o}mez}, {Grux}, {Gueguen}, {Heyrovsky}, {Hoar}, {Iannicola}, {Isasi Parache}, {Janotto}, {Joliet}, {Jonckheere}, {Keil}, {Kim}, {Klagyivik}, {Klar}, {Knude}, {Kochukhov}, {Kolka}, {Kos}, {Kutka}, {Lainey}, {LeBouquin}, {Liu}, {Loreggia}, {Makarov}, {Marseille}, {Martayan}, {Martinez-Rubi}, {Massart}, {Meynadier}, {Mignot}, {Munari}, {Nguyen}, {Nordlander}, {Ocvirk}, {O'Flaherty}, {Olias Sanz}, {Ortiz}, {Osorio}, {Oszkiewicz}, {Ouzounis}, {Palmer}, {Park}, {Pasquato}, {Peltzer}, {Peralta}, {P{\'e}turaud}, {Pieniluoma}, {Pigozzi}, {Poels}, {Prat}, {Prod'homme}, {Raison}, {Rebordao}, {Risquez}, {Rocca-Volmerange}, {Rosen},
  {Ruiz-Fuertes}, {Russo}, {Sembay}, {Serraller Vizcaino}, {Short}, {Siebert}, {Silva}, {Sinachopoulos}, {Slezak}, {Soffel}, {Sosnowska}, {Strai{\v{z}}ys}, {ter Linden}, {Terrell}, {Theil}, {Tiede}, {Troisi}, {Tsalmantza}, {Tur}, {Vaccari}, {Vachier}, {Valles}, {Van Hamme}, {Veltz}, {Virtanen}, {Wallut}, {Wichmann}, {Wilkinson}, {Ziaeepour}, \& {Zschocke}}]{gaiamission}
{Gaia Collaboration}, {Prusti}, T., {de Bruijne}, J.~H.~J., {et~al.} 2016, \aap, 595, A1, \dodoi{10.1051/0004-6361/201629272}

\bibitem[{{Gaia Collaboration} {et~al.}(2023){Gaia Collaboration}, {Vallenari}, {Brown}, {Prusti}, {de Bruijne}, {Arenou}, {Babusiaux}, {Biermann}, {Creevey}, {Ducourant}, {Evans}, {Eyer}, {Guerra}, {Hutton}, {Jordi}, {Klioner}, {Lammers}, {Lindegren}, {Luri}, {Mignard}, {Panem}, {Pourbaix}, {Randich}, {Sartoretti}, {Soubiran}, {Tanga}, {Walton}, {Bailer-Jones}, {Bastian}, {Drimmel}, {Jansen}, {Katz}, {Lattanzi}, {van Leeuwen}, {Bakker}, {Cacciari}, {Casta{\~n}eda}, {De Angeli}, {Fabricius}, {Fouesneau}, {Fr{\'e}mat}, {Galluccio}, {Guerrier}, {Heiter}, {Masana}, {Messineo}, {Mowlavi}, {Nicolas}, {Nienartowicz}, {Pailler}, {Panuzzo}, {Riclet}, {Roux}, {Seabroke}, {Sordo}, {Th{\'e}venin}, {Gracia-Abril}, {Portell}, {Teyssier}, {Altmann}, {Andrae}, {Audard}, {Bellas-Velidis}, {Benson}, {Berthier}, {Blomme}, {Burgess}, {Busonero}, {Busso}, {C{\'a}novas}, {Carry}, {Cellino}, {Cheek}, {Clementini}, {Damerdji}, {Davidson}, {de Teodoro}, {Nu{\~n}ez Campos}, {Delchambre}, {Dell'Oro}, {Esquej},
  {Fern{\'a}ndez-Hern{\'a}ndez}, {Fraile}, {Garabato}, {Garc{\'\i}a-Lario}, {Gosset}, {Haigron}, {Halbwachs}, {Hambly}, {Harrison}, {Hern{\'a}ndez}, {Hestroffer}, {Hodgkin}, {Holl}, {Jan{\ss}en}, {Jevardat de Fombelle}, {Jordan}, {Krone-Martins}, {Lanzafame}, {L{\"o}ffler}, {Marchal}, {Marrese}, {Moitinho}, {Muinonen}, {Osborne}, {Pancino}, {Pauwels}, {Recio-Blanco}, {Reyl{\'e}}, {Riello}, {Rimoldini}, {Roegiers}, {Rybizki}, {Sarro}, {Siopis}, {Smith}, {Sozzetti}, {Utrilla}, {van Leeuwen}, {Abbas}, {{\'A}brah{\'a}m}, {Abreu Aramburu}, {Aerts}, {Aguado}, {Ajaj}, {Aldea-Montero}, {Altavilla}, {{\'A}lvarez}, {Alves}, {Anders}, {Anderson}, {Anglada Varela}, {Antoja}, {Baines}, {Baker}, {Balaguer-N{\'u}{\~n}ez}, {Balbinot}, {Balog}, {Barache}, {Barbato}, {Barros}, {Barstow}, {Bartolom{\'e}}, {Bassilana}, {Bauchet}, {Becciani}, {Bellazzini}, {Berihuete}, {Bernet}, {Bertone}, {Bianchi}, {Binnenfeld}, {Blanco-Cuaresma}, {Blazere}, {Boch}, {Bombrun}, {Bossini}, {Bouquillon}, {Bragaglia}, {Bramante}, {Breedt},
  {Bressan}, {Brouillet}, {Brugaletta}, {Bucciarelli}, {Burlacu}, {Butkevich}, {Buzzi}, {Caffau}, {Cancelliere}, {Cantat-Gaudin}, {Carballo}, {Carlucci}, {Carnerero}, {Carrasco}, {Casamiquela}, {Castellani}, {Castro-Ginard}, {Chaoul}, {Charlot}, {Chemin}, {Chiaramida}, {Chiavassa}, {Chornay}, {Comoretto}, {Contursi}, {Cooper}, {Cornez}, {Cowell}, {Crifo}, {Cropper}, {Crosta}, {Crowley}, {Dafonte}, {Dapergolas}, {David}, {David}, {de Laverny}, {De Luise}, {De March}, {De Ridder}, {de Souza}, {de Torres}, {del Peloso}, {del Pozo}, {Delbo}, {Delgado}, {Delisle}, {Demouchy}, {Dharmawardena}, {Di Matteo}, {Diakite}, {Diener}, {Distefano}, {Dolding}, {Edvardsson}, {Enke}, {Fabre}, {Fabrizio}, {Faigler}, {Fedorets}, {Fernique}, {Fienga}, {Figueras}, {Fournier}, {Fouron}, {Fragkoudi}, {Gai}, {Garcia-Gutierrez}, {Garcia-Reinaldos}, {Garc{\'\i}a-Torres}, {Garofalo}, {Gavel}, {Gavras}, {Gerlach}, {Geyer}, {Giacobbe}, {Gilmore}, {Girona}, {Giuffrida}, {Gomel}, {Gomez}, {Gonz{\'a}lez-N{\'u}{\~n}ez},
  {Gonz{\'a}lez-Santamar{\'\i}a}, {Gonz{\'a}lez-Vidal}, {Granvik}, {Guillout}, {Guiraud}, {Guti{\'e}rrez-S{\'a}nchez}, {Guy}, {Hatzidimitriou}, {Hauser}, {Haywood}, {Helmer}, {Helmi}, {Sarmiento}, {Hidalgo}, {Hilger}, {H{\l}adczuk}, {Hobbs}, {Holland}, {Huckle}, {Jardine}, {Jasniewicz}, {Jean-Antoine Piccolo}, {Jim{\'e}nez-Arranz}, {Jorissen}, {Juaristi Campillo}, {Julbe}, {Karbevska}, {Kervella}, {Khanna}, {Kontizas}, {Kordopatis}, {Korn}, {K{\'o}sp{\'a}l}, {Kostrzewa-Rutkowska}, {Kruszy{\'n}ska}, {Kun}, {Laizeau}, {Lambert}, {Lanza}, {Lasne}, {Le Campion}, {Lebreton}, {Lebzelter}, {Leccia}, {Leclerc}, {Lecoeur-Taibi}, {Liao}, {Licata}, {Lindstr{\o}m}, {Lister}, {Livanou}, {Lobel}, {Lorca}, {Loup}, {Madrero Pardo}, {Magdaleno Romeo}, {Managau}, {Mann}, {Manteiga}, {Marchant}, {Marconi}, {Marcos}, {Marcos Santos}, {Mar{\'\i}n Pina}, {Marinoni}, {Marocco}, {Marshall}, {Martin Polo}, {Mart{\'\i}n-Fleitas}, {Marton}, {Mary}, {Masip}, {Massari}, {Mastrobuono-Battisti}, {Mazeh}, {McMillan}, {Messina}, {Michalik},
  {Millar}, {Mints}, {Molina}, {Molinaro}, {Moln{\'a}r}, {Monari}, {Mongui{\'o}}, {Montegriffo}, {Montero}, {Mor}, {Mora}, {Morbidelli}, {Morel}, {Morris}, {Muraveva}, {Murphy}, {Musella}, {Nagy}, {Noval}, {Oca{\~n}a}, {Ogden}, {Ordenovic}, {Osinde}, {Pagani}, {Pagano}, {Palaversa}, {Palicio}, {Pallas-Quintela}, {Panahi}, {Payne-Wardenaar}, {Pe{\~n}alosa Esteller}, {Penttil{\"a}}, {Pichon}, {Piersimoni}, {Pineau}, {Plachy}, {Plum}, {Poggio}, {Pr{\v{s}}a}, {Pulone}, {Racero}, {Ragaini}, {Rainer}, {Raiteri}, {Rambaux}, {Ramos}, {Ramos-Lerate}, {Re Fiorentin}, {Regibo}, {Richards}, {Rios Diaz}, {Ripepi}, {Riva}, {Rix}, {Rixon}, {Robichon}, {Robin}, {Robin}, {Roelens}, {Rogues}, {Rohrbasser}, {Romero-G{\'o}mez}, {Rowell}, {Royer}, {Ruz Mieres}, {Rybicki}, {Sadowski}, {S{\'a}ez N{\'u}{\~n}ez}, {Sagrist{\`a} Sell{\'e}s}, {Sahlmann}, {Salguero}, {Samaras}, {Sanchez Gimenez}, {Sanna}, {Santove{\~n}a}, {Sarasso}, {Schultheis}, {Sciacca}, {Segol}, {Segovia}, {S{\'e}gransan}, {Semeux}, {Shahaf}, {Siddiqui}, {Siebert},
  {Siltala}, {Silvelo}, {Slezak}, {Slezak}, {Smart}, {Snaith}, {Solano}, {Solitro}, {Souami}, {Souchay}, {Spagna}, {Spina}, {Spoto}, {Steele}, {Steidelm{\"u}ller}, {Stephenson}, {S{\"u}veges}, {Surdej}, {Szabados}, {Szegedi-Elek}, {Taris}, {Taylor}, {Teixeira}, {Tolomei}, {Tonello}, {Torra}, {Torra}, {Torralba Elipe}, {Trabucchi}, {Tsounis}, {Turon}, {Ulla}, {Unger}, {Vaillant}, {van Dillen}, {van Reeven}, {Vanel}, {Vecchiato}, {Viala}, {Vicente}, {Voutsinas}, {Weiler}, {Wevers}, {Wyrzykowski}, {Yoldas}, {Yvard}, {Zhao}, {Zorec}, {Zucker}, \& {Zwitter}}]{gaiddr3}
{Gaia Collaboration}, {Vallenari}, A., {Brown}, A.~G.~A., {et~al.} 2023, \aap, 674, A1, \dodoi{10.1051/0004-6361/202243940}

\bibitem[{{Gaudin} {et~al.}(2024){Gaudin}, {Kennea}, {Coe}, {Monageng}, {Udalski}, {Townsend}, {Buckley}, \& {Evans}}]{Gaudin2024Bexray}
{Gaudin}, T.~M., {Kennea}, J.~A., {Coe}, M.~J., {et~al.} 2024, \apjl, 965, L10, \dodoi{10.3847/2041-8213/ad354a}

\bibitem[{{Georgy} {et~al.}(2013){Georgy}, {Ekstr{\"o}m}, {Eggenberger}, {Meynet}, {Haemmerl{\'e}}, {Maeder}, {Granada}, {Groh}, {Hirschi}, {Mowlavi}, {Yusof}, {Charbonnel}, {Decressin}, \& {Barblan}}]{georgy2013grids}
{Georgy}, C., {Ekstr{\"o}m}, S., {Eggenberger}, P., {et~al.} 2013, \aap, 558, A103, \dodoi{10.1051/0004-6361/201322178}

\bibitem[{{Gies} {et~al.}(1998){Gies}, {Bagnuolo}, {Ferrara}, {Kaye}, {Thaller}, {Penny}, \& {Peters}}]{Gies1998phiper}
{Gies}, D.~R., {Bagnuolo}, Jr., W.~G., {Ferrara}, E.~C., {et~al.} 1998, \apj, 493, 440, \dodoi{10.1086/305113}

\bibitem[{{Gies} {et~al.}(2020){Gies}, {Lester}, {Wang}, {Couperus}, {Shepard}, {Neiner}, {Wade}, {Dunham}, \& {Dunham}}]{gies2020spectroscopic}
{Gies}, D.~R., {Lester}, K.~V., {Wang}, L., {et~al.} 2020, \apj, 902, 25, \dodoi{10.3847/1538-4357/abb372}

\bibitem[{{Gieseking}(1981)}]{Gieseking1981v658Car}
{Gieseking}, F. 1981, \aaps, 43, 33

\bibitem[{{Goodricke}(1783)}]{Goodricke1783BetPer}
{Goodricke}, J. 1783, Philosophical Transactions of the Royal Society of London Series I, 73, 474

\bibitem[{{Halm}(1911)}]{Halm1911binaries}
{Halm}, J. 1911, \mnras, 71, 610, \dodoi{10.1093/mnras/71.8.610}

\bibitem[{{Hanuschik} {et~al.}(1996){Hanuschik}, {Hummel}, {Sutorius}, {Dietle}, \& {Thimm}}]{hanuschik1996}
{Hanuschik}, R.~W., {Hummel}, W., {Sutorius}, E., {Dietle}, O., \& {Thimm}, G. 1996, \aaps, 116, 309

\bibitem[{{Haubois} {et~al.}(2012){Haubois}, {Carciofi}, {Rivinius}, {Okazaki}, \& {Bjorkman}}]{haubois2012buildup}
{Haubois}, X., {Carciofi}, A.~C., {Rivinius}, T., {Okazaki}, A.~T., \& {Bjorkman}, J.~E. 2012, \apj, 756, 156, \dodoi{10.1088/0004-637X/756/2/156}

\bibitem[{{Hauck}(2018)}]{hauck2018eclipsing}
{Hauck}, N. 2018, BAV Rundbrief - Mitteilungsblatt der Berliner Arbeits-gemeinschaft fuer Veraenderliche Sterne, 67, 41, \dodoi{10.48550/arXiv.1805.04318}

\bibitem[{{Hirata}(2007)}]{hirata2007}
{Hirata}, R. 2007, in Astronomical Society of the Pacific Conference Series, Vol. 361, Active OB-Stars: Laboratories for Stellare and Circumstellar Physics, ed. A.~T. {Okazaki}, S.~P. {Owocki}, \& S.~{Stefl}, 267

\bibitem[{{Huang} {et~al.}(2010){Huang}, {Gies}, \& {McSwain}}]{Huang2010wparameter}
{Huang}, W., {Gies}, D.~R., \& {McSwain}, M.~V. 2010, \apj, 722, 605, \dodoi{10.1088/0004-637X/722/1/605}

\bibitem[{{Kervella} {et~al.}(2022){Kervella}, {Arenou}, \& {Th{\'e}venin}}]{kervella2022gaia}
{Kervella}, P., {Arenou}, F., \& {Th{\'e}venin}, F. 2022, \aap, 657, A7, \dodoi{10.1051/0004-6361/202142146}

\bibitem[{{Klement} {et~al.}(2015){Klement}, {Carciofi}, {Rivinius}, {Panoglou}, {Vieira}, {Bjorkman}, {{\v{S}}tefl}, {Tycner}, {Faes}, {Kor{\v{c}}{\'a}kov{\'a}}, {M{\"u}ller}, {Zavala}, \& {Cur{\'e}}}]{klement2015betacmi}
{Klement}, R., {Carciofi}, A.~C., {Rivinius}, T., {et~al.} 2015, \aap, 584, A85, \dodoi{10.1051/0004-6361/201526535}

\bibitem[{{Klement} {et~al.}(2019){Klement}, {Carciofi}, {Rivinius}, {Ignace}, {Matthews}, {Torstensson}, {Gies}, {Vieira}, {Richardson}, {Domiciano de Souza}, {Bjorkman}, {Hallinan}, {Faes}, {Mota}, {Gullingsrud}, {de Breuck}, {Kervella}, {Cur{\'e}}, \& {Gunawan}}]{Klement2019sedturndown}
---. 2019, \apj, 885, 147, \dodoi{10.3847/1538-4357/ab48e7}

\bibitem[{{Klement} {et~al.}(2022){Klement}, {Baade}, {Rivinius}, {Gies}, {Wang}, {Labadie-Bartz}, {Ticiani dos Santos}, {Monnier}, {Carciofi}, {M{\'e}rand}, {Anugu}, {Schaefer}, {Le Bouquin}, {Davies}, {Ennis}, {Gardner}, {Kraus}, {Setterholm}, \& {Labdon}}]{Klement2022dynamical}
{Klement}, R., {Baade}, D., {Rivinius}, T., {et~al.} 2022, \apj, 940, 86, \dodoi{10.3847/1538-4357/ac98b8}

\bibitem[{{Klement} {et~al.}(2024){Klement}, {Rivinius}, {Gies}, {Baade}, {M{\'e}rand}, {Monnier}, {Schaefer}, {Lanthermann}, {Anugu}, {Kraus}, \& {Gardner}}]{klement2024interfsdo}
{Klement}, R., {Rivinius}, T., {Gies}, D.~R., {et~al.} 2024, \apj, 962, 70, \dodoi{10.3847/1538-4357/ad13ec}

\bibitem[{{Kriz} \& {Harmanec}(1975)}]{kriz1975hypothesis}
{Kriz}, S., \& {Harmanec}, P. 1975, Bulletin of the Astronomical Institutes of Czechoslovakia, 26, 65

\bibitem[{{Labadie-Bartz} {et~al.}(2022){Labadie-Bartz}, {Carciofi}, {Henrique de Amorim}, {Rubio}, {Luiz Figueiredo}, {Ticiani dos Santos}, \& {Thomson-Paressant}}]{labadie-bartz2022classifying}
{Labadie-Bartz}, J., {Carciofi}, A.~C., {Henrique de Amorim}, T., {et~al.} 2022, \aj, 163, 226, \dodoi{10.3847/1538-3881/ac5abd}

\bibitem[{{Lanz} \& {Hubeny}(2007)}]{Lanz2007Tlusty}
{Lanz}, T., \& {Hubeny}, I. 2007, \apjs, 169, 83, \dodoi{10.1086/511270}

\bibitem[{{Lee} {et~al.}(1991){Lee}, {Osaki}, \& {Saio}}]{lee1991VDD}
{Lee}, U., {Osaki}, Y., \& {Saio}, H. 1991, \mnras, 250, 432, \dodoi{10.1093/mnras/250.2.432}

\bibitem[{{Lindegren} {et~al.}(2021){Lindegren}, {Klioner}, {Hern{\'a}ndez}, {Bombrun}, {Ramos-Lerate}, {Steidelm{\"u}ller}, {Bastian}, {Biermann}, {de Torres}, {Gerlach}, {Geyer}, {Hilger}, {Hobbs}, {Lammers}, {McMillan}, {Stephenson}, {Casta{\~n}eda}, {Davidson}, {Fabricius}, {Gracia-Abril}, {Portell}, {Rowell}, {Teyssier}, {Torra}, {Bartolom{\'e}}, {Clotet}, {Garralda}, {Gonz{\'a}lez-Vidal}, {Torra}, {Abbas}, {Altmann}, {Anglada Varela}, {Balaguer-N{\'u}{\~n}ez}, {Balog}, {Barache}, {Becciani}, {Bernet}, {Bertone}, {Bianchi}, {Bouquillon}, {Brown}, {Bucciarelli}, {Busonero}, {Butkevich}, {Buzzi}, {Cancelliere}, {Carlucci}, {Charlot}, {Cioni}, {Crosta}, {Crowley}, {del Peloso}, {del Pozo}, {Drimmel}, {Esquej}, {Fienga}, {Fraile}, {Gai}, {Garcia-Reinaldos}, {Guerra}, {Hambly}, {Hauser}, {Jan{\ss}en}, {Jordan}, {Kostrzewa-Rutkowska}, {Lattanzi}, {Liao}, {Licata}, {Lister}, {L{\"o}ffler}, {Marchant}, {Masip}, {Mignard}, {Mints}, {Molina}, {Mora}, {Morbidelli}, {Murphy}, {Pagani}, {Panuzzo}, {Pe{\~n}alosa
  Esteller}, {Poggio}, {Re Fiorentin}, {Riva}, {Sagrist{\`a} Sell{\'e}s}, {Sanchez Gimenez}, {Sarasso}, {Sciacca}, {Siddiqui}, {Smart}, {Souami}, {Spagna}, {Steele}, {Taris}, {Utrilla}, {van Reeven}, \& {Vecchiato}}]{Lindegren2021gaia}
{Lindegren}, L., {Klioner}, S.~A., {Hern{\'a}ndez}, J., {et~al.} 2021, \aap, 649, A2, \dodoi{10.1051/0004-6361/202039709}

\bibitem[{{Marr} {et~al.}(2022){Marr}, {Jones}, {Tycner}, {Carciofi}, \& {Silva}}]{marr2022}
{Marr}, K.~C., {Jones}, C.~E., {Tycner}, C., {Carciofi}, A.~C., \& {Silva}, A.~C.~F. 2022, \apj, 928, 145, \dodoi{10.3847/1538-4357/ac551b}

\bibitem[{{Martayan} {et~al.}(2006){Martayan}, {Fr{\'e}mat}, {Hubert}, {Floquet}, {Zorec}, \& {Neiner}}]{martayan2006effects}
{Martayan}, C., {Fr{\'e}mat}, Y., {Hubert}, A.~M., {et~al.} 2006, \aap, 452, 273, \dodoi{10.1051/0004-6361:20053859}

\bibitem[{{Martin} {et~al.}(1992){Martin}, {Adamson}, {Whittet}, {Hough}, {Bailey}, {Kim}, {Sato}, {Tamura}, \& {Yamashita}}]{martin}
{Martin}, P.~G., {Adamson}, A.~J., {Whittet}, D. C.~B., {et~al.} 1992, \apj, 392, 691

\bibitem[{{Martin} \& {Lepp}(2022)}]{martin2022}
{Martin}, R.~G., \& {Lepp}, S. 2022, \mnras, 516, L86, \dodoi{10.1093/mnrasl/slac090}

\bibitem[{{McSwain} \& {Gies}(2005)}]{mcswain2005}
{McSwain}, M.~V., \& {Gies}, D.~R. 2005, \apjs, 161, 118, \dodoi{10.1086/432757}

\bibitem[{{Milson} {et~al.}(2020){Milson}, {Barton}, \& {Bennett}}]{Milson2020pythonbinaries}
{Milson}, N., {Barton}, C., \& {Bennett}, P.~D. 2020, arXiv e-prints, arXiv:2011.13914, \dodoi{10.48550/arXiv.2011.13914}

\bibitem[{{Miroshnichenko} {et~al.}(2023){Miroshnichenko}, {Chari}, {Danford}, {Prendergast}, {Aarnio}, {Andronov}, {Chinarova}, {Lytle}, {Amantayeva}, {Gabitova}, {Vaidman}, {Baktybayev}, \& {Khokhlov}}]{Miroshnichenko2023SB1}
{Miroshnichenko}, A.~S., {Chari}, R., {Danford}, S., {et~al.} 2023, Galaxies, 11, 83, \dodoi{10.3390/galaxies11040083}

\bibitem[{{Mourard} {et~al.}(2015){Mourard}, {Monnier}, {Meilland}, {Gies}, {Millour}, {Benisty}, {Che}, {Grundstrom}, {Ligi}, {Schaefer}, {Baron}, {Kraus}, {Zhao}, {Pedretti}, {Berio}, {Clausse}, {Nardetto}, {Perraut}, {Spang}, {Stee}, {Tallon-Bosc}, {McAlister}, {ten Brummelaar}, {Ridgway}, {Sturmann}, {Sturmann}, {Turner}, \& {Farrington}}]{Mourard2015phiper}
{Mourard}, D., {Monnier}, J.~D., {Meilland}, A., {et~al.} 2015, \aap, 577, A51, \dodoi{10.1051/0004-6361/201425141}

\bibitem[{{Neiner} {et~al.}(2011){Neiner}, {de Batz}, {Cochard}, {Floquet}, {Mekkas}, \& {Desnoux}}]{neiner2011bess}
{Neiner}, C., {de Batz}, B., {Cochard}, F., {et~al.} 2011, \aj, 142, 149, \dodoi{10.1088/0004-6256/142/5/149}

\bibitem[{{Offner} {et~al.}(2023){Offner}, {Moe}, {Kratter}, {Sadavoy}, {Jensen}, \& {Tobin}}]{offner2023binaries}
{Offner}, S.~S.~R., {Moe}, M., {Kratter}, K.~M., {et~al.} 2023, in Astronomical Society of the Pacific Conference Series, Vol. 534, Protostars and Planets VII, ed. S.~{Inutsuka}, Y.~{Aikawa}, T.~{Muto}, K.~{Tomida}, \& M.~{Tamura}, 275, \dodoi{10.48550/arXiv.2203.10066}

\bibitem[{{Okazaki}(1996)}]{Okazaki1996m1waves}
{Okazaki}, A.~T. 1996, \pasj, 48, 305, \dodoi{10.1093/pasj/48.2.305}

\bibitem[{{Okazaki} {et~al.}(2002){Okazaki}, {Bate}, {Ogilvie}, \& {Pringle}}]{okazaki2002}
{Okazaki}, A.~T., {Bate}, M.~R., {Ogilvie}, G.~I., \& {Pringle}, J.~E. 2002, \mnras, 337, 967, \dodoi{10.1046/j.1365-8711.2002.05960.x}

\bibitem[{{Panoglou} {et~al.}(2016){Panoglou}, {Carciofi}, {Vieira}, {Cyr}, {Jones}, {Okazaki}, \& {Rivinius}}]{panoglou2016SPH}
{Panoglou}, D., {Carciofi}, A.~C., {Vieira}, R.~G., {et~al.} 2016, \mnras, 461, 2616, \dodoi{10.1093/mnras/stw1508}

\bibitem[{{Panoglou} {et~al.}(2018){Panoglou}, {Faes}, {Carciofi}, {Okazaki}, {Baade}, {Rivinius}, \& {Borges Fernandes}}]{panoglou2018discs}
{Panoglou}, D., {Faes}, D.~M., {Carciofi}, A.~C., {et~al.} 2018, \mnras, 473, 3039, \dodoi{10.1093/mnras/stx2497}

\bibitem[{{Pereyra} {et~al.}(2009){Pereyra}, {de Ara{\'u}jo}, {Magalh{\~a}es}, {Borges Fernandes}, \& {Domiciano de Souza}}]{Pereyra2009GGCarinae}
{Pereyra}, A., {de Ara{\'u}jo}, F.~X., {Magalh{\~a}es}, A.~M., {Borges Fernandes}, M., \& {Domiciano de Souza}, A. 2009, \aap, 508, 1337, \dodoi{10.1051/0004-6361/200913250}

\bibitem[{{Peters} {et~al.}(2008){Peters}, {Gies}, {Grundstrom}, \& {McSwain}}]{Peters2008fycma}
{Peters}, G.~J., {Gies}, D.~R., {Grundstrom}, E.~D., \& {McSwain}, M.~V. 2008, \apj, 686, 1280, \dodoi{10.1086/591145}

\bibitem[{{Peters} {et~al.}(2013){Peters}, {Pewett}, {Gies}, {Touhami}, \& {Grundstrom}}]{Peters201359cyg}
{Peters}, G.~J., {Pewett}, T.~D., {Gies}, D.~R., {Touhami}, Y.~N., \& {Grundstrom}, E.~D. 2013, \apj, 765, 2, \dodoi{10.1088/0004-637X/765/1/2}

\bibitem[{{Peters} {et~al.}(2016){Peters}, {Wang}, {Gies}, \& {Grundstrom}}]{Peters2016hr2142}
{Peters}, G.~J., {Wang}, L., {Gies}, D.~R., \& {Grundstrom}, E.~D. 2016, \apj, 828, 47, \dodoi{10.3847/0004-637X/828/1/47}

\bibitem[{{Poeckert}(1981)}]{phiper1981}
{Poeckert}, R. 1981, \pasp, 93, 297, \dodoi{10.1086/130828}

\bibitem[{{Pollmann} {et~al.}(2018){Pollmann}, {Bennett}, {Vollmann}, \& {Somogyi}}]{Pollmann2018VVcep}
{Pollmann}, E., {Bennett}, P.~D., {Vollmann}, W., \& {Somogyi}, P. 2018, Information Bulletin on Variable Stars, 6249, 1, \dodoi{10.22444/IBVS.6249}

\bibitem[{{Pols} {et~al.}(1991){Pols}, {Cote}, {Waters}, \& {Heise}}]{pols1991formation}
{Pols}, O.~R., {Cote}, J., {Waters}, L.~B.~F.~M., \& {Heise}, J. 1991, \aap, 241, 419

\bibitem[{{Rappaport} {et~al.}(2009){Rappaport}, {Podsiadlowski}, \& {Horev}}]{rappaport2009past}
{Rappaport}, S., {Podsiadlowski}, P., \& {Horev}, I. 2009, \apj, 698, 666, \dodoi{10.1088/0004-637X/698/1/666}

\bibitem[{{Rappaport} \& {van den Heuvel}(1982)}]{Rappaport1982xray}
{Rappaport}, S., \& {van den Heuvel}, E.~P.~J. 1982, in Be Stars, ed. M.~{Jaschek} \& H.~G. {Groth}, Vol.~98, 327--344

\bibitem[{{Ricker} {et~al.}(2015){Ricker}, {Winn}, {Vanderspek}, {Latham}, {Bakos}, {Bean}, {Berta-Thompson}, {Brown}, {Buchhave}, {Butler}, {Butler}, {Chaplin}, {Charbonneau}, {Christensen-Dalsgaard}, {Clampin}, {Deming}, {Doty}, {De Lee}, {Dressing}, {Dunham}, {Endl}, {Fressin}, {Ge}, {Henning}, {Holman}, {Howard}, {Ida}, {Jenkins}, {Jernigan}, {Johnson}, {Kaltenegger}, {Kawai}, {Kjeldsen}, {Laughlin}, {Levine}, {Lin}, {Lissauer}, {MacQueen}, {Marcy}, {McCullough}, {Morton}, {Narita}, {Paegert}, {Palle}, {Pepe}, {Pepper}, {Quirrenbach}, {Rinehart}, {Sasselov}, {Sato}, {Seager}, {Sozzetti}, {Stassun}, {Sullivan}, {Szentgyorgyi}, {Torres}, {Udry}, \& {Villasenor}}]{ricker2015TESS}
{Ricker}, G.~R., {Winn}, J.~N., {Vanderspek}, R., {et~al.} 2015, Journal of Astronomical Telescopes, Instruments, and Systems, 1, 014003, \dodoi{10.1117/1.JATIS.1.1.014003}

\bibitem[{{Riello} {et~al.}(2021){Riello}, {De Angeli}, {Evans}, {Montegriffo}, {Carrasco}, {Busso}, {Palaversa}, {Burgess}, {Diener}, {Davidson}, {Rowell}, {Fabricius}, {Jordi}, {Bellazzini}, {Pancino}, {Harrison}, {Cacciari}, {van Leeuwen}, {Hambly}, {Hodgkin}, {Osborne}, {Altavilla}, {Barstow}, {Brown}, {Castellani}, {Cowell}, {De Luise}, {Gilmore}, {Giuffrida}, {Hidalgo}, {Holland}, {Marinoni}, {Pagani}, {Piersimoni}, {Pulone}, {Ragaini}, {Rainer}, {Richards}, {Sanna}, {Walton}, {Weiler}, \& {Yoldas}}]{gaia2021rpbp}
{Riello}, M., {De Angeli}, F., {Evans}, D.~W., {et~al.} 2021, \aap, 649, A3, \dodoi{10.1051/0004-6361/202039587}

\bibitem[{{Rivinius} {et~al.}(2003){Rivinius}, {Baade}, \& {{\v{S}}tefl}}]{rivi2003NRP}
{Rivinius}, T., {Baade}, D., \& {{\v{S}}tefl}, S. 2003, \aap, 411, 229, \dodoi{10.1051/0004-6361:20031285}

\bibitem[{{Rivinius} {et~al.}(2013){Rivinius}, {Carciofi}, \& {Martayan}}]{rivinius2013classical}
{Rivinius}, T., {Carciofi}, A.~C., \& {Martayan}, C. 2013, \aapr, 21, 69, \dodoi{10.1007/s00159-013-0069-0}

\bibitem[{{Rivinius} {et~al.}(2025){Rivinius}, {Klement}, {Chojnowski}, {Baade}, {Abdul-Masih}, {Przybilla}, {Guarro Fl{\'o}}, {Heathcote}, {Hadrava}, {Gies}, {Shepard}, {Buil}, {Garde}, {Thizy}, {Monnier}, {Anugu}, {Lanthermann}, {Schaefer}, {Davies}, {Kraus}, {Ennis}, {Setterholm}, {Gardner}, {Ibrahim}, {Chhabra}, {Gutierrez}, \& {Codron}}]{rivi2025bloated}
{Rivinius}, T., {Klement}, R., {Chojnowski}, S.~D., {et~al.} 2025, \aap, 694, A172, \dodoi{10.1051/0004-6361/202347275}

\bibitem[{{Rosales} {et~al.}(2023){Rosales}, {Mennickent}, {Djura{\v{s}}evi{\'c}}, {Araya}, {Cur{\'e}}, {Schleicher}, \& {Petrovi{\'c}}}]{rosales2023V4142Sgr}
{Rosales}, J.~A., {Mennickent}, R.~E., {Djura{\v{s}}evi{\'c}}, G., {et~al.} 2023, \aap, 670, A94, \dodoi{10.1051/0004-6361/202244046}

\bibitem[{{Rubio} {et~al.}(2025){Rubio}, {Carciofi}, {Bjorkman}, {de Amorim}, {Okazaki}, {Suffak}, {Jones}, \& {Candido}}]{rubio2025}
{Rubio}, A.~C., {Carciofi}, A.~C., {Bjorkman}, J.~E., {et~al.} 2025, \aap, 698, A309, \dodoi{10.1051/0004-6361/202452724}

\bibitem[{{Sana} {et~al.}(2012){Sana}, {de Mink}, {de Koter}, {Langer}, {Evans}, {Gieles}, {Gosset}, {Izzard}, {Le Bouquin}, \& {Schneider}}]{sana2012binaries}
{Sana}, H., {de Mink}, S.~E., {de Koter}, A., {et~al.} 2012, Science, 337, 444, \dodoi{10.1126/science.1223344}

\bibitem[{{Sana} {et~al.}(2013){Sana}, {de Koter}, {de Mink}, {Dunstall}, {Evans}, {H{\'e}nault-Brunet}, {Ma{\'\i}z Apell{\'a}niz}, {Ram{\'\i}rez-Agudelo}, {Taylor}, {Walborn}, {Clark}, {Crowther}, {Herrero}, {Gieles}, {Langer}, {Lennon}, \& {Vink}}]{Sana2013}
{Sana}, H., {de Koter}, A., {de Mink}, S.~E., {et~al.} 2013, \aap, 550, A107, \dodoi{10.1051/0004-6361/201219621}

\bibitem[{{Schlawin} {et~al.}(2014){Schlawin}, {Herter}, {Henderson}, {Wilson}, {Probst}, {Sprayberry}, {Bonati}, {Schurter}, {James}, {Warner}, {Tighe}, {Adams}, \& {Mart{\'\i}nez}}]{Schlawin2014triplespec}
{Schlawin}, E., {Herter}, T.~L., {Henderson}, C., {et~al.} 2014, in Society of Photo-Optical Instrumentation Engineers (SPIE) Conference Series, Vol. 9147, Ground-based and Airborne Instrumentation for Astronomy V, ed. S.~K. {Ramsay}, I.~S. {McLean}, \& H.~{Takami}, 91472H, \dodoi{10.1117/12.2055233}

\bibitem[{{Serkowski} {et~al.}(1975){Serkowski}, {Mathewson}, \& {Ford}}]{Serkowski}
{Serkowski}, K., {Mathewson}, D.~S., \& {Ford}, V.~L. 1975, \apj, 196, 261, \dodoi{10.1086/153410}

\bibitem[{{Shafter} \& {Harkness}(1986)}]{Shafter1986bisector}
{Shafter}, A.~W., \& {Harkness}, R.~P. 1986, \aj, 92, 658, \dodoi{10.1086/114198}

\bibitem[{{Shakura} \& {Sunyaev}(1973)}]{Shakura1973alpha}
{Shakura}, N.~I., \& {Sunyaev}, R.~A. 1973, \aap, 24, 337

\bibitem[{{Shao} \& {Li}(2014)}]{shao2014formation}
{Shao}, Y., \& {Li}, X.-D. 2014, \apj, 796, 37, \dodoi{10.1088/0004-637X/796/1/37}

\bibitem[{{Siverd} {et~al.}(2018){Siverd}, {Brown}, {Barnes}, {Bowman}, {De Vera}, {Foale}, {Harbeck}, {Henderson}, {Hygelund}, {Kirby}, {McCully}, {Nation}, {Smith}, {Taylor}, \& {Tufts}}]{siverd2018nres}
{Siverd}, R.~J., {Brown}, T.~M., {Barnes}, S., {et~al.} 2018, in Society of Photo-Optical Instrumentation Engineers (SPIE) Conference Series, Vol. 10702, Ground-based and Airborne Instrumentation for Astronomy VII, ed. C.~J. {Evans}, L.~{Simard}, \& H.~{Takami}, 107026C, \dodoi{10.1117/12.2312800}

\bibitem[{{Staritsin}(2024)}]{Staritsin2024binaries}
{Staritsin}, E. 2024, Research in Astronomy and Astrophysics, 24, 015001, \dodoi{10.1088/1674-4527/ad089e}

\bibitem[{{Suffak} {et~al.}(2022){Suffak}, {Jones}, \& {Carciofi}}]{suffak2002}
{Suffak}, M., {Jones}, C.~E., \& {Carciofi}, A.~C. 2022, \mnras, 509, 931, \dodoi{10.1093/mnras/stab3024}

\bibitem[{{Suffak} {et~al.}(2024){Suffak}, {Jones}, \& {Carciofi}}]{suffak2024}
{Suffak}, M.~W., {Jones}, C.~E., \& {Carciofi}, A.~C. 2024, \mnras, 527, 7515, \dodoi{10.1093/mnras/stad3659}

\bibitem[{{van Bever} \& {Vanbeveren}(1997)}]{vanbever1997number}
{van Bever}, J., \& {Vanbeveren}, D. 1997, \aap, 322, 116

\bibitem[{{van Dam} {et~al.}(2020){van Dam}, {Kenworthy}, {David}, {Mamajek}, {Hillenbrand}, {Cody}, {Howard}, {Isaacson}, {Ciardi}, {Rebull}, {Stauffer}, {Patel}, {Cameron + WASP Collaborators}, {Rodriguez}, {Pojma{\'n}ski}, {Gonzales}, {Schlieder}, {Hambsch}, {Dufoer}, {Vanmunster}, {Dubois}, {Vanaverbeke}, {Logie}, \& {Rau}}]{vanDam2020Ttauri}
{van Dam}, D.~M., {Kenworthy}, M.~A., {David}, T.~J., {et~al.} 2020, \aj, 160, 285, \dodoi{10.3847/1538-3881/abc259}

\bibitem[{{Vieira} {et~al.}(2017){Vieira}, {Carciofi}, {Bjorkman}, {Rivinius}, {Baade}, \& {R{\'\i}mulo}}]{Vieira2017VDD}
{Vieira}, R.~G., {Carciofi}, A.~C., {Bjorkman}, J.~E., {et~al.} 2017, \mnras, 464, 3071, \dodoi{10.1093/mnras/stw2542}

\bibitem[{{von Zeipel}(1924)}]{vonZeipel1924}
{von Zeipel}, H. 1924, \mnras, 84, 684, \dodoi{10.1093/mnras/84.9.684}

\bibitem[{{{\v{S}}tefl} {et~al.}(2009){{\v{S}}tefl}, {Rivinius}, {Carciofi}, {Le Bouquin}, {Baade}, {Bjorkman}, {Hesselbach}, {Hummel}, {Okazaki}, {Pollmann}, {Rantakyr{\"o}}, \& {Wisniewski}}]{stefl2009m1wave}
{{\v{S}}tefl}, S., {Rivinius}, T., {Carciofi}, A.~C., {et~al.} 2009, \aap, 504, 929, \dodoi{10.1051/0004-6361/200811573}

\bibitem[{{Wang} {et~al.}(2017){Wang}, {Gies}, \& {Peters}}]{Wang201760cyg}
{Wang}, L., {Gies}, D.~R., \& {Peters}, G.~J. 2017, \apj, 843, 60, \dodoi{10.3847/1538-4357/aa740a}

\bibitem[{{Wang} {et~al.}(2021){Wang}, {Gies}, {Peters}, {G{\"o}tberg}, {Chojnowski}, {Lester}, \& {Howell}}]{wang2021detection}
{Wang}, L., {Gies}, D.~R., {Peters}, G.~J., {et~al.} 2021, \aj, 161, 248, \dodoi{10.3847/1538-3881/abf144}

\bibitem[{{Wang} \& {Chen}(2019)}]{wang2019gaia}
{Wang}, S., \& {Chen}, X. 2019, \apj, 877, 116, \dodoi{10.3847/1538-4357/ab1c61}

\bibitem[{{Whittet} {et~al.}(1992){Whittet}, {Martin}, {Hough}, {Rouse}, {Bailey}, \& {Axon}}]{whittet}
{Whittet}, D. C.~B., {Martin}, P.~G., {Hough}, J.~H., {et~al.} 1992, \apj, 386, 562

\bibitem[{{Zhang} {et~al.}(2025){Zhang}, {Navarete}, {G{\'a}lvez-Ortiz}, {Jones}, {Burgasser}, {Cruz}, {Marocco}, {Lodieu}, {Shan}, {Gauza}, {Raddi}, {Huang}, {Smart}, {Baig}, {Cheng}, \& {Pinfield}}]{navarete2025quadruple}
{Zhang}, Z.~H., {Navarete}, F., {G{\'a}lvez-Ortiz}, M.~C., {et~al.} 2025, \mnras, \dodoi{10.1093/mnras/staf895}

\bibitem[{{Zharikov} {et~al.}(2013){Zharikov}, {Miroshnichenko}, {Pollmann}, {Danford}, {Bjorkman}, {Morrison}, {Favaro}, {Guarro Fl{\'o}}, {Terry}, {Desnoux}, {Garrel}, {Martineau}, {Buchet}, {Ubaud}, {Mauclaire}, {Kalbermatten}, {Buil}, {Sawicki}, {Blank}, \& {Garde}}]{Zharikov2013}
{Zharikov}, S.~V., {Miroshnichenko}, A.~S., {Pollmann}, E., {et~al.} 2013, \aap, 560, A30, \dodoi{10.1051/0004-6361/201322114}

\bibitem[{{Zhou} {et~al.}(2018){Zhou}, {Rappaport}, {Nelson}, {Huang}, {Senhadji}, {Rodriguez}, {Vanderburg}, {Quinn}, {Johnson}, {Latham}, {Torres}, {Gary}, {Tan}, {Johnson}, {Burt}, {Kristiansen}, {Jacobs}, {LaCourse}, {Schwengeler}, {Terentev}, {Bieryla}, {Esquerdo}, {Berlind}, {Calkins}, {Bento}, {Cochran}, {Karjalainen}, {Hatzes}, {Karjalainen}, {Holden}, \& {Butler}}]{Zhou2018Sauleclipse}
{Zhou}, G., {Rappaport}, S., {Nelson}, L., {et~al.} 2018, \apj, 854, 109, \dodoi{10.3847/1538-4357/aaa9b9}

\end{thebibliography}
\bibliographystyle{aasjournal}

\end{document}